\documentclass[nofootinbib, superscriptaddress, twocolumn]{revtex4-1}

\usepackage{amsmath}
\usepackage{amssymb}
\usepackage{braket}
\usepackage{color}
\usepackage{xcolor}
\usepackage{graphicx}
\usepackage{latexsym}
\usepackage{listings}
\usepackage{mathrsfs}
\usepackage{pdfpages}
\usepackage{verbatim}
\usepackage{tabularx}
\usepackage{ragged2e}
\usepackage{multirow}
\usepackage{amsbsy}
\usepackage{siunitx}
\usepackage{numprint}
\usepackage{makecell}
\usepackage[utf8]{inputenc}

\usepackage[colorlinks,citecolor=blue]{hyperref}

\newcommand{\R}{\mathbb{R}}
\newcommand{\eval}{\biggr\rvert} %evaluated at
\newcommand{\N}{\mathbb{N}} % Integers
\newcommand{\myvec}[1]{\pmb{\mathbf{#1}}}

\newcommand{\dd}[1]{\frac{d}{d #1}}
\newcommand{\ddII }[1]{\frac{\partial^2}{\partial #1^2}}
\newcommand{\pd}[2]{\frac{\partial #1}{\partial #2}} 
\newcommand{\pdd}[1]{\frac{\partial}{\partial #1}} 
\newcommand{\half}{\frac{1}{2}}

\newcommand{\paren}[1]{\left( #1 \right)}
\newcommand{\sqrbrace}[1]{\left[ #1 \right]}

\renewcommand{\braket}[1]{\left\langle #1 \right\rangle}

\newcommand{\V}{\mathcal{V}}

\newcommand{\F}{\mathcal{F}}

\newcommand{\Lie}{\mathcal{L}}

\newcommand{\Ord}[1]{\mathcal{O}\paren{#1}}

\newcommand{\mT}{\mathcal{T}}

\newcommand{\mI}{\mathcal{I}}

\newcommand{\Ma}{\mathcal{M}}
\newcommand{\St}{\mathcal{S}}

\newcommand{\ME}{\mathcal{E}}

\newcommand{\MH}{\mathcal{H}}

\newcommand{\NF}{\text{NF}}

\newcommand{\OH}{\text{OH}}

\newcommand{\MO}{\text{MO}}
\newcommand{\floor}[1]{\left\lfloor #1 \right\rfloor}

\newcommand{\FD}{\text{FD}}
\newcommand{\DG}{\text{DG}}

\newcommand{\BSSN}{\text{BSSN}}

\newcommand{\B}{\text{B}}

\newcommand{\Lo}{\mathcal{L}}

\npdecimalsign{.}
\npthousandsep{,}

\newcolumntype{Y}{>{\RaggedRight\arraybackslash}X}

\begin{document}

\title{An Operator-Based Local Discontinuous Galerkin Method
  Compatible With the BSSN Formulation of the Einstein Equations}

\author{Jonah M. Miller}
\affiliation{Department of Physics, University of Guelph, Guelph, ON,
  Canada}
\affiliation{Perimeter Institute for Theoretical Physics, Waterloo,
  ON, Canada}
\email{jmiller@perimeterinstitute.ca}

\author{Erik Schnetter}
\affiliation{Perimeter Institute for Theoretical Physics, Waterloo,
  ON, Canada}
\affiliation{Department of Physics, University of Guelph, Guelph, ON,
  Canada}
\affiliation{Center for Computation \& Technology, Louisiana State
  University, Baton Rouge, LA, USA}
\email{eschnetter@perimeterinstitute.ca}

\date{2016-04-14}

\begin{abstract}

  Discontinuous Galerkin Finite Element (DGFE) methods offer a
  mathematically beautiful, computationally efficient, and efficiently
  parallelizable way to solve partial differential
  equations (PDEs). These properties make them highly desirable for
  numerical calculations in relativistic astrophysics and many other
  fields.  The BSSN formulation of the Einstein equations has
  repeatedly demonstrated its robustness. The formulation is not only
  stable but allows for puncture-type evolutions of black hole
  systems.  To-date no one has been able to solve the full
  (3+1)-dimensional BSSN equations using DGFE methods. This is partly
  because DGFE discretization often occurs at the level of the
  equations, not the derivative operator, and partly because DGFE
  methods are traditionally formulated for manifestly
  flux-conservative systems. By discretizing the derivative operator,
  we generalize a particular flavor of DGFE methods, Local DG methods,
  to solve arbitrary second-order hyperbolic equations. Because we
  discretize at the level of the derivative operator, our method can
  be interpreted as either a DGFE method or as a finite differences
  stencil with non-constant coefficients.

\end{abstract}

\maketitle

\section{Introduction}
\label{sec:intro}

In numerical relativity, Einstein's equations are typically decomposed
in one of several ways. Drawing on the constraint damping proposed by
Gundlach et al. \cite{GundlachConstraintDamping}, the generalized
harmonic (GH) formulation was originally developed in second-order form by
Pretorious \cite{PretoriuousGH}. Pretorious used it with finite
differences to provide the first successful evolution and merger of a
binary black hole system \cite{pretorius2005evolution}. Lindblom et
al. \cite{LindblomGH} rewrote the generalized harmonic
formulation in first-order form. Using pseudospectral methods, this
formulation has successfully been used to accurately describe a wide
variety of astrophysical situations. The literature is very extensive,
but the interested reader can find much of the relevant work in
\cite{BoyleBBH,ScheelWaveforms,SPeC,BAMPS} and references therein.

The Baumgarte-Shapiro-Shibata-Nakamura (BSSN) formulation of the
Einstein equations
\cite{ShibataBSSN,BaumgarteBSSN,BrownBSSN,AlcubierreConformalDecomp,AlcubierreGammaDriver}
is a second-order formulation. Some key ingredients of the BSSN
formulation are the conformal re-scaling of geometric quantities,
treating the trace of the connection coefficients as independent
variables, and the separation of the trace of the extrinsic curvature
tensor from its other components. These ingredients not only make the
formulation well-posed \cite{SarbachBSSNHyperbolicity} but allow for
so-called ``moving puncture'' evolutions, where the
singularity within a black hole is not resolved on the computational
grid, and where the thus non-physical interior of the black hole
can be safely evolved thanks to the characteristic
structure of the system \cite{CampanelliPunctures,BakerPunctures}. In
three dimensions, puncture solutions are very desirable because they
are significantly easier to implement than the other techniques for
avoiding singularities.

For these reasons, the BSSN formulation of the Einstein equations is
used by many relativity groups and a great deal of expertise has been
acquired. This is strong motivation for the development of efficient
numerical methods for evolving the BSSN equations. For smooth
problems, such as the Einstein equations, pseudospectral methods
converge exponentially. They are therefore a very appealing approach
to solving the Einstein equations. They have been successfully used
with the GH formulation in a number of contexts, especially for
compact binary mergers of all flavors
\cite{BoyleBBH,ScheelWaveforms,SPeC}. They have also been enormously
successful in generating initial data for numerical relativity
\cite{LORENEPaper,LORENE,SpecInitialData,TwoPunctures,SopuertaToyModelFE,PhysRevD.73.044028,aksoylu2008FEEinstein,KorobkinFE,CaoTwoPunctureFE}.\footnote{We
  note that the calculation of initial data involves solving an
  elliptic differential system, not a hyperbolic one. This requires
  different considerations than those we discuss here.} If one imposes
appropriately flux-conservative penalty-type boundary conditions and
uses many small spectral domains, these techniques become nodal
discontinuous Galerkin finite element (DGFE) methods
\cite{gottlieb2001spectralReview}.

DGFE methods combine the high-order accuracy of spectral methods with
the flexibility and parallelizability of finite volume type methods
\cite{hesthaven2008nodal}. In smooth regions they provide spectral
accuracy and in non-smooth regions they can be combined with
high-resolution shock capturing (HRSC) techniques to accurately
resolve discontinuities. (See
\cite{Dumbser20096991,radice2011discontinuous,Zhao2013138,zanotti2015solving,BugnerDGWENO,kidder2016spectre}
for some recent applications of DGFE combined with HRSC for
relativistic hydrodynamics.)  Importantly, DGFE methods allow for a
domain decomposition which requires only a single layer of ghost
points.

There are several extensions of DGFE methods for second-order and non
flux-conservative systems. Using distributional theory, Vol'pert
\cite{VolpertBV}, LeFloch and collaborators
\cite{maso1995definition,lefloch1988entropy,lefloch1989shock,hou1994nonconservative},
and Colombeau and coworkers
\cite{colombeau2000new,colombeau1988multiplications} have all
developed different techniques to define shock wave solutions for
hyperbolic systems of the form
$$\partial_t \myvec{\psi} + g(\myvec{\psi}) \partial_i\myvec{\psi} = 0\ \forall\ i\in\{1,2,3\},$$
where $\myvec{\psi}$ is a collection of variables, each of which may
be discontinuous. These techniques have been applied numerically first
in a finite volume context
\cite{cauret1989discontinuous,TOUMI1992360,toumi1996approximate,castro2006high,pares2006numerical}
and later in a discontinuous Galerkin setting
\cite{rhebergen2008discontinuous,tassi2008discontinuous}.

In \cite{teukolsky2015formulation}, Teukolsky develops a formalism for
DGFE methods in arbitrary curved spacetimes for both conservative and
non-conservative first-order systems. In \cite{TaylorPenaltyWave},
Taylor et al. derive a penalty method for the wave equation based on
energy methods.  \textit{Interior Penalty discontinuous Galerkin}
(IPDG) methods
\cite{arnold1982interior,ShahbaziInterior,cheng2008discontinuous,grote2006discontinuous,GroteMaxwellIPDG,HesthavenNodal1,GroteThesis}
discretize a second-order system by imposing additional penalty
boundary terms. \textit{Local discontinuous Galerkin} (LDG) methods,
developed by Shu and collaborators \cite{LDG1,LDG2,LDG3} and based on
the early work by Bassi et al. \cite{bassiDu1,bassiDu2}, introduce
auxiliary variables to facilitate second-differentiation. These
variables are evaluated at each time step but not evolved and allow
penalty boundaries to be imposed as for a first-order system. For a
review, see \cite{LDGReview}.

DGFE methods exist within a rich ecosystem of penalty methods,
including but not limited to: spectral penalty methods
\cite{funaro1988new,funaro1991convergence,hesthaven2000SpectralPenalty},
spectral finite volume methods
\cite{spectralFV1,spectralFV2,spectralFV3,spectralFV4,spectralFV5,spectralFV6},
and spectral difference methods
\cite{spectralDifference1,spectralDifference2}. Indeed, nodal DGFE
methods (and likely pseudospectral penalty methods in general) can be
cast as multi-domain summation-by-parts finite differences methods
\cite{Gassner-DG-SBP,fernandez2014generalized}. In this formalism, the
penalty boundary terms are called \textit{simultaneous approximation
  terms}
\cite{carpenter1993stability,svardSBPReview,fernandez2014review}.

In \cite{tichy2006} and \cite{tichy2009}, Tichy evolved a static black
hole on a single spectral domain using the BSSN system and a
pseudospectral scheme. Using a variational principle, Zumbusch
developed a DGFE discretization in both space and time for the
second-order GH formulation of the Einstein equations
\cite{zumbuschHarmonicDGFE}. Field et al. developed a DGFE method for
the second-order BSSN equations in spherical symmetry
\cite{field2010discontinuous}. In \cite{radice2011discontinuous},
Radice and Rezzola developed a DGFE formulation for fluids in a
general relativistic setting. In the process, they use a DGFE method
to solve the Einstein equations in spherical symmetry with maximal
slicing and areal coordinates.\footnote{In this gauge, the reduction
  of the Einstein equations to spherical symmetry is an elliptic,
  rather than hyperbolic, system.} In \cite{BugnerDGWENO}, Bugner et
al. build on this work to combine DGFE methods with an HRSC scheme
based on weighted essentially nonoscillatory (WENO) algorithms for
fluids on a fixed, curved, spacetime background.  Motivated by the
first-order-in-space nature of DGFE methods, Brown et al. developed a
fully first-order version of the BSSN system and evolved a binary
black hole in-spiral using finite differences. They also evolved a
reduction of the system to spherical symmetry using a DGFE scheme
\cite{BrownFOBSSN}.

It is desirable to evolve the second-order BSSN equations in full 3+1
dimensions via DGFE methods. Unfortunately, to-date this has not been
possible.

DGFE methods are typically formulated for manifestly flux-conservative
systems of the form
\begin{equation}
  \label{eq:flux:conservative}
  \partial_t \myvec{\psi} + \partial_j f^j(\myvec{\psi}) = 0,
\end{equation}
where $\myvec{\psi}$ is a collection of variables and $\myvec{f}$ is a
nonlinear flux, which is a function of $\myvec{\psi}$. In these
systems, differentiation of the flux is the most natural
operation. Therefore, one may discretize $\partial_j f^j$ all at once.

In contrast, the BSSN system is roughly of the form
\begin{equation}
  \label{eq:general:nonlinear:operator}
  \partial_t \myvec{\psi} = \Lo\sqrbrace{\myvec{\psi},\partial_j\myvec{\psi},\partial_j\partial_k\myvec{\psi}},
\end{equation}
where $\Lo$ is a nonlinear operator that acts on $\myvec{\psi}$ and its
derivatives. In this case, the natural operation is to differentiate
$\myvec{\psi}$ directly. Therefore, there is no reason to discretize the
entire operator $\Lo$, which may be very cumbersome. Instead, it may
be cleaner to discretize the differential operators $\partial_i$.

There are then two related difficulties in evolving the BSSN equation
using DGFE methods. First DGFE methods are usually formulated for
manifestly flux-conservative systems, which are first-order in space
and time. The penalty-type boundary conditions imposed in DGFE methods
must therefore be generalized to second-order systems. Second, DGFE
methods are usually formulated at the level of the equations, not the
level of the derivative operator. Given the complexity of the BSSN
system, this is a serious impediment to the development of a working
scheme.

In this work we develop a new generalization of DGFE methods,
\textit{operator-based local discontinuous Galerkin} (OLDG) methods,
which address these problems and allows us to evolve the BSSN
equations. We then subject our approach to a battery of
community-developed tests for numerical relativity: the
Apples-with-Apples tests \cite{AwA1,AwA2}. So that we can handle
second-order systems, we base our scheme off of LDG methods. We draw
particular inspiration from the work of Xing et
al. \cite{LDGTruncate}, where they develop a superconvergent
energy-conserving LDG method for the wave equation.

To avoid the complications of discretizing the BSSN equations, we
perform our discretization at the level of the derivative operator,
rather than at the level of the equations. This requires a different
formalism for describing the piecewise polynomial space. Our formalism
uses distributional theory and is inspired by a pedagogical exercise
in \cite{hesthaven2008nodal}, which we make rigorous. Because we focus
our discretization at the level of the differential operator and not
the equations, our method provides a \textit{drop-in} solution for
working relativity codes. All that is necessary to convert a finite
differences code to a DGFE code is to replace the derivative operator
with ours.

Discretizing at the level of the derivative requires special care with
respect to the stability of our scheme. Integration by parts is often
an integral step in proofs of the well-posedness of a continuum system
of initial-value problems \cite{Alcubierre}. Essentially, one finds an
energy norm and shows that it is non-increasing. Hyperbolic systems
for which an energy norm exists are called \textit{symmetrizable
  hyperbolic}.
In their pioneering work, Sarbach et al. showed that
the second-order BSSN system is a second-order version of a
first-order strongly hyperbolic system
\cite{SarbachBSSNHyperbolicity}. Strong hyperbolicity is weaker than
symmetrizable hyperbolicity, but both imply stability.

The discrete analog of integration by parts is \textit{summation by
  parts}, developed by Kreiss and Scherer
\cite{Kreiss1974,Kreiss1977}, and it tremendously simplifies proofs of
numerical stability. Indeed, a summation-by-parts operator, combined
with strong or symmetrizable hyperbolicity and appropriate conditions
on initial and boundary data, is often enough to demonstrate
stability.

Given the complicated nature of the BSSN equations and their
discretizations, we do not seek to prove the stability of our
scheme. Rather we insist that our discretized derivative operator
satisfy summation by parts. Given the strongly hyperbolic nature of
the BSSN system, we expect this restriction to provide linear
stability. Nonlinear stability is enforced both by a
\textit{truncation} scheme we develop and by more traditional
filtering techniques as needed.

Our paper is organized as follows. In section \ref{sec:methods}, we
develop the formalism for OLDG methods and define the OLDG
operator. For brevity, we skip the details of our results regarding
summation-by-parts, stability, and convergence. The interested reader
can find these in appendices \ref{sec:sbp:proof},
\ref{sec:stability:wave:equation} and \ref{sec:convergence}
respectively. In section \ref{sec:asymptotics}, we describe some of
the computational properties of interest, such as computation,
communication, and memory access costs. In section \ref{sec:tests}, we
describe the numerical tests we perform and their results. Finally, in
section \ref{sec:conclusion}, we offer some concluding remarks.

\section{Methods}
\label{sec:methods}

\begin{table}[!tb]
  \centering
  \begin{tabularx}{\columnwidth}{| l | Y |}
    \toprule
    \textbf{Symbol} & \textbf{Meaning}\\
    \colrule
    $\Omega = [X_l,X_r]$ & Our domain of interest\\ 
    [1ex]
    $\tilde{\Omega} = [X_l-\varepsilon,X_r+\varepsilon]$ & Our extended domain\\ 
    [1ex]
    $\Omega^k = [x_l^k,x_r^k]$ & An element within $\Omega$\\ 
    [1ex]
    $K$ & The number of elements in domain $\Omega$\\
    [1ex]
    $P^k$ & The maximum polynomial order of the test functions $\{\Phi^k_i\}$ 
            used to represent a function within element $\Omega^k$\\
    [1ex]
    $h^k$ & The width of an element, $x^k_r - x^k_l$\\
    [1ex]
    $\psi,\phi$ & Piecewise smooth functions living on our broken domain\\
%     [1ex]
%     $\nabla\psi$ & The derivative of $\psi$ in space with respect to $x$ \\
    \colrule
    $\chi^k$ & The characteristic function for $\Omega^k$\\
    [1ex]
    $\Theta^k_l$,$\Theta^k_r$ & The Heaviside function centred 
                               on $x^k_l$ and $x^k_r$ respectively\\
    [1ex]
    $\delta^k_l$,$\delta^k_r$ & The Dirac delta function centred
                               on $x^k_l$ and $x^k_r$ respectively\\
    [1ex]
    $\psi^k$ & The restriction of $\psi$ onto $\Omega^k$\\
    [1ex]
    $\psi^k_l,\psi^k_r$ & $\psi^k$ evaluated at $x^k_l$ and $x^k_r$ 
                         respectively\\
    [1ex]
    $\Phi^k_i$ & The $i^{\text{th}}$ test function in $\Omega^k$\\
    \colrule
    $(\psi^*)^k_l, (\psi^*)^k_r$ & Weak boundary terms\\
    [1ex]
    $\tilde{\psi}_s^k,\tilde{\psi}_d^k$  & The smooth and discontinuous extensions 
                                          of $\psi^k$ outside its element respectively\\
    \colrule
    $x^k_i$ & The $i^{\text{th}}$ collocation point ($0 \le i \le P^k$)
              on discretized element $\Omega^k$\\
    [1ex]
    $w^k_i$ & Weights for the discrete inner product within an element\\
    [1ex]
    $\psi^k_i$ & $\psi^k$ evaluated at $x^k_i$\\
    [1ex]
    $\hat{\psi}^k_i$ & The projection of $\psi^k$ onto $\Phi^k_i$ \\
    [1ex]
    $\hat{d}^k$ & The change of basis matrix that maps the 
                      spectral coefficients for $\psi^k$ to
                      those for its derivative\\
    [1ex]
    $\V^k$ & The Vandermonde matrix, which transforms between the
             modal and nodal bases\\
    [1ex]
    $d^k$ & The narrow nodal derivative operator \\
    \hline
    $w^k$ & The discrete weights as a matrix\\
    [1ex]
    $b^k$ & The boundary operator, the discretization of a Dirac delta
            function over the boundary of an element $\Omega^k$\\
    [1ex]
    $F^k$ & The ``fetch'' operator, which pulls information from 
            elements neighbouring element $\Omega^k$\\
    [1ex]
    $D^k$ & The wide derivative operator, which takes
            neighbouring elements into account\\
    \botrule    
  \end{tabularx}
  \caption{Notation used in the construction of our discontinuous
    Galerkin scheme. All symbols are explained in more detail in the
    main text.}
  \label{tab:notation}
\end{table}

In the usual formulation of DGFE methods one replaces the conserved
flux through the boundary of a fixed volume with a \textit{numerical
  flux}, which takes information from within the volume and from the
boundaries of neighbouring volumes \cite{hesthaven2008nodal}. Here we
take a different approach. We use distributional theory to replace the
derivative of a smooth function with the \textit{weak} derivative of a
\textit{piecewise smooth} function, appropriately chosen to recover
the small communication overhead characteristic of these types of
methods.

This approach was first proposed by Hesthaven and Warburton
\cite{hesthaven2008nodal}, who use it pedagogically to argue that the
strong form of the canonical DGFE operator is just an encoding of the
notion of a weak derivative.  We make this assertion rigorous and
argue that this weak derivative formulation provides a generic way to
place arbitrary nonlinear hyperbolic equations into the DGFE
framework, even if they are not manifestly flux-conservative.

This focus on the derivative operator has very practical consequences:
It allows computer programmes that currently employ finite differences
methods to replace the finite differences stencil with our OLDG
stencil, converting a finite differences code to a DGFE code. This
transition requires: implementing loop tiling (for efficiency) so that
the band-diagonal finite differences operator can be replaced by our
block-diagonal discontinuous Galerkin operator, implementing the
truncation scheme described in section \ref{sec:truncation}, and
replacing the Kreiss-Oliger dissipation operator with the
right-hand-side filter operator discussed in section
\ref{sec:filtering}. We discuss those issues in this section
below. This change from a finite difference to a DGFE method should
improve the parallel efficiency significantly, as we discuss in the
next section \ref{sec:asymptotics}.

% one partitions---in the set
%theoretic sense \cite{halmos2013naive}---
In a DGFE method, there are two levels of discretization. At the top
level, one breaks the domain of interest $\Omega$ into many
subdomains, or \textit{elements}, $\{\Omega_k\}_{k=1}^K$ which overlap
on a set of measure zero (see figure \ref{fig:broken:domain}).  One
must choose how to approximate the derivatives of a function that
lives on this broken domain. At a lower level, one must choose how to
approximate the piece of the global function that lives on each
subdomain. From the perspective of a single element, one can think of
the former as a choice of boundary conditions for the piece of the
function living on the element and the latter as an ansatz for the
types of functions that can live on the element. These two levels of
discretization can be lumped into a single discretization
step. However, making them explicit allows us to develop our
discretization approach formally.

For the reader's convenience we provide a reference for the notation used
in our construction in table \ref{tab:notation}.

\subsection{The Main Idea}
\label{sec:main:idea}

Before we proceed with our construction, we present a toy problem
which encapsulates some of the core ideas of our method. Consider two
smooth functions:
\begin{eqnarray}
  \label{eq:def:toy:phi}
  \phi : [0,1] &\to& \R\\
  \label{eq:def:toy:pi}
  \text{and }\pi : [0,1] &\to& \R,
\end{eqnarray}
each of which is defined on the interval $[0,1]$. From these two
functions, we can construct a third function
$\psi : [0,1]\to \R$ defined by
\begin{equation}
  \label{eq:def:toy:psi}
  \psi(x) = \begin{cases}
    \phi(x)&\text{if }0\leq x\leq x_0\\
    \pi(x)&\text{if }x_0 < x \leq 1
  \end{cases}
\end{equation}
for some $0 < x_0 < 1$. 

We wish to differentiate $\psi$. However, generically, $\psi$ has a
discontinuity at $x_0$ and is not a differentiable function. If we
treat $\psi$ as a \textit{distribution} then its derivative can be
defined \textit{weakly}. To make this manifest, we write $\psi$ as the
sum of two distributions:
\begin{equation}
  \label{eq:phi:toy:sum}
  \psi(x) = \phi(x)\Theta(x_0-x) + \pi(x)\Theta(x-x_0),
\end{equation}
where 
\begin{equation}
  \label{eq:def:heaviside}
  \Theta(x) = \begin{cases}0&\text{if }x < 0\\ 1&\text{if }x\geq 0\end{cases}
\end{equation}
is the Heaviside function. Then the \textit{weak}, or
\textit{distributional} derivative of $\psi$ is given by
\begin{widetext}
\begin{eqnarray}
  \partial_x \psi(x) &=& \sqrbrace{\partial_x\phi(x)}\Theta(x_0-x) - \phi(x)\delta(x-x_0) + \sqrbrace{\partial_x\pi(x)}\Theta(x-x_0) + \pi(x)\delta(x-x_0)\nonumber\\
  \label{eq:toy:dphi}
  &=& \sqrbrace{\partial_x\phi(x)}\Theta(x_0-x) + \sqrbrace{\partial_x\pi(x)}\Theta(x-x_0)+\sqrbrace{\pi(x)-\psi(x)}\delta(x-x_0),
\end{eqnarray}
\end{widetext}
where we have used the fact that the distributional derivative of the
Heaviside function $\Theta(x)$ is the Dirac delta function $\delta(x)$
\cite{duistermaat2010distributions}.

Equation \eqref{eq:toy:dphi} is well defined under integration with
any smooth test function that has compact support over the interval
$[0,1]$. In other words,
\begin{displaymath}
  \label{eq:toy:integral:dphi}
  \int_0^1 \Phi\partial_x\psi dx = \int_0^{x_0}\Phi\partial_x\phi dx + \int_{x_0}^1 \Phi\partial_x\pi dx + \sqrbrace{\pi - \phi}\eval_{x_0}
\end{displaymath}
for all smooth functions $\Phi$ such that $\Phi(0)=\Phi(1)=0$. 

In the following sections, we will use a decomposition much like that
given by equation \eqref{eq:phi:toy:sum}. In this toy example, our
decomposition relied on the existence of functions $\phi$ and $\pi$
which were defined on the \textit{entire} interval $[0,1]$. More
generally, these functions may not be given to us \textit{a priori},
but they can be constructed. 

% This construction injects an ambiguity
% into our method, which we will first make explicit and then resolve by
% insisting our discrete approximation of a derivative satisfy summation
% by parts.
  
\subsection{The Broken Domain}
\label{sec:methods:breakup}

\begin{figure}[!tb]
  \centering
  \includegraphics[width=0.6\columnwidth]{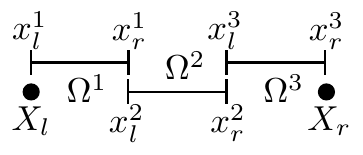}
  \caption{A broken domain with $K=3$.}
  \label{fig:broken:domain}
\end{figure}

We now proceed with the main construction. For simplicity suppose that
the domain of interest $\Omega$ is the real interval $[X_l, X_r]$,
$X_l < X_r\in\R$. For higher dimensions, we simply assume a Cartesian
product topology. We break our domain $\Omega$ into $K$ interior
elements
\begin{equation}
  \label{eq:def:omega:k}
  \Omega^k := [x_l^k, x_r^k],\ x^k_l < x^k_r \in\R
\end{equation}
for all $k=1,\ldots,K$ and two boundary elements
\begin{equation}
  \label{eq:boundary:elements}
  \Omega^0 := \{x^0_r\},\ \Omega^{K+1} := \{x^{K+1}_l\}
\end{equation}
with $x^0_r=X_l$ and $x^{K+1}_l=X_r$ such that,
\begin{equation}
  \label{eq:element:boundary:condition}
  x_r^{k-1} = x_l^k \text{ and } x_r^k = x_l^{k+1}
\end{equation}
for all $k=1,\ldots,K$. (Note that these boundary elements are
singleton sets.) We also demand that the union of all $K+2$ elements
comprises the whole domain:
\begin{equation}
  \label{eq:domain:union}
  \bigcup_{k=0}^{K+1} \Omega^k = \Omega.
\end{equation}
For convenience, we define the element width
\begin{equation}
  \label{eq:def:hk}
  h^k = x^k_r - x^k_l.
\end{equation}
Figure \ref{fig:broken:domain} shows the structure of $\Omega$ for three
elements, e.g., when $K=3$.

On our domain $\Omega$, we wish to represent the arbitrary function
$\psi(x)$. $\psi$ may be either a scalar or
vector quantity. For simplicity, we will assume here that it is a
scalar. We call the restriction of $\psi$ onto a given
element $\Omega^k$, $\psi^k$. We demand that each $\psi^k$ be smooth
but we allow $\psi$ to have jump discontinuities at the domain
boundaries $\{x_r^k = x_l^{k+1}\}$. Notice that in this description,
each $\Omega^k$ overlaps with its neighbours $\Omega^{k-1}$ and
$\Omega^{k+1}$ at exactly one point. These points of overlap are
exactly the points where $\psi^k$ is allowed to be discontinuous and
\begin{equation}
  \label{eq:def:psi:k:lr}
  \psi^k_r := \psi^k(x_r^k) \text{ and } \psi^{k+1}_l := \psi^{k+1}(x_l^{k+1})
\end{equation}
can be thought of as the left- and right-hand limits of $\psi$
respectively. These conditions are equivalent to the standard choices
one makes for a typical one-dimensional DGFE method.

For clarity, we define the following convention. Functions living on
the domain $\Omega$ will be represented by lower-case Greek
letters. Elements will be indexed by a superscript, and positions
within an element will be denoted by a subscript.

We can formalize the restriction of $\psi$ to $\psi^k$ by defining the
\textit{characteristic function}
\begin{equation}
  \label{eq:def:selection:distribution}
  \chi^k(x) := \begin{cases}1&\text{if }x\in\Omega^k\\0&\text{else}\end{cases}
\end{equation}
such that
\begin{equation}
  \label{eq:restricted:function}
  \psi(x)\chi^k(x) = \begin{cases}\psi^k(x)&\text{if }x\in\Omega^k\\0&\text{else}\end{cases}.
\end{equation}
Note that product \eqref{eq:restricted:function} is defined pointwise
as a function. However, since the product of two distributions is in
general not well-defined, it is \textit{not} a proper
distribution. The techniques Colombeau and coworkers
\cite{colombeau2000new} can be used to define such a
distribution. However, we will not need to do so.

We also introduce two inner products, one local to a subdomain
$\Omega^k$ and one for the entire domain. If $\psi^k$ and
$\phi^k$ are functions on $\Omega^k$, then the subdomain inner
product is
\begin{equation}
  \label{eq:subdomain:inner:product}
  \braket{\psi^k, \phi^k}_{\Omega^k} = \int_{x_l^k}^{x_r^k} \psi^k(x) w^k(x) \phi^k(x) dx,
\end{equation}
where $w^k(x)$ is an as-of-yet undecided weight function. If $\psi$
and $\phi$ are instead functions on the whole domain $\Omega$, then
the overall inner product is:
\begin{equation}
  \label{eq:total:domain:inner:product}
  \braket{\psi,\phi}_\Omega = \int_{X_l}^{X_r} \psi(x)w(x)\phi(x) dx= \sum_{k=1}^K \braket{\psi^k,\phi^k}_{\Omega^k},
\end{equation}
where $w(x)$ is the weight function for the whole domain and $\psi^k$
and $\phi^k$ are the appropriate restrictions of $\psi$ and $\phi$.

We will sometimes be interested in a slightly extended integral. Let
\begin{equation}
  \label{eq:varepsilon:introduction}
  \varepsilon > 0,\ \varepsilon\in\R. 
\end{equation}
Then we define the \textit{extended domain} 
\begin{equation}
  \label{eq:def:omega:tilde}
  \tilde{\Omega} := [X_l-\varepsilon,X_r+\varepsilon]
\end{equation}
and \textit{extended inner product}
\begin{equation}
  \label{eq:def:extended:product}
  \braket{\psi,\phi}_{\tilde{\Omega}} := \int_{X_l-\varepsilon}^{X_r+\varepsilon} \psi(x)w(x)\phi(x) dx.
\end{equation}
This extension is useful for handling the discontinuities at
$\Omega^0$ and $\Omega^{K+1}$.

\begin{figure}[!tb]
  \centering
  \includegraphics[width=\columnwidth]{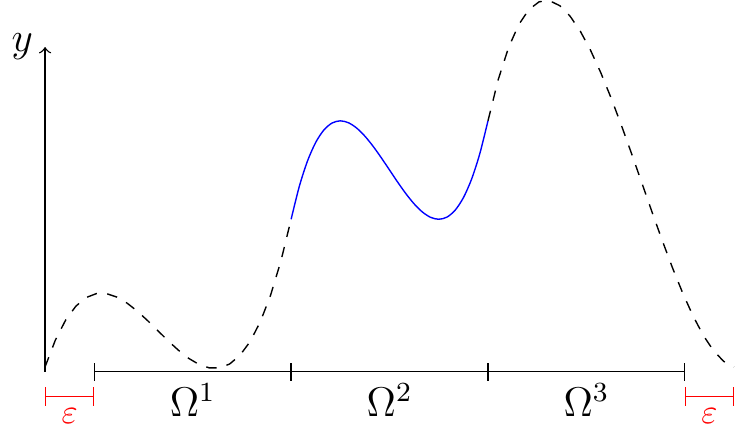}
  \caption{An example test function $\Phi^2_i$ for the domain
    $\Omega^2$ with $K=3$. When restricted to $\Omega^2$, the set
    $\{\chi^2 \Phi^2_i\}_{i=0}^\infty$ forms an orthonormal basis for
    $C^\infty(\Omega^2)$. However, each $\Phi^2_i$ is smooth over and
    has compact support on the \textit{whole} interval
    $[X_l-\varepsilon,X_r+\varepsilon]$. The restriction $\chi^2
    \Phi^2_i$ (for some $i$) is shown as a solid line. The remainder
    of the function is shown as a dashed line.}
  \label{fig:test-functions}
\end{figure}

Since we allow functions on $\Omega$ to be piecewise smooth and
therefore not everywhere differentiable, we are interested in their
properties in a \textit{weak} or \textit{distributional}
sense. Therefore for each $k=1,\ldots,K$ and each element $\Omega^k$,
we define a set of smooth test functions, $\{\Phi_i^k\}_{i=0}^\infty$,
each defined on the \textit{whole} extended interval
$\tilde{\Omega}=[X_l-\varepsilon,X_r+\varepsilon]$.

We demand that, for all $i\in\N$ and $k=1,\ldots,K$,
\begin{equation}
  \label{eq:compact:support}
  \Phi_i^k(X_l-\varepsilon) = \Phi_i^k(X_r+\varepsilon) = 0
\end{equation}
and that, for all $\phi^k\in C^\infty(\Omega^k)$, there exists a set
of spectral coefficients $\{\hat{\phi}^k_i\}_{i=0}^\infty$,
$\hat{\phi}^k_i\in\R$, such that
\begin{equation}
  \label{eq:span}
  \phi^k(x) = \sum_{i=0}^\infty \hat{\phi}^k_i \chi^k(x)\Phi^k_i(x).
\end{equation}
In other words, every $\Phi_i^k$ has compact support on the interval
$[X_l-\varepsilon,X_r+\varepsilon]$ and, for each $k$, the set of
restrictions $\{\chi^k \Phi^k_i\}_{i=0}^\infty$ forms an orthonormal
basis for $C^\infty(\Omega^k)$. Other than insisting on smoothness and
compact support, we do not need to constrain our test functions
outside of their respective elements. (From now on, we will represent
spectral coefficients with hats.)

The choice of weight function $w^k$ is tied to the choice of basis
functions $\Phi^k_i$. We may use it to ensure that our test functions
are orthogonal. The weight function for the whole domain, $w(x)$ must
be chosen for compatibility with $w^k$.

Note that our basis functions are quite different from those defined
in a standard discontinuous Galerkin method. In the usual DGFE
approach one can define the test functions as a set of piecewise
smooth functions on $\Omega$ with discontinuities at element
boundaries. However, if the test functions themselves are
discontinuous, we cannot rigorously apply distribution theory.

In our case we define a set of test functions on each
element. However, each test function is defined outside the element,
with compact support on an appropriately extended domain, as shown in
figure \ref{fig:test-functions}. This construction ensures that all
our test functions are smooth and that the standard results from
distributional theory hold.

In particular, if $\Theta(x)$ and $\delta(x)$ are the Heaviside and
Dirac delta functions respectively, then
% \begin{equation}
%   \label{eq:def:heaviside}
%   \Theta(x) = \begin{cases}0&\text{if }x < 0\\ 1&\text{if }x\geq 0\end{cases}
% \end{equation}
\begin{equation}
  \label{eq:dheaviside:1}
  \braket{\partial_x\Theta,\Phi_i^k}_{\tilde{\Omega}} = \braket{\delta,\Phi_i^k}_{\tilde{\Omega}}
%   \label{eq:dheaviside:2}
%   \braket{\ddII{x}\Theta,\Phi_i^k}_\Omega &=& \braket{\delta',\Phi_i^k}_\Omega
\end{equation}
for all $i\in\N$ and $k=1,2,\ldots,K$
\cite{duistermaat2010distributions}. 

For convenience, we define the
\textit{shifted} Heaviside and Dirac delta functions
\begin{eqnarray}
  \label{eq:shifted:heaviside}
  \Theta^k_l (x):= \Theta(x^k_l-x),&&\Theta^k_r(x) := \Theta(x-x^k_r),\\
  \label{eq:shifted:dirac}
  \delta^k_l (x):= \delta(x-x^k_l),&\text{ and }&\delta^k_r(x) := \delta(x-x^k_r),
\end{eqnarray}
which are centred on the element boundaries
$\{x^k_l,x^k_r\}_{k=1}^{K}$. (Note that $\Theta^k_l$ is inverted
in $x$ so it is nonzero for $x<x^k_l$ and zero for $x> x^k_l$.)
In this language, we can write the distributional derivative of the
characteristic function as 
\begin{equation}
  \label{eq:def:dchi}
  \braket{\partial_x\chi^j,\Phi^k_i}_{\tilde{\Omega}} = \braket{\delta^j_l,\Phi^k_i}_{\tilde{\Omega}} - \braket{\delta^j_r,\Phi^k_i}_{\tilde{\Omega}}
\end{equation}
for all $i\in\N$ and all $j,k = 1,2,\ldots,K$.

For the remainder of this paper, the restrictions $\chi^k \Phi^k_i$ of
our test functions $\Phi^k_i$ to the interior of an element $\Omega^k$
are assumed to be the Legendre polynomials, defined in appendix
\ref{sec:collocation}. (It doesn't matter what they are outside
$\Omega^k$, as long as they are smooth and have compact support.) This
means our element-wise weight function is $w^k = 1$. 

We make this assumption partly to make contact with traditional nodal
DGFE methods, where this is the norm, and partly for
simplicity. Although we did not explore them, other choices such as
Chebyshev polynomials are certainly possible and potentially
desirable. Choice of (orthonormal) basis function and associated
collocation points affect both the timestep and the conditioning of
the linear operator that transforms between nodal and modal bases
\cite{hesthaven2008nodal} (see section \ref{sec:movng:discrete}). For
more details on some of the effects varying the basis or collocation
points can have, see \cite{hesthaven2008nodal} or
\cite{press2007numerical}.

As long as the collocation points include the boundaries of the
element, as described appendix \ref{sec:collocation}, our main results
are unchanged by the choice of basis or collocation points.

\subsection{The Modified Derivative Operator}
\label{sec:modified:derivative}

To take the derivative of an arbitrary distribution $\phi$ defined on
$\Omega$, we will decompose it into the sum of several distributions,
just as in section \ref{sec:main:idea}. However, to follow this
procedure, each $\phi^k$ must be smoothly defined on the whole
extended domain $\tilde{\Omega}$. Therefore, for each $\psi^k$, we
define a \textit{smooth continuation}
\begin{equation}
  \label{eq:def:psi:smooth}
  \tilde{\psi}^k_s \in C^\infty(\tilde{\Omega}) : \tilde{\Omega} \to \R
\end{equation}
such that
\begin{equation}
  \label{eq:extension:restriction}
  \chi^k\tilde{\psi}^k_s = \chi^k\psi.
\end{equation}
In other words, within $\Omega^k$, $\psi$ and $\tilde{\psi}^k_s$ must
agree. However, outside of $\Omega^k$ they generically do not.

% The machinery described in section \ref{sec:methods:breakup} is
% sufficient to define 
%However, the natural
% derivative is not quite suitable for our needs. To retain the spirit
% of DGFE methods, we would like our derivative operator to depend on a
% quantity analogous to the flux through the boundary of an element.

Armed with this machinery, we can perform a procedure analogous to
that in section \ref{sec:main:idea} to compute the the derivative of a
piecewise smooth function $\phi$ on $\Omega$. To this end, consider
the following \textit{discontinuous continuation} of the restriction
of $\psi^k$:
\begin{equation}
  \label{eq:modified:restriction}
  \tilde{\psi}^k_d(x) := \tilde{\psi}^k_s(x)\chi^k(x) 
  + (\psi^*)^k_l\Theta^k_l(x)
  + (\psi^*)_r^k\Theta^k_r(x),
\end{equation}
where $(\psi^*)_l^k$ and $(\psi^*)_r^k$ are chosen to incorporate
information about $\psi^{k-1}_r$ and $\psi^{k+1}_l$ respectively. Note
that since $\tilde{\psi}^k_s$ is smooth, $\tilde{\psi}^k_d$ is a
proper distribution with no ambiguities other than the choice of
$\psi^*$, which is analogous to the choice of numerical flux in a
traditional DGFE scheme.

% The
% choice of $\psi^*$ is analogous to the choice of numerical flux in a
% traditional DGFE scheme. 

We now approximate the derivative of $\psi$ as the weak derivative of
$\tilde{\psi}_d^k(x)$ for $x\in\Omega^k$. To calculate this weak
derivative, we differentiate and take the inner product with an
arbitrary test function $\Phi_i^k$:
\begin{widetext}
\begin{eqnarray}
  \braket{\Phi_i^k, \partial_x \tilde{\psi}_d^k}_{\tilde{\Omega}} 
  &=& \braket{\Phi_i^k,\partial_x\sqrbrace{\chi^k(x)\tilde{\psi}^k_s(x)}}_{\tilde{\Omega}}
  + (\psi^*)^k_l \braket{\Phi_i^k,\partial_x\Theta^k_l(x)}_{\tilde{\Omega}}
  + (\psi^*)_r^k\braket{\Phi_i^k,\partial_x\Theta^k_r(x)}_{\tilde{\Omega}}\nonumber\\
  &=& \braket{\Phi_i^k,\chi^k\partial_x\tilde{\psi}_s^k}_{\tilde{\Omega}} 
  + \braket{\Phi^k_i,\tilde{\psi}^k_s\partial_x\chi^k}_{\tilde{\Omega}}
  + (\psi^*)^k_l \braket{\Phi_i^k,\partial_x\Theta^k_l(x)}_{\tilde{\Omega}}
  + (\psi^*)_r^k\braket{\Phi_i^k,\partial_x\Theta^k_r(x)}_{\tilde{\Omega}}\nonumber\\
  \label{eq:weak:derivative:intermediate}
  &=& \braket{\Phi_i^k,\partial_x\psi^k}_{\Omega^k} 
  - \sqrbrace{(\psi^*)^k_l-\psi^k_l}\braket{\Phi_i^k,\delta_l^k}_{\tilde{\Omega}}
  + \sqrbrace{(\psi^*)_r^k-\psi_r^k}\braket{\Phi_i^k,\delta_r^k}_{\tilde{\Omega}}\nonumber\\
  &=& \braket{\Phi_i^k,\partial_x\psi^k}_{\Omega^k} + \Phi_i^k(x) \sqrbrace{(\psi^*)^k-\psi^k}\eval_{x^k_l}^{x^k_r}\nonumber\\
  \label{eq:weak:derivative}
  &=& \braket{\Phi_i^k,\partial_x\psi^k}_{\Omega^k} + \braket{\Phi_i^k,(\psi^*)^k-\psi^k}_{\partial\Omega^k},
\end{eqnarray}
\end{widetext}
where in the first step we employ the product rule; in the second, we
utilize equations \eqref{eq:def:selection:distribution},
\eqref{eq:def:dchi}, and \eqref{eq:dheaviside:1}; in the third, we
utilize the definition of the Dirac delta function; and in the final
step, we recognize $\braket{\phi,\psi}_{\partial\Omega^k}$ as the
integral over the boundary of $\Omega^k$, defined in the usual way.

Note that, although $\partial_x \tilde{\psi}^k_d$ is well-defined as
distribution, the product of distributions
$g(\psi)\partial_x\tilde{\psi}^k_d$, for an arbitrary nonlinear
function $g$, may not be. Therefore, although we have a distributional
derivative, we may not be able to use it to weakly define a system of
equations. This difficulty can be overcome on a system-by-system basis
via the work of, e.g., LeFloch et
al. \cite{maso1995definition,lefloch1988entropy,lefloch1989shock,hou1994nonconservative}
or Colombeau et
al. \cite{colombeau2000new,colombeau1988multiplications}. We are more
interested in a general framework which may be used with any
sufficiently well-behaved hyperbolic system. Therefore we do not
address this issue here.

\subsection{Moving to the Discrete}
\label{sec:movng:discrete}

Our description of the derivative so far assumes that the restriction
$\psi^k$ is an arbitrary smooth function. Therefore, to obtain a discrete
scheme, we must make an ansatz about $\psi^k$. We choose a
pseudospectral ansatz. We briefly discuss some of the details of this
ansatz in appendix \ref{sec:collocation}. For a review of pseudospectral
methods, see \cite{Grandclement2009}.

To make our method a Galerkin method, for each element $\Omega^k$, we
choose some $P^k\in\N$ and some subset $\{\Phi_i^k\}_{i=0}^{P^k}$ of
the test functions $\Phi_i^k$ and demand that any function $\psi^k$ is
a linear combination of the restriction of those test functions onto
$\Omega^k$:
\begin{equation}
  \label{eq:uk:ansatz}
  \psi^k(x) = \sum_{i=0}^{P^k} \hat{\psi}^k_i \Phi_i^k(x)\ \forall\ x\in\Omega^k,
\end{equation}
where $\hat{\psi}^k_i\in\R$ are spectral coefficients. Note that since
we have chosen our test functions to be the Legendre polynomials,
$P^k$ is also the highest-order polynomial which can be represented
within an element.

Then, we demand that the weak derivative relations
\eqref{eq:dheaviside:1} and \eqref{eq:weak:derivative} only hold for
all $0 \leq i \leq P^k$, rather than for all $k\in\N$. In this modal
representation, we can relate $\psi^k(x)$ to its derivative
$\partial_x\psi^k(x)$ in the standard way,
\begin{equation}
  \label{eq:ukprime}
  \partial_x\psi^k(x) = \sum_{i=0}^{P^k} \paren{\widehat{\partial_x\psi}}^k_i \Phi_i^k(x),
\end{equation}
where the spectral coefficients are given by
\begin{equation}
  \label{eq:modal:derivative}
  \paren{\widehat{\partial_x\psi}}^k_i = \sum_{j=0}^{P^k}\braket{\partial_x\Phi^k_i(x), \Phi^k_j(x)}_{\Omega^k} \hat{\psi}^k_j,
\end{equation}
which can be thought of as a change-of-basis operation. For
convenience, we define the matrix $\hat{d}^k$, whose components are
given by
\begin{equation}
  \label{eq:def:dmodal}
  \hat{d}^k_{ij} := \braket{\partial_x\Phi^k_i(x), \Phi^k_j(x)}_{\Omega^k}.
\end{equation}

We also construct a nodal representation. Within each element, we
assume a discrete set of points $\{x_i^k\}_{i=0}^{P^k}$ such that
$x_0^k=x_l^k$ and $x_{(P^k)}^k = x_r^k$. One good choice for
$\{x_i^k\}_{i=0}^{P^k}$ is the Gauss-Lobatto points of
$\{\Phi_i^k\}_{k=0}^{P^k}$. Given a set of values $\psi^k_i$,
$\psi^k(x)$ is assumed to interpolate the values such that
$\psi^k(x_i^k) = \psi^k_i$. In this nodal representation, the
element-wise inner product is approximated by the Gauss-Lobatto
quadrature rule
\begin{equation}
  \label{eq:def:quadrature:rule}
  \braket{\phi^k,\psi^k}_{\Omega^k} \approx \sum_{i=0}^{P^k} \phi^k_i \psi^k_i w_i^k,
\end{equation}
where $w_i^k$ are the discrete weights of the inner product. 

We emphasize that even when the continuum weights are trivial, as they
are for the Legendre polynomials, the discrete weights $w^k_i$ will
not be. Given a set of test functions and continuum weights, they are
determined by the locations of collocation points, as described in
appendix \ref{sec:collocation}. Figure \ref{fig:weights:abcissas}
shows the approximate values of the weights and locations of the
collocation points for an element of order $P^k=4$ and weight $h^k$.

\begin{figure}[tb!]
  \centering
  \begin{displaymath}
    \myvec{w}^k \approx \frac{h^k}{2} \left(\begin{array}{S[table-format=1.2]}0.1\\0.54\\0.71\\0.54\\0.1\end{array}\right)
    \text{ and }
    \myvec{x}^k \approx \frac{x^k_l + x^k_r}{2}
    + \frac{h^k}{2}\left(\begin{array}{S[table-format=2.2]}-1\\-0.65\\0\\0.65\\1\end{array}\right)
  \end{displaymath}
  \caption{The approximate weights $w^k_i$ and collocation points $x^k_i$ for an
    element of order $P^k=4$ and width $h^k$.}
  \label{fig:weights:abcissas}
\end{figure}

If we collect the $\phi^k_i$ and $\psi^k_i$ into vectors
$\myvec{\phi}^k$ and $\myvec{\psi}^k$ respectively, we can define the
quadrature rule as a matrix operation:
\begin{equation}
  \label{eq:quadrature:matrix}
  \braket{\phi^k,\psi^k}_{\Omega^k} = (\myvec{\phi}^k)^T w^k \myvec{\psi}^k,
\end{equation}
where $w^k_{ij} = w^k_i \delta_{ij}$. The inner product over the whole
domain remains the sum over the element-wise inner products as in
equation \eqref{eq:total:domain:inner:product}.

To move between the modal and nodal representations, we introduce the
generalized Vandermonde matrix,
\begin{equation}
  \label{eq:def:vandermonde}
  \V^k_{ij} := \Phi^k_j(x^k_i),
\end{equation}
such that the vector of nodal coefficients is obtained by applying the
Vandermonde matrix to the vector of modal coefficients:
\begin{equation}
  \label{eq:vandermonde:use}
  \myvec{\psi}^k = \V^k \myvec{\hat{\psi}}^k.
\end{equation}
This also gives us a nodal representation of element-wise
differentiation. We define the element-wise derivative operator as
\begin{equation}
  \label{eq:elementwise:derivative}
  d^k:= \V^k \hat{d}^k (\V^k)^{-1},
\end{equation}
which is nothing more than the standard pseudospectral derivative
operator for the domain $\Omega^k$. For reasons that will soon become
clear, we call $d^k$ the \textit{narrow} derivative operator.

At this point, we must make a choice about how to represent an inner
product that integrates over the boundary of a domain. As a guiding
principle, we will use the discrete analogue of integration by parts,
summation by parts, for the inner product within a single
element. In other words, we want the following to hold:
\begin{equation}
  \label{eq:integration:by:parts}
  \braket{\phi^k,\partial_x\psi^k}_{\Omega^k} + \braket{\partial_x \phi^k,\psi^k}_{\Omega^k} = \braket{\phi^k,\psi^k}_{\partial\Omega^k}.
\end{equation}
In the discrete case, we can write an inner product over the boundary
of an element as a standard inner product where one of the operands is
multiplied by a special matrix, $b^k$, which we call the
\textit{boundary operator}:
\begin{equation}
  \label{eq:boundary:inner:product}
  \braket{\phi^k,\psi^k}_{\partial\Omega^k} = (\myvec{\phi}^k)^T w^k b^k \myvec{\psi}^k = \paren{b^k \myvec{\phi}^k}^Tw^k\myvec{\psi}^k.
\end{equation}
The boundary operator is determined by equations
\eqref{eq:quadrature:matrix} and \eqref{eq:integration:by:parts} to be
given by
\begin{equation}
  \label{eq:def:b:operator}
  w^k b^k = w^k d^k + \paren{w^k d^k}^T,
\end{equation}
where $w^k$ and $d^k$ are the element-wise weight and differentiation
matrices respectively.

The matrix $b^k$ is essentially a discretization of the Dirac delta
function centred on the boundary of $\Omega^k$. It has components
given by
\begin{equation}
  \label{eq:form:bk}
  b^k_{ij} = \frac{2 b^k_{0}}{h^k} \paren{\delta_{i,P^k}\delta_{j,P^k} - \delta_{i,0}\delta_{j,0}},
\end{equation}
where $b_{0} > 0\in\R$ depends on $P^k$. The product $w^kb^k$ has the particularly simple form
\begin{equation}
  \label{eq:bk:wk:properties}
  \paren{w^k b^k}_{ij} = \delta_{i,P^k}\delta_{j,P^k} - \delta_{i,0}\delta_{j,0}
\end{equation}
so that the integral over the boundaries of an element matches the
continuum result,
\begin{displaymath}
  \braket{\phi^k,\psi^k}_{\partial \Omega^k} = \phi^k_r\psi^k_r - \phi^k_l \psi^k_l.
\end{displaymath}
Figure \ref{fig:bk:example} provides an example of $b^k$ in matrix
form for an element of width $h^k$ and order $P^k=4$.

\begin{figure}[!tb]
  \centering
  \begin{displaymath}
  b^k = \frac{1}{h^k}\left(\begin{array}{r r r r r}
      -20 & 0 & 0 & 0 & 0\\
      0 & 0 & 0 & 0 & 0\\
      0 & 0 & 0 & 0 & 0\\
      0 & 0 & 0 & 0 & 0\\
      0 & 0 & 0 & 0 & 20
      \end{array}\right)
  \end{displaymath}
  \caption{The matrix $b^k$ for an element of width $h^k$ and
    order $P^k = 4$.}
  \label{fig:bk:example}
\end{figure}

We now set about the task of defining the discrete analog of equation
\eqref{eq:weak:derivative} which, in our scheme, will replace the
pseudospectral derivative \eqref{eq:elementwise:derivative}. We must
first choose a definition of $\psi^*$, the quantity in equation
\eqref{eq:modified:restriction} that represents the information we
pull from a neighbouring element. There are a number of choices one
can make. However, we make the following, relatively simple,
choice. Let the suggestively named $(b^k)^{-1}F^k$ be any operator
that maps
\begin{eqnarray}
  \label{eq:def:b:inverse:J:1}
  \psi^k_l &\to& \psi^{k-1}_r\\
  \label{eq:def:b:inverse:J:2}
  \text{and }\psi^k_r&\to&\psi^{k+1}_l.
\end{eqnarray}
Note that the name $(b^k)^{-1}F^k$ is an abuse of notation, as $b^k$
has no inverse. We then define the \textit{fetch operator}
\begin{equation}
  \label{eq:def:J}
  F^k := b^k \sqrbrace{(b^k)^{-1}F^k},
\end{equation}
which performs the same role as $(b^k)^{-1}F^k$, but selects only
boundary terms for integration, making any non-boundary properties of
$(b^k)^{-1}F^k$ irrelevant. 

Like the boundary operator $b^k$, the fetch operator $F^k$ is
essentially a discretization of a delta function centred at the
boundary of $\Omega^k$. However, since functions are allowed to be
discontinuous at element boundaries, there is an ambiguity. The
boundary operator selects for the values of a function within an
element, the ``inside limit,'' while the fetch operator selects for
values of a function outside an element, the ``outside limit.'' If we
introduce the shorthand
\begin{equation}
  \label{eq:shorthand:indexing}
  x^{k-1}_r = x^k_{-1}\text{ and } x^{k+1}_l = x^k_{P^k+1},
\end{equation}
then the fetch operator has components
\begin{equation}
  \label{eq:jump:operator:explicit:form}
  F^k_{ij} = \frac{2 b^k_0}{h^k}\paren{\delta_{i,P^k}\delta_{j,P^k+1} - \delta_{i,0}\delta_{j,-1}}
\end{equation}
for all $i=0,1,\ldots,P^k$ and $j=-1,0,\ldots,P^k,P^k+1$.

Now let
\begin{equation}
  \label{eq:def:psi:star}
  (\psi^*)^k = \half\xi \sqrbrace{(b^k)^{-1}F^k}\psi^k - \half \psi^k
\end{equation}
so that
\begin{equation}
  \label{eq:def:b:psi:star}
  b^k (\psi^*)^k = \half\sqrbrace{\xi F^k\psi^k - b^k\psi^k},
\end{equation}
where $\xi\in\R$. Combining this choice of $(\psi^*)^k$ with equation
\eqref{eq:weak:derivative} results in the following discrete
element-wise, \textit{weak} derivative operator:
\begin{equation}
  \label{eq:def:Dk}
  \partial_x\tilde{\psi}^k = D^k \psi^k := \sqrbrace{d^k - \half b^k + \half\xi F^k}\psi^k,
\end{equation}
where we call $D^k$ the element-wise \textit{wide derivative} operator
because it takes information from neighbouring elements. $D^k$ is the
differential operator for OLDG methods.

In general we define the following convention. Any operator that takes
information from a \textit{single} element we call
\textit{narrow}. Any operator that takes information from an element
and its nearest neighbours, we call \textit{wide}. Narrow operators
will be represented by lower-case Latin symbols while wide operators
will be represented by upper-case Latin symbols. In the functional
notation of equations \eqref{eq:def:b:inverse:J:1} and
\eqref{eq:def:b:inverse:J:2}, the addition of wide and narrow
operators is unambiguous. In matrix notation, we simply pad the narrow
operator with columns of zeros so that the matrices are the same
shape.

\subsection{Summation By Parts}
\label{sec:methods:sbp}

We insist that our operator $D^k$ from equation \eqref{eq:def:Dk}
satisfy summation by parts. Given the strongly hyperbolic nature of
the BSSN system \cite{SarbachBSSNHyperbolicity}, we expect this
restriction to provide linear stability.

By construction, summation-by-parts is satisfied within each element
$\Omega^k$. However, stability proofs require integration over the
\textit{whole domain} $\Omega$. We define the wide derivative operator
over the whole domain $D$ such that, for all $\psi$ on $\Omega$,
\begin{equation}
  \label{eq:def:D}
  D\psi(x) = (D^k\psi^k)(x)\ \forall\ x\in \Omega^k\ \forall\ 1\leq k \leq K.
\end{equation}
We then seek a value of $\xi$, defined in equation
\eqref{eq:def:Dk}, such that, for all $\psi,\phi$,
\begin{equation}
  \label{eq:def:summation:by:parts:omega}
  \braket{\psi, D\phi}_\Omega + \braket{D\psi, \phi}_\Omega = \braket{\psi,\phi}_{\partial\Omega},
\end{equation}
where 
\begin{equation}
    \braket{\psi,\phi}_{\partial\Omega} = \psi\phi\eval_{x=X_r} - \psi\phi\eval_{x=X_l}.
\end{equation}
Note that there are two collocation points at the physical position
$X_l$: $x^0_r$ and $x^1_l$, and similarly for $X_r$. Therefore, we can
reasonably expect summation by parts to average over these two values
in some way. For example, we might accept the relationship
\begin{displaymath}
  \braket{\psi,\phi}_{\partial\Omega}  = 
  \half \sqrbrace{ \paren{\psi^K_r\phi^{K+1}_l+\psi^{K+1}_l\phi^K_r}
    -\paren{\psi^1_l\phi^0_r + \psi^0_r\phi^1_l }},
\end{displaymath}
which is the mixed average of left- and right-hand limits of $\phi$
and $\psi$ at the boundary. (This combination is arbitrary. Other
relationships might also be acceptable.)

We find that the only value of $\xi$ that satisfies summation by parts
is $\xi =1$. (See appendix \ref{sec:sbp:proof} for a proof.)
Therefore, the final version of the derivative operator is
\begin{eqnarray}
  \label{eq:def:Dk:final}
    D^k  &=& d^k - \half \sqrbrace{b^k -  F^k}.
\end{eqnarray}
Note that $\xi = 1$ may not be the only stable choice. Depending on
the continuum differential system, other values may result in a scheme
that is dissipative at element boundaries so that the energy norm is
non-increasing. Indeed, dissipation at element boundaries is typical
of DGFE schemes \cite{Cockburn2003}. However, we did not explore this
possibility.

In appendix \ref{sec:stability:wave:equation}, we provide an example
of how summation-by-parts can be used to demonstrate the stability of
an OLDG discretization of the wave equation in second-order form.

\subsection{Properties}
\label{sec:methods:properties}

Given a set of collocation points $x^k_i$, $D^k$ maps the $P^k+3$
points $\{x_r^{k-1}\}\cup\{x_i^k\}_{i=0}^{P^k}\cup\{x_l^{k+1}\}$ to
the $P^k+1$ points $\{x_i^k\}_{i=0}^{P^k}$. Physically, it maps a
function on the exterior faces and interior of an element $\Omega^k$
to a function defined only on the interior. In this picture the OLDG
stencil can be thought of as a finite differences stencil with
non-constant coefficients and a special, weak, boundary operator. If
we use our wide derivative to discretize the linear wave equation in
first-order form, we recover a standard DGFE method in the strong
formulation with a simple central flux. We demonstrate this in
appendix \ref{sec:lax-friedrich}. Therefore OLDG methods are truly a
generalization of current DGFE methods.

Figures \ref{fig:Dk:order2}, \ref{fig:Dk:order3}, and
\ref{fig:Dk:order4} show examples of $D^k$ for elements of width
$h^k$ and order $P^k=2$, 3, and 4 respectively. Here we use the
shorthand described in equation \eqref{eq:shorthand:indexing} such that
$i$ ranges from $0$ to $P^k$ and $j$ ranges from $-1$ to
$P^k+1$. The first and last columns pull information from the
neighbouring elements. Because of space constraints, with the
exception of the $P^k=2$ case, we show only the approximate numeric
values of the coefficients of $D^k$. The full values are available in
a public repository containing our supplemental materials
\cite{dgfeSupp}.

In the first column, only the first row is nonzero. And likewise in
the last column, only the last row is nonzero. This indicates that
collocation points in the neighbouring elements affect only the face
of the element $\Omega^k$. The internal diagonal of the matrix
vanishes, which indicates that the derivative of a function $\phi$ at
a collocation point $x$ is independent of the value of $\phi$ at
$x$. Finally, because $D^k$ represents a first derivative with respect
to $x$, $D^k\phi \to -D^k\phi$ as $x\to -x$. This is reflected in the
symmetry properties of the matrix, which obeys the relationship:
\begin{equation}
  \label{eq:Dk:symmetry:properties}
  D^k_{ij} = -D^k_{(P^k-i)(P^k-j)}
\end{equation}
for all $i=0,1,2,\ldots,P^k$ and $j=-1,1,2,\ldots,P^k+1$. These properties
are generic in our scheme.

\begin{figure}[tb]
  \centering
  \begin{displaymath}
   D^k = \frac{1}{h^k}
   \left(\begin{array}{r r r r r}
      -3&0&4&-1&0\\
      0&-1&0&1&0\\
      0&1&-4&0&3
      \end{array}\right)
  \end{displaymath}
  \caption{The wide derivative operator $D^k$ for elements with a
    Legendre basis, polynomial order $P^k=2$, and element width $h^k$.}
  \label{fig:Dk:order2}
\end{figure}

\begin{figure}[tb]
  \centering
  \begin{displaymath}
    D^k \approx \frac{1}{h^k}
    \left(\begin{array}{r S[table-format=3.2] S[table-format=3.2]
        S[table-format=3.2] S[table-format=3.2] r}
      -6&0&8.09&-3.09&1&0\\
      0&-1.62&0&2.24&-0.62&0\\
      0&0.62&-2.24&0&1.62&0\\
      0&-1&3.09&-8.09&0&6
    \end{array}\right)
  \end{displaymath}
  \caption{The wide derivative operator $D^k$ for elements with a
    Legendre basis, polynomial order $P^k=3$, and element width $h^k$. }
  \label{fig:Dk:order3}
\end{figure}

\begin{figure}[t!]
  \centering
  \begin{displaymath}
    D^k \approx \frac{1}{h^k}
    \left(\begin{array}{r S[table-format=3.2] S[table-format=3.2] S[table-format=3.2] S[table-format=3.2] S[table-format=3.2] r}
      -10&0&13.51&-5.33&2.82&-1&0\\
      0&-2.48&0&3.49&-1.53&0.52&0\\
      0&0.75&-2.67&0&2.67&-0.75&0\\
      0&-0.52&1.53&-3.49&0&2.48&0\\
      0&1&-2.82&5.33&-13.51&0&10
    \end{array}\right)
  \end{displaymath}
  \caption{The wide derivative operator $D^k$ for elements with a
    Legendre basis, polynomial order $P^k=4$, and element width $h^k$. }
  \label{fig:Dk:order4}
\end{figure}

\subsection{Consistency}
\label{sec:consistency}

It is important to check that, if we differentiate a smooth function,
our OLDG derivative converges to the continuum derivative in the
separate limits of $h^k\to 0$ and $P^k\to\infty$. We call this
property \textit{consistency}.

To check consistency, suppose we seek to approximately differentiate a
smooth function $\phi$. Then
$$D^k \phi^k = d^k \phi^k$$
and our OLDG derivative reduces to a pseudospectral derivative within
the domain $\Omega^k$. Therefore the standard results for spectral
methods apply and, as long as the physical solution is smooth,
equation \eqref{eq:def:Dk:final} provides a consistent approximation
of a derivative
\cite{bernardi1992polynomial,schwab1998p,hesthaven2008nodal}.  More
generally, if the continuum function $\phi$ is continuous at element
boundaries, the spectral consistency results hold. If the continuum
solution is \textit{discontinuous} at element boundaries, then the
wide component of our operator contributes and the issue is more
delicate. We do not address it here.

% For non-smooth problems, consistency is a more delicate issue. Our
% distributional construction is only valid when each $\phi^k$ is
% differentiable. In this setting, it may be possible to establish
% consistency by utilizing the work of LeFloch et
% al. \cite{maso1995definition,lefloch1988entropy,lefloch1989shock} or
% by relying on a mapping between OLDG methods and more traditional DGFE
% methods. However, we do not address this problem here.

\subsection{Higher Derivatives}
\label{sec:higher:derivatives}

So far, we have only described how to approximate the continuum
operator $\partial_x$. However, we'd also like to approximate higher
derivatives. To approximate the second derivative $\partial_x^2 \psi$ of
a function $\psi$, we introduce an auxiliary variable
\begin{equation}
  \label{eq:def:du}
  \nabla\psi := D\psi,
\end{equation}
where $D$ is the wide derivative operator defined in equation
\eqref{eq:def:D}, which is not evolved, but calculated globally
whenever a second derivative is required.  We then calculate the
second derivative as
\begin{equation}
  \label{eq:def:d2u:dx2}
  \ddII{x} \psi \approx D \nabla\psi .
\end{equation}
This may not be the most efficient way of calculating higher
derivatives. However, it was the only generic approach we could find that
produced second derivatives compatible with the first derivatives
defined in section \ref{sec:methods:sbp} in the summation by parts
sense:
\begin{equation}
  \label{eq:second:order:sbp}
  \braket{\psi,D^2 \phi}_\Omega + \braket{D\psi, D\phi}_\Omega = \braket{\psi,D\phi}_{\partial\Omega},
\end{equation}
where $D^2$ refers to a wide second derivative operator.

We note that this approach to higher derivatives does come at a
price. Since we must know $\nabla\phi$ on neighbouring elements before
we can calculate $\partial_x^2\phi$, we must wait for this calculation
to finish before proceeding with our calculation of the
right-hand-side. This reduces the parallelism of the scheme.

\subsection{Truncation}
\label{sec:truncation}

In this section, we make a connection to a recent development in the
LDG methods of Shu and collaborators. Consider the linear first-order
in time, second-order in space wave equation defined on the interval
$[X_l,X_r]$,
\begin{equation}
  \label{eq:linear:wave:equation}
  \begin{aligned}
  \pd{\phi}{t} &= \psi\\
  \pd{\psi}{t} &= c^2 \partial_x^2 \phi,
  \end{aligned}
\end{equation}
where $\phi,\psi\in L_2(\Omega)$ are subject to appropriate initial
and boundary conditions. We seek to discretize equations
\eqref{eq:linear:wave:equation} using our DGFE scheme. This means
replacing $\phi$ and $\psi$ by approximate representations $\phi^k$
and $\psi^k$ made up of the basis functions $\Phi_i^k$. We abuse
notation here and denote by $\psi^k$ not only the restriction to
element $\Omega^k$ but also the discretization.

Naively, one would use the same number of basis functions to represent
$\phi$ and $\psi$ within each element. And indeed this is precisely
the procedure Cockburn and Shu propose in \cite{LDG1}. In
\cite{LDGTruncate}, Xing et al. show that if $\psi$ is represented as
a piecewise polynomial of order $P^k=p\in\N$ in each element
$\Omega^k$, but $\phi$ is represented only as a piecewise polynomial
of order $P^{k}=p-1$, it is possible to obtain an energy conserving
method that is superconvergent such that the error in the solution is
of order $\Ord{(h^k)^{P_k+3/2}}$. Whether or not a DGFE method is
superconvergent depends strongly on the choice of numerical flux. For
linear first-order systems, superconvergence is expected
\cite{hesthaven2008nodal} but it is not always achieved in LDG methods
\cite{LDGReview}.

Motivated by this choice, we represent $\phi$ as a polynomial of order
$P^k=p$ but we represent $\psi$ as a polynomial of order $P^k=p-1$.
This matches with the observation due to Richardson
\cite{richardson1911approximate} that in the numerical solution of a
hyperbolic PDE, the $(h^k)^{-n}$ term in the solution introduced by
differentiation $n$ times is exactly cancelled by multiplications by
$h^k$ due to time integration (with a CFL factor). $\phi$ is $\psi$
integrated, so it is multiplied by $h^k$, reducing the error. We do
not obtain superconvergence, but we have found this procedure to
significantly improve convergence and stability. 

% We implement this scheme by a ``truncation'' operation. Both $\psi$
% and $\phi$ live on the same grid of collocation points. But at each
% timestep and in each element $\Omega^k$, we multiply $\psi$ by our
% truncation operator
% \begin{equation}
%   \label{eq:def:truncation:operator}
%   \mathcal{T}^k = (\V^k)^{-1} \hat{\mathcal{T}}^k\V^k,
% \end{equation}
% where $\V^k$ is the Vandermonde matrix and 
% \begin{equation}
%   \label{eq:def:truncation:modal}
%   \hat{\mathcal{T}}^k_{ij} = \delta_{ij} - \delta_{iP^k}\delta_{jP^k},
% \end{equation}
% where $P^k$ is the polynomial order within the $k^{th}$ element. 

More generally, given a system of equations, 
\begin{displaymath}
  \partial_t \myvec{\psi} = \Lo\sqrbrace{\myvec{\psi},\partial_j \myvec{\psi},\partial_j\partial_k \myvec{\psi}},
\end{displaymath}
we call all variables with no spatial derivatives in their
right-hand-sides \textit{primary variables} and all other variables
\textit{conjugate variables}. Within an element, we represent all
variables as linear combinations of the $P^k$ basis
functions. However, at each integrator substep and for each conjugate
variable, we set the spectral coefficient associated with the
$(P^k)^{\text{th}}$ mode of a conjugate variable to zero. We call this
procedure \textit{truncation}. The process of truncation acts as
artificial dissipation, improving the nonlinear stability of our
scheme. (OLDG methods are linearly stable, so the stability cannot be
improved.)  Truncation also eliminates terms of order
$\Ord{(h^k)^{P^k+1}}$ in our pointwise error estimates, providing
cleaner convergence results.

\subsection{Time Integration}
\label{sec:time:integration}

With our derivatives defined, we integrate our scheme via the method
of lines and an explicit time integrator such as the fourth-order
total variation diminishing Runge-Kutta scheme proposed by Gottlieb
and Shu \cite{gottlieb1998total} or the 8(7) scheme due to Dormand and
Prince \cite{dormand1980dupri,prince1981high}.

Unfortunately, the timesteps for a DGFE method must
be smaller than for a finite differences method. Although there is no
proof for high-order elements \cite{cockburn1999discontinuous}, the
time step size for DGFE methods for global timestepping is believed to obey
\begin{equation}
  \label{eq:CFL:factor}
  \Delta t \leq \min_{0< k \leq K}\Delta t^k
\end{equation}
where 
\begin{equation}
  \label{eq:def:delta:t:k}
  \Delta t^k := C_{\text{CFL}} \frac{h^k}{\paren{P^k+1}^2}
\end{equation}
where $C_{\text{CFL}}>0\in\R $ is a constant that depends on the system being solved
\cite{hesthaven2008nodal}. (The bound was proven for polynomial order
$P^k=1$ \cite{cockburn1991runge}.) In our simulations we choose 
\begin{equation}
  \label{eq:our:delta:t}
  C_{\text{CFL}} \le 0.45
\end{equation}
however larger timesteps may be possible.  

There exist several techniques for increasing the CFL factor beyond
that implied by equations \eqref{eq:CFL:factor} and
\eqref{eq:def:delta:t:k}, which we do not explore but which may be of
use to the interested reader. For more details on these techniques,
see
\cite{kosloff1993modified,warburton2008taming,cockburn1999discontinuous}
and references therein. Local time-stepping in the context of DGFE
methods is discussed in \cite{qiu2005lotalTimeStepping}. We do not
implement local time-stepping, however the implementation for OLDG
methods should be the same as for standard DGFE methods.

\subsection{Convergence}

We do not prove convergence for the BSSN system. However, we know from
section \ref{sec:consistency} and appendix
\ref{sec:stability:wave:equation} that an OLDG discretization of the
linear wave equation is both consistent and stable. Therefore by the
Lax-Richtmyer theorem \cite{lax1956survey}, it is convergent, with
error bounded by the standard pseudospectral consistency bounds
\cite{bernardi1992polynomial,schwab1998p,hesthaven2008nodal}.

We can also obtain stronger, pointwise bounds. In appendix
\ref{sec:convergence} we show that, for the linear, second-order wave
equation \eqref{eq:linear:wave:equation} with arbitrary initial
conditions, we have
\begin{equation}
  \label{eq:wavetoy:convergence:formula}
  \phi^k_i = \phi(t,x^k_i) + \ME^k(t,x^k_i) \paren{h^k}^{P^k},
\end{equation}
where $\phi^k_i$ is the numerical solution in element $\Omega^k$ with
element width $h^k$ and element order $P^k$ at collocation point
$x^k_i$. $\phi$ is the true solution, and $\ME^k$ is a function, which
may depend on the location of the collocation points $x^k_i$ but is
independent of $h^k$ and $P^k$, that determines the error. In this
proof we show that convergence does not depend on whether or not we
truncate as described in section \ref{sec:truncation}.

We argue that these convergence results provide analytic evidence that
discretizations of more general systems are convergent to the
appropriate order when discretized with our OLDG stencil. The
numerical experiments discussed in section \ref{sec:tests} support our
claim.

\subsection{Filtering}
\label{sec:filtering}

It is well known that a numerical scheme that is linearly stable may
become unstable when used to solve a nonlinear system of equations,
even when those equations are well-posed. In spectral and DGFE
methods, we can interpret the loss of stability as emerging from the
fact that the interpolating polynomial representing the derivative of
a function is not the same as the derivative of an interpolating
polynomial. This is usually called \textit{aliasing error}. Stability
can often be restored by \textit{filtering} the spectral coefficients
to remove energy from the short-wavelength modes
\cite{kreiss1979stability}.

These filtering techniques were originally motivated by the need to
capture shocks in nonlinear flux-conservative systems and, more
generally, to efficiently represent discontinuous functions spectrally
\cite{Vandeven1991a}. However, they are often required even when the
solution is smooth. If the system is nonlinear, aliasing error can
drive an instability. The BSSN equations are no exception
\cite{tichy2006,tichy2009}. 

One can think of the truncation scheme described in section
\ref{sec:truncation} as a type of filter and we have found that, in
many situations it provides all the required dissipation. However,
since truncation is a projection-type operation, it is not tunable. We
therefore develop a more traditional, tunable, filtering scheme which
can be utilized if necessary.

We choose a modification of the spectral viscosity technique developed
by Tadmor and collaborators for pseudospectral methods
\cite{tadmor1989convergence,maday1989analysis,tadmor1990shock,schochet1990rate}. Consider
the semi-discrete system of the form
\begin{equation}
  \label{eq:nonlinear:PDE}
  \pdd{t} \myvec{\phi} = \Lo\sqrbrace{\myvec{\phi},D\myvec{\phi},D^2\myvec{\phi}},\ x\in [X_l,X_r],
\end{equation}
where $\Lo$ is some nonlinear operator that acts on $\myvec{\phi}$ and
its first two wide OLDG derivatives, as defined in equation
\eqref{eq:def:Dk:final}, subject to appropriate initial and boundary
data. To make contact with traditional pseudospectral methods, we
initially assume that $K=1$ such that there is a single element, of
width $h^k=h$ and order $P^k=p$. In this limit, our discontinuous
Galerkin scheme becomes a pseudospectral method with weak boundary
conditions. Once we have developed filtering in this setting, we will
generalize to the full case.

Tadmor modifies equation \eqref{eq:nonlinear:PDE} by including an
artificial dissipation term which vanishes in the continuum limit:
\begin{eqnarray}
  \label{eq:nonlinear:PDE:artificial:viscosity}
  \pdd{t} \myvec{\phi} &=& \mathcal{L}\sqrbrace{\myvec{\phi},D\myvec{\phi},D^2\myvec{\phi}}\\
  &&+ \epsilon_p (-1)^{s+1}\sqrbrace{\partial_x(1-x^2)\partial_x}^s \myvec{\phi},\nonumber
\end{eqnarray}
where $s$ is the so-called order of the dissipation and strength of
dissipation, and $\epsilon_p$ varies as
\begin{equation}
  \label{eq:epsilon_p:def}
  \epsilon_p \sim \frac{C_s h}{p^{2s-1}}.
\end{equation}
The constant $C_s$ depends on the regularity of the solution
$\myvec{\phi}$. Roughly, it should be
$$C_s\sim \max_{0\leq k\leq s}\|\myvec{\phi}\|_\infty^k,$$
where $\|\myvec{\phi}\|_\infty$ is the infinity-norm of
$\myvec{\phi}$. Under these conditions, Tadmor shows that equation
\eqref{eq:nonlinear:PDE:artificial:viscosity} converges spectrally to
the true solution. By inspection, we can see that Tadmor's spectral
viscosity technique is the spectral analog to the artificial
dissipation proposed by Kreiss and Oliger \cite{kreiss1973methods}.

Tadmor's spectral viscosity technique is roughly equivalent to
filtering the modes $\hat{\myvec{\phi}}$ of $\myvec{\phi}$ via the exponential
filter first proposed by Vandeven \cite{Vandeven1991a}. In this case,
the spectral representation of $\phi$
$$\phi(x) = \sum_{i=0}^p \hat{\phi}_i \Phi_i(x),$$
where $\Phi_i$ are the test functions, becomes
\begin{equation}
  \label{eq:filtered:phi}
  \phi^\sigma(x) = \sum_{i=0}^p \sigma\paren{\frac{i}{p}}\hat{\phi}_i \Phi_i(x),
\end{equation}
where
\begin{equation}
  \label{eq:exponential:filter}
  \sigma(\eta) := \exp\sqrbrace{-(C_s p \Delta t) \eta^s},
\end{equation}
where $\Delta t$ is the discrete time step used in evolution.

On the other hand, applying an exponential filter to the modes
$\hat{\phi}$ is equivalent to solving equation
\eqref{eq:nonlinear:PDE} but also solving the ordinary differential
equation
\begin{equation}
  \label{eq:mode:damping:equation}
  \frac{d}{dt} \hat{\phi}^i = - \epsilon_p i^{2s} \hat{\phi}^i,\ t\in [0,\Delta t]
\end{equation}
for all $0 \leq i \leq p$ at each time step. We claim that equations
\eqref{eq:nonlinear:PDE} and \eqref{eq:mode:damping:equation} need not
be solved in separate steps and that the artificial viscosity
formulation of Tadmor \eqref{eq:nonlinear:PDE:artificial:viscosity}
can be well approximated by ``filtering the right-hand-side'' as
\begin{equation}
  \label{eq:filtering:rhs}
    \pdd{t} \myvec{\phi} = \mathcal{L}\sqrbrace{\myvec{\phi},D\myvec{\phi},D^2\myvec{\phi}}  - C_s p \V \F \V^{-1} \myvec{\phi},
\end{equation}
where $\V$ is the Vandermonde matrix defined in equation
\eqref{eq:def:vandermonde} and
\begin{equation}
  \label{eq:def:F}
  \F_{ij} := \paren{\frac{i}{p}}^{2s}\delta_{ij}
\end{equation}
is a diagonal matrix defining the decay coefficients of the modal
representation of $\phi$. 

Alternatively, we can use a modified version of $\F$:
\begin{equation}
  \label{eq:def:F:mod}
  \F_{ij} := \begin{cases} 0&\text{if } i/p \leq \eta_\mathrm{crit}\\
    \delta_{ij}\sqrbrace{\frac{i/p - \eta_\mathrm{crit}}{1 - \eta_\mathrm{crit}}}^{2s}&\text{ else}
    \end{cases},
\end{equation}
where $0 \leq \eta_\mathrm{crit} < 1$
\cite{hesthaven2008nodal}. Equation \eqref{eq:def:F:mod} does not
precisely correspond to the viscosity term provided in equation
\eqref{eq:nonlinear:PDE:artificial:viscosity}. Rather, it filters only
the higher-order modes. Ideally, this is less destructive to the
accuracy of the solution. We are currently using equation
\eqref{eq:def:F:mod} in our implementation, but further investigation
is necessary to determine what approach is best.

Of course, a DGFE method usually has more than one
element, and we would like the dissipation term in equation
\eqref{eq:filtering:rhs} to scale appropriately with the number of
elements. We therefore modify our dissipation term to the final form
\begin{eqnarray}
  \label{eq:filtering:rhs:final}
    \pdd{t} \myvec{\phi}^k &=& \mathcal{L}\sqrbrace{\myvec{\phi}^k,D^k\myvec{\phi}^k,(D^k)^2\myvec{\phi}^k}  \\
    &&\qquad - C_s\frac{P^k}{h^k} \V^k \F^k \paren{\V^k}^{-1}\myvec{\phi}^k,\nonumber
\end{eqnarray}
where we have re-introduced the indexing for the element
$\Omega^k$. In the full DGFE scheme, $C_s$ remains a global quantity,
independent of the grid and $\F^k$ generalizes to multiple elements in
the obvious way. Since the average distance between nodes is
approximately $\Delta x^k := h^k/P^k$, the right-hand-side of equation
\eqref{eq:filtering:rhs:final} manifestly has the appropriate units of
$1/\Delta x^k$ for a hyperbolic problem.

% Although we have no formal proof that this
% formulation is stable and convergent, our numerical experiments have
% indicated that it is flexible and robust.

\section{Asymptotic Properties}
\label{sec:asymptotics}

In this section we analyze the computational costs of OLDG methods
and compare them to finite differences.

\subsection{Floating Point Operations for First Derivatives}
\label{sec:asymptotics:flops}

Here we ask how many floating point operations are required to
approximate $\partial_x\phi$ for some function $\phi$. For simplicity,
suppose a three-dimensional domain with $K$ elements on a side, for
$K^3$ elements total. To make contact with finite differences, each
element is of the same order $P^k=p$, $p$ even, such that the number
of collocation points on a side is $n = (p+1)K$. Further suppose that
we are evolving only one variable.

Our in-element wide derivative operator is dense with vanishing
diagonal. Therefore, in one dimension, a first derivative requires
$$2\sqrbrace{(p+1)^2  - (p+1) + 2} = 2 \sqrbrace{(p+1) p + 2}$$
floating point operations per element for the multiplication of the
length $p+3$ in-element state vector by our differentiation
matrix. (The overall factor of 2 comes from the fact that add and
multiply are separate operations.) In three dimensions and over the
whole domain, this translates to
\begin{eqnarray}
  \label{eq:nflops:dg:1}
  \NF_{\DG}^{(1)} &:=& 2K^3 (p+1)^2 \sqrbrace{(p+1)p+2}\nonumber\\
  &=& 2\frac{n^3}{p+1}\sqrbrace{(p+1)p+2}\nonumber\\
  &=& 2n^3 \sqrbrace{p + \frac{2}{1+p}}
\end{eqnarray}
floating point operations. In contrast, a $p^{th}$-order finite
differences stencil requires
\begin{equation}
  \label{eq:nflops:fd:1}
  \NF^{(1)}_{\FD} := 2n^3 p
\end{equation}
floating point operations for a first derivative. 

Figure \ref{fig:flops:overhead} plots $\NF_{\DG}^{(1)}/\NF_{\FD}^{(1)}$,
which tells us how much more a DGFE derivative costs compared to a
finite differences derivative. To leading order, both DGFE and finite
differences stencils require a number of operations equal to
$\Ord{n^3p}$. However, the DGFE method has sub-leading terms which
will contribute significantly when $p$ is small and which become
negligible when $p$ is large.

\subsection{Floating Point Operations for First and Second
  Derivatives}
\label{sec:asymptotics:flops:2nd}

For most wave-like systems such as the BSSN system, we need both the
first and second derivatives of variables in the state vector. We
therefore ask how many floating point operations are required to
approximate both
\begin{equation}
  \label{eq:first:and:second:derivs}
  \partial_i \phi \text{ and } \partial_i\partial_j \phi,\ i,j=1,2,3
\end{equation}
for some continuum variable $\phi$. We make the same
assumptions here as in section \ref{sec:asymptotics:flops}. 

In the OLDG case, we take a first derivative, store
it, and calculate a second derivative. Therefore the cost to
approximate quantity \eqref{eq:first:and:second:derivs} is just the
cost of calculating the three first derivatives of $\phi$
and then the cost of differentiating each of those quantities for a
total of
\begin{eqnarray}
  \label{eq:flops:dg:2}
  \NF_{\DG}^{(1,2)} &:=& 3 \NF_{\DG}^{(1)} + 6 \NF_{\DG}^{(1)}\nonumber\\
  &=& 18n^3 \sqrbrace{p+\frac{2}{1+p}} = \Ord{n^3 p}
\end{eqnarray}
floating point operations.

Finite differences differentiation could be performed the same way,
but it is typically not done. Usually $\partial^2_i \phi$
and $\partial_i\partial_j \phi,\ i\neq j$ are calculated
independently as full stencils without any intermediate steps or
storage. With this approach, approximating
$\partial_i\partial_j\phi,\ i \neq j$ costs $2n^3p^2$ operations
for each combination of $i$ and $j$. Approximating
$\partial^2_i\phi$ to the same order of accuracy requires
an extra two operations for the additional non-zero stencil
point. So, for
all three directions, the cost of approximating quantity
\eqref{eq:first:and:second:derivs} is
\begin{eqnarray}
  \label{eq:flops:fd:2}
  \NF_{\FD}^{(1,2)} &:=& 3 \NF_{\FD}^{(1)} + 6 n^3 (p+1) + 6 n^3 p^2\nonumber\\
  &=& 6n^3\sqrbrace{p+ (p+1) + p^2}\nonumber\\
  &=& 6 n^3 \paren{p + 1}^2 = \Ord{n^3 p^2}
\end{eqnarray}
operations.

\begin{figure}[tb]
  \centering
  \includegraphics[width=\columnwidth]{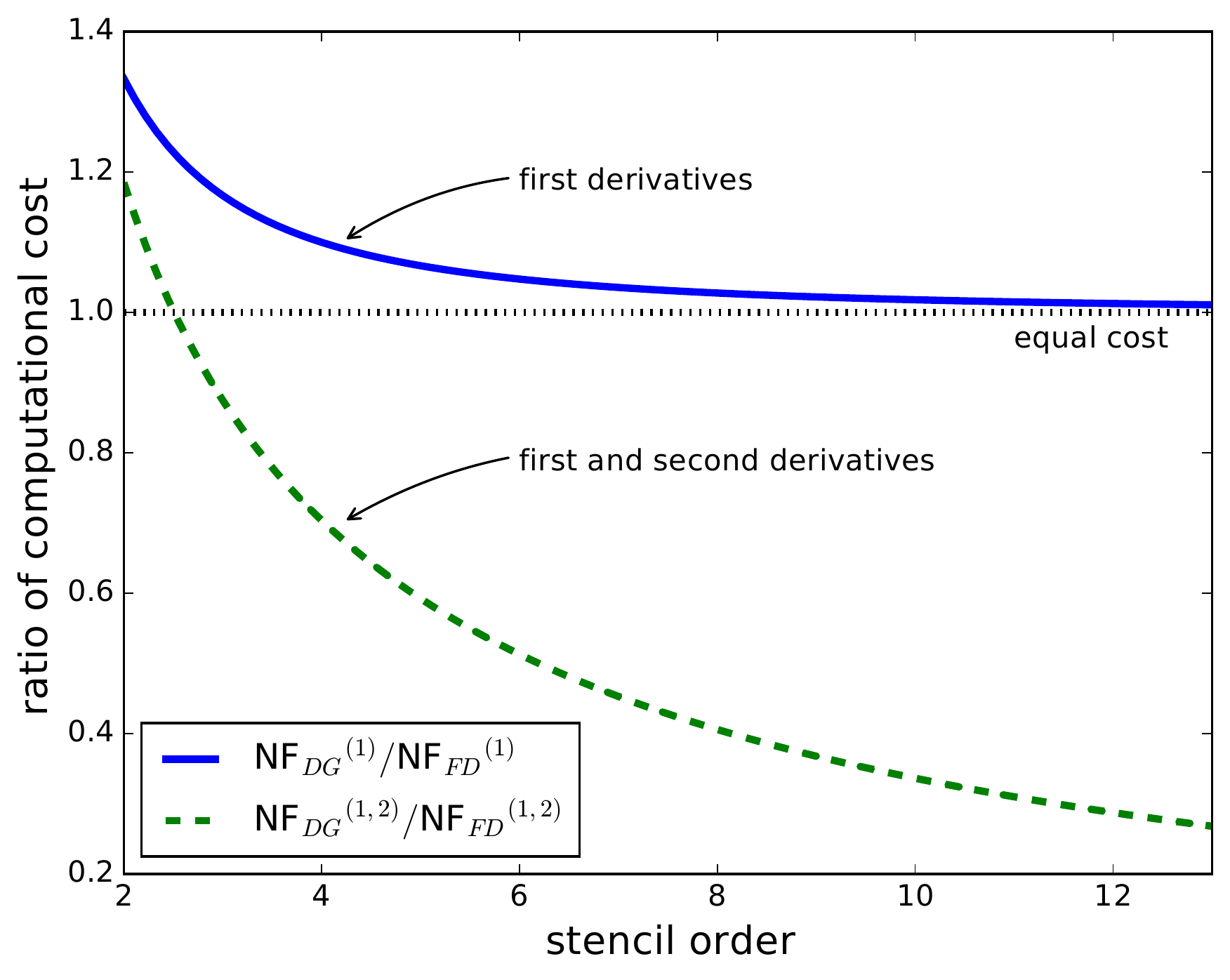}
  \caption{The ratio of the computation cost in floating point
    operations of our DGFE stencil compared to
    finite differences for the approximation of a first derivative
    (blue) and both first and second derivatives (green). The black
    line shows equal cost.}
  \label{fig:flops:overhead}
\end{figure}

Figure \ref{fig:flops:overhead} plots
$\NF_{\DG}^{(1,2)}/\NF_{\FD}^{(1,2)}$, which tells us how much more
(or less) it costs to calculate both first and second derivatives
using our OLDG stencil compared to finite differences. For $p\geq 3$,
the cost of calculating all first and second derivatives for a
function is larger for finite differences as usually implemented than
for OLDG methods.

The standard finite differences implementation trades
computational cost for memory and communication overhead.
We note that finite differences implementations could calculate
first derivatives and store them, just as we do in OLDG methods.
See e.g. \cite{hu2015model} for much more advanced FD
stencil algorithms that greatly reduce memory access cost.
We are
unable to make this trade-off in our OLDG scheme because we need to
respect both the weak boundary conditions between elements and
summation-by-parts over the whole domain.

\subsection{Communication Cost}
\label{sec:communication:cost}

We now investigate the communication costs in a distributed memory
environment. We are interested in \textit{strong scaling}: given a
three-dimensional problem of fixed size $S^3$, where $S$ is the number
of collocation points in a single dimension, across how many
separate memory domains can we efficiently distribute the calculation?

\begin{figure}[tb]
  \centering
  \includegraphics[width=0.87\linewidth]{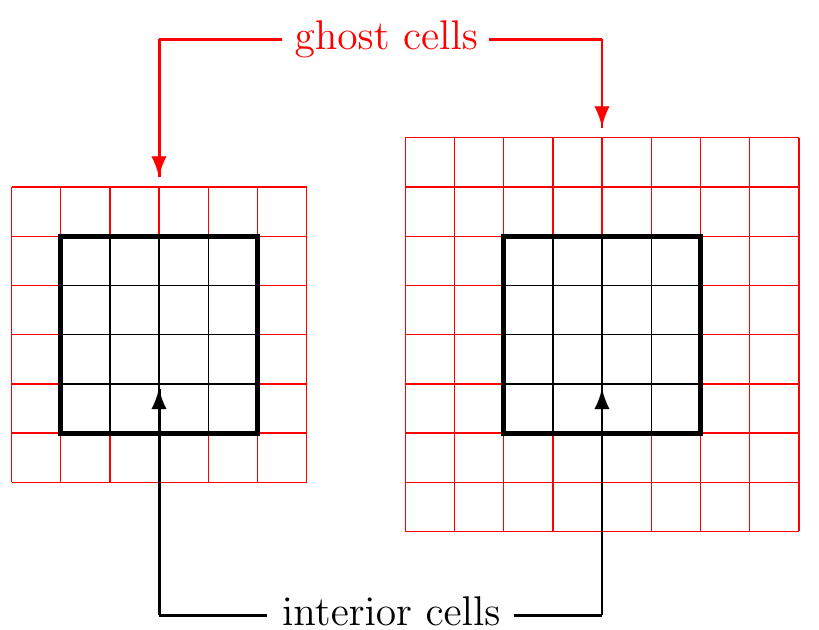}
  \caption{A two-dimensional slice of the collocation cells on a
    single cpu.  \textbf{Left:} DGFE methods of all orders have only
    one layer of ghost cells. \textbf{Right:} fourth-order finite
    differences methods use two layers of ghost cells.}
  \label{fig:cells:plot}
\end{figure}

To differentiate at a collocation point $g$, a finite differences
stencil of order $p$, $p$ even, needs $p/2$ collocation points on each
side. For distributed memory, this translates to a layer of unevolved
``ghost cells,'' $p$ cells deep around the border of the memory
domain, as shown in figure \ref{fig:cells:plot}. These cells are
synchronized whenever differentiation is required and cause a
bottleneck for parallelization.

In contrast, a DGFE method requires a layer of ghost
cells only \textit{one} cell deep, no matter the order of the method,
as shown in figure \ref{fig:cells:plot}. This stems from the fact
that, to differentiate within an element, one only needs data from the
\textit{boundaries} of neighbouring elements. The distributional
derivative of the DGFE approximation couples the
elements only weakly.

We quantify communication overhead by calculating the ratio
\begin{equation}
  \label{eq:def:overhead}
  \OH := \frac{N_{\text{ghost cells}}}{N_{\text{interior cells}}},
\end{equation}
or the number of ghost cells divided by the number of interior
cells. This depends on the total problem size $S^3$ and the number of
memory domains $D$ across which we want to distribute our
problem. Additionally, in the finite differences case, it depends on
the order of the method.

The number of interior cells is always $\floor{S^3/D}$. (It is
impossible to have a fractional number of interior cells. However, for
brevity of notation, we will suppress the floor term from now on.)
This translates to an overhead of
\begin{equation}
  \label{eq:fd:overhead:comm:1}
  \OH_{\FD} = \frac{D}{S^3}\paren{\frac{S}{D^{1/3}}+p}^3-1
\end{equation}
for finite differences. At scale, $S^3$ and $D$ are of the same order,
so even a moderate $p$ such as $p=4$ can produce very large
overheads. In contrast, DGFE methods have an overhead of
\begin{equation}
  \label{eq:dg:overhead:comm:1}
    \OH_{\DG} = \frac{D}{S^3}\paren{\frac{S}{D^{1/3}}+2}^3-1
\end{equation}
independent of the order of the element. 

Since our method is an LDG method, we introduce extra communication
steps. We communicate when we calculate both first and second
derivatives and when we perform the truncation operation described in
section \ref{sec:truncation}. This does not change the ratio of ghost
to interior cells, but is an additional communication cost.

\begin{figure}[tb]
\begin{center}
  \includegraphics[width=\linewidth]{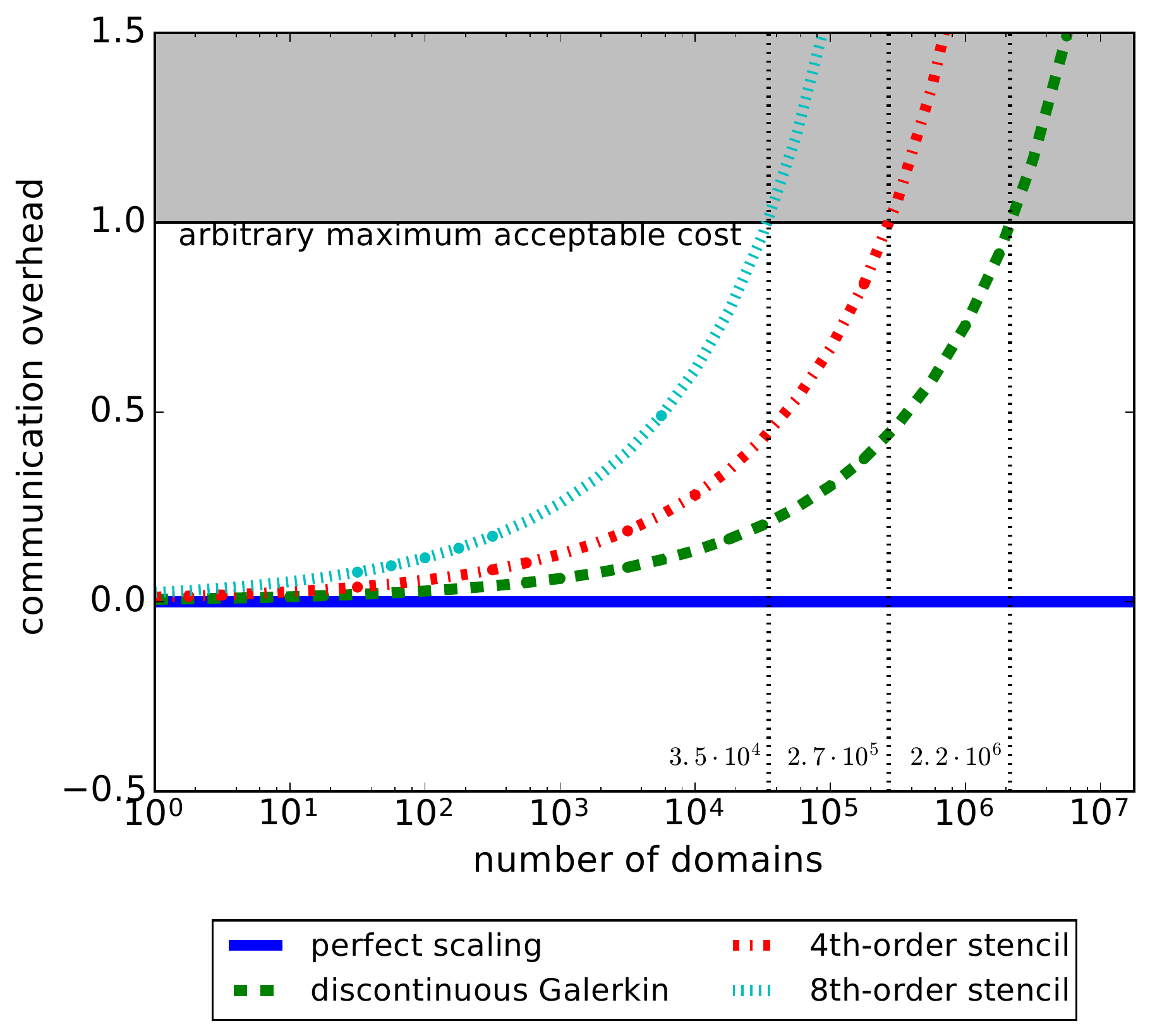}
\end{center}
\caption{The communication overhead for computing the solution to a 3D
  problem of fixed size $1000^3$ cells compared between DGFE methods
  and fourth- and eighth-order finite differences methods. For
  comparison, we also include the unrealistic cost for an approach
  that scales perfectly.}
\label{fig:cost:plot}
\end{figure}

As a concrete example we plot in figure \ref{fig:cost:plot} the
overhead associated with a fixed problem size of $S^3=1000^3$
collocation points and for different values of $D$. We compare fourth-
and eighth-order finite differences stencils with discontinuous
Galerkin methods of any order. Perfect strong scaling has a constant
overhead of 0 (blue line). If one arbitrarily assumes a maximum
acceptable overhead of 1.0, meaning we have as many ghost cells as
interior cells, then the eighth-order stencil scales to about $D =
3.5\times 10^4$ domains, fourth-order to $D = 2.7\times 10^5$, and
DGFE to $D = 2.2\times 10^6$. In this particular situation, DGFE
stencils of any order scale about ten times further than fourth-order
finite differences.

\subsection{Memory Access Cost}
\label{sec:memory:access}

Loading values from memory is a costly operation; accessing memory has
on today's systems a latency more than a hundred times larger than a
floating point operation. It is thus important that as many memory
load operations as possible can be served from a cache. To allow this,
one simple optimization method arranges loops in such a way that one
first loads a block of collocation points into the cache, and then
performs as many operations as possible on this block without
requiring additional memory accesses. This is called \emph{loop
  blocking}.

Here we assume ideal loop blocking, and then calculate how many memory
accesses are necessary to calculate a derivative. The small number of
ghost zones required by DGFE methods also serves to improve
performance.

To calculate a derivative at a collocation point $g$, its respective
neighbours must also be present. This means that when we calculate a
derivative within a block of collocation points, we once again have
interior and ghost cells. To quantify the additional memory access
cost due to ghost cells, we define a \textit{memory access overhead}
\begin{equation}
  \label{eq:def:cache:overhead}
  \MO := \frac{N_{\text{ghost cells}}}{N_{\text{interior cells}}}.
\end{equation}
Here we calculate the memory access overhead for an idealized L3 cache
of fixed size $C$ per process for OLDG methods and for finite
differences.

A double-precision number requires $8\,\text{Bytes}$, so a cache of
size $C$ can contain
\begin{equation}
  \label{eq:cache:grid:points:on:a:side}
  N_{\text{total}} := N_{\text{ghost cells}}+N_{\text{interior cells}} =
  \frac{C}{V\; (8\,\B)}
\end{equation}
total points, where $V$ is the total number of variables required. If
we define $l$ such that
\begin{equation}
  \label{eq:def:l:memory:access}
  N_{\text{interior cells}} =: l^3,
\end{equation}
then the total number of cells is 
\begin{equation}
  \label{eq:def:ghost:for:fd}
  N^{\FD}_{\text{total}} = (l+p)^3
\end{equation}
for a finite differences stencil of order $p$ and
\begin{equation}
  \label{eq:def:ghost:for:dg}
  N^{\DG}_{\text{total}} = (l+2)^3
\end{equation}
for DGFE methods at any order. If we solve for $l$ we find
that
\begin{equation}
  \label{eq:l:fd}
  N^{\FD}_{\text{interior cells}} = \paren{N_{\text{total}}^{1/3} - p}^3
\end{equation}
for finite differences and 
\begin{equation}
  \label{eq:l:dg}
  N^{DG}_{\text{interior cells}} = \paren{N_{\text{total}}^{1/3} - 2}^3
\end{equation}
for DGFE methods. This gives us a memory access overhead of 
\begin{equation}
  \label{eq:cache:overhead:fd}
  \MO_{\FD} = \frac{N_{\text{total}}^{\FD}}{\sqrbrace{\paren{N_{\text{total}}^{\FD}}^{1/3}-p}^3}-1
\end{equation}
for finite differences and
\begin{equation}
  \label{eq:cache:overhead:dg}
  \MO_{\DG} = \frac{N_{\text{total}}^{\FD}}{\sqrbrace{\paren{N_{\text{total}}^{\FD}}^{1/3}-2}^3}-1
\end{equation}
for DGFE methods.

As a concrete example, we calculate the memory access overhead for the
BSSN system and a realistic cache size of $C=1.5\,\text{MByte}$ per
core. Here we ignore details of a realistic cache and assume a simple
$1.5\,\text{MB}$ ``container.'' The Einstein Toolkit
\cite{loffler2012einstein,EinsteinToolkit:web,EinsteinToolkit:ascl}
implementation of the BSSN system has 24 evolved variables
\cite{AlcubierreConformalDecomp,AlcubierreGammaDriver}. The cache must
also contain the right-hand-side and everything we need to calculate
it. In the case of finite differences, this is just the state vector
and the right-hand-side. In the case of our discontinuous Galerkin
scheme, this includes both first derivatives and temporary variables
for the truncation operation described in section
\ref{sec:truncation}. Therefore we find that the BSSN system requires
\begin{equation}
  \label{eq:V:BSSN}
  V^{\BSSN}_{\FD} = 48\text{ and }V_{\DG}^{\BSSN} = 92
\end{equation}
for finite differences and OLDG stencils respectively. 

\begin{figure}[tb]
  \centering
  \includegraphics[width=\columnwidth]{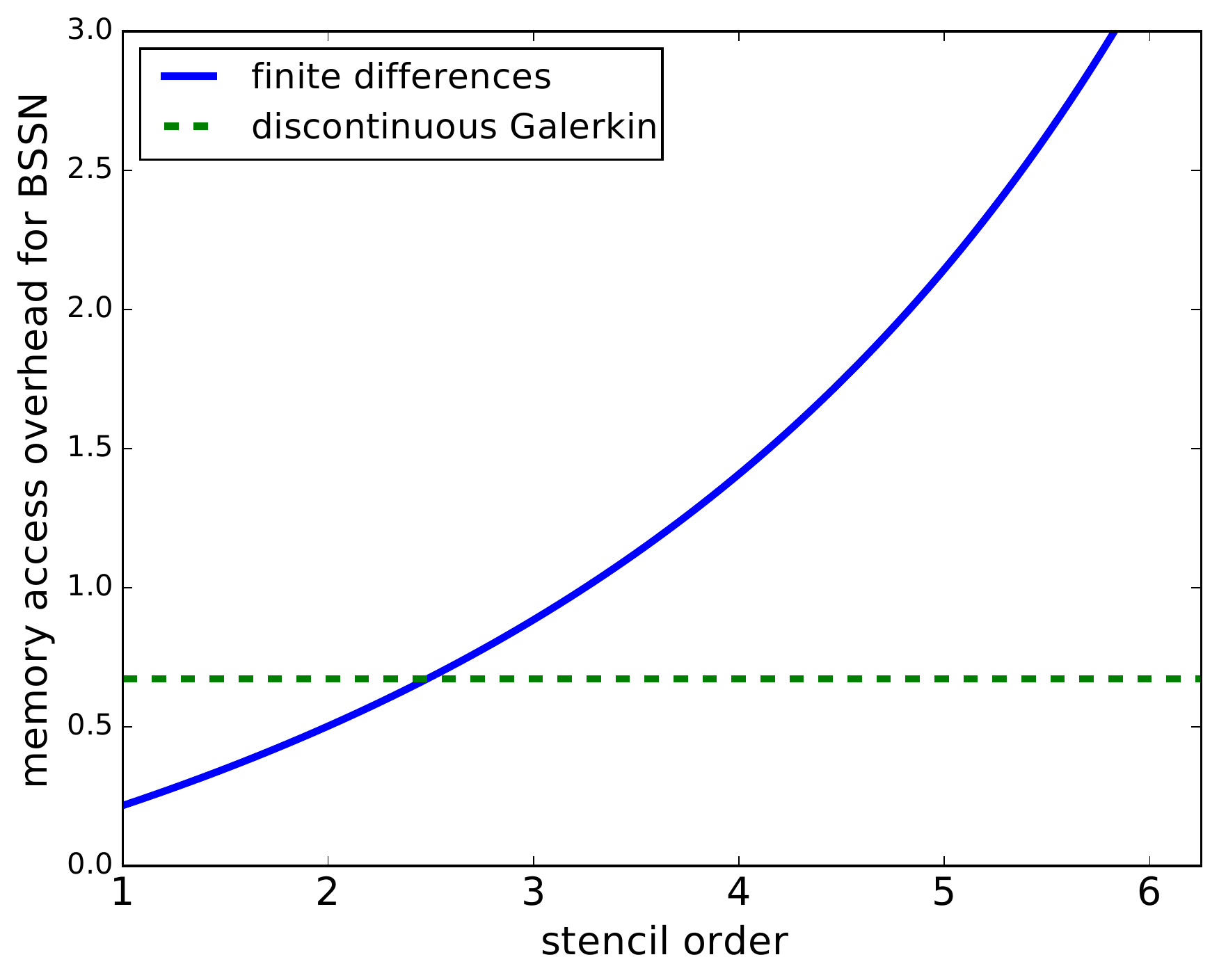}
  \caption{The memory access overhead for the BSSN system with a fixed
    cache size of $C=1.5$ MB for both finite differences and DGFE
    stencils as a function of the stencil order. We assume the BSSN
    system requires 48 variables per collocation point for finite
    differences and 92 variables per collocation point for DGFE}
  \label{fig:cache:plot}
\end{figure}

Given these assumptions, we plot the memory access overhead points as
a function of the stencil order for both finite differences and DGFE
stencils in figure \ref{fig:cache:plot}. At secnd-order, and for this
cache size, finite differences utilizes the cache slightly better. But
at higher order, our dicontinuous Galerkin scheme becomes
significantly more efficient. 

% We note that we have ignored the fact
% that, in the case of DGFE methods, interior points must come in
% $(p+1)^3$ chunks, where $p$ is the order, as elements cannot be broken
% up.

% 
% Here we investigate memory usage in our desired use-case: the BSSN
% equations. We calculate the number of grid points that can fit in an
% L3 cache of fixed size per core $C_{L3}$ given the BSSN state vector
% for both DGFE and finite differences stencils. Once
% again assume that each element is of the same order $P^k=p$, $p$
% even. Further suppose that the total number of elements that fit in
% the L3 cache is $K$ such that the total number of grid points is $n =
% (p+1)K$.
% 

\section{Numerical Tests}
\label{sec:tests}

We have implemented our OLDG scheme as a thorn in the Einstein Toolkit
\cite{loffler2012einstein,EinsteinToolkit:web,EinsteinToolkit:ascl},
using Kranc for code generation \cite{Husa:2004ip,Kranc:web}.  We
provide our implementation, which is based on the McLachlan thorn
\cite{Brown:2008sb,McLachlan:web}, in a public repository
\cite{MclachlanDGFE2}.  We emphasize that our implementation is a
proof-of-concept implementation and has not yet been optimized for
performance. We are currently testing a more efficient implementation.

To establish the basic numerical properties of our method, we first
investigate its applicability in the context of the second-order wave
equation \eqref{eq:linear:wave:equation} of section
\ref{sec:tests:we}. We then investigate in the context of the BSSN
equations by performing some of the standard Apples-With-Apples tests
\cite{AwA1,AwA2}, which we discuss in sections
\ref{sec:tests:robust:stability}, \ref{sec:tests:gauge:wave}, and
\ref{sec:tests:gamma:driver}. 

The original Apples With Apples tests were intended to test not only
the stability and convergence of a code, but also the formulation of
the Einstein equations on which that code is based. In this work, we
are interested only in establishing the numerical properties of our
scheme. Thus we only perform those Apples with Apples tests which probe
the numerical scheme rather than the formulation.

We discuss both stability and convergence. All our numerical tests are
performed with truncation as described in section \ref{sec:truncation}
but \textit{without} the filtering described in section
\ref{sec:filtering}.

\subsection{The Second-Order Equation}
\label{sec:tests:we}

Recall the second-order wave equation \eqref{eq:linear:wave:equation}:
\begin{eqnarray}
  \pd{\phi}{t} &=& \psi\nonumber\\
  \pd{\psi}{t} &=& c^2 \partial_x^2\phi.\nonumber
\end{eqnarray}
A DGFE method has two types of resolution: the order of the polynomial
interpolant within each element and the total number of elements in
the domain $K$. In appendix \ref{sec:convergence}, we show that the
pointwise error obeys equation \eqref{eq:wavetoy:convergence:formula}:
\begin{displaymath}
  \phi^k_i = \phi(t,x^k_i) + \ME^k(t,x^k_i) \paren{h^k}^{P^k},
\end{displaymath}
however, because DGFE methods are defined only in a
weak sense, and because the positions of collocation points change with
resolution, equation \eqref{eq:wavetoy:convergence:formula} best
translates to the following statement over the whole domain
\begin{equation}
  \label{eq:wavetoy:convergence:formula:norm2}
  \|E[\phi]\|_2 := \|\phi^k-\phi\|_2 \leq \|\ME^k\|_2 \paren{h^k}^{P^k},
\end{equation}
where $\|\phi\|_2$ is the 2-norm of $\phi$ over the domain. (Here we
abuse notation and allow $\phi^k$ to not only represent the
restriction of $\phi$ onto the element $\Omega^k$ but the numerical
solution.)

To investigate convergence of OLDG methods, we numerically solve
equation \eqref{eq:linear:wave:equation} with different numbers of
elements and different in-element orders (the order stays fixed to
$P^k=p$ over the whole grid.) The former is called
$h$-\textit{refinement}. The latter is called
$p$-\textit{refinement}. We solve equation
\eqref{eq:linear:wave:equation} in 1D over the domain
$x\in\sqrbrace{-\frac{1}{2},\frac{1}{2}}$ with periodic boundary
conditions and initial conditions of the form
\begin{eqnarray}
  \label{eq:wave:equation:initial:data}
  \phi(t=0,x) &=& A \sin(2\pi k x)\\
  \psi(t=0,x) &=& -A \omega \cos(k x),
\end{eqnarray}
where $\omega = \sqrt{k^2 c^2}$ for some amplitude $A$ and wavenumber
$k$. For simplicity, we fix $c=k=A=1$. We plot our error at $t=0.75$.

\begin{figure}[tb]
  \centering
  \includegraphics[width=\columnwidth]{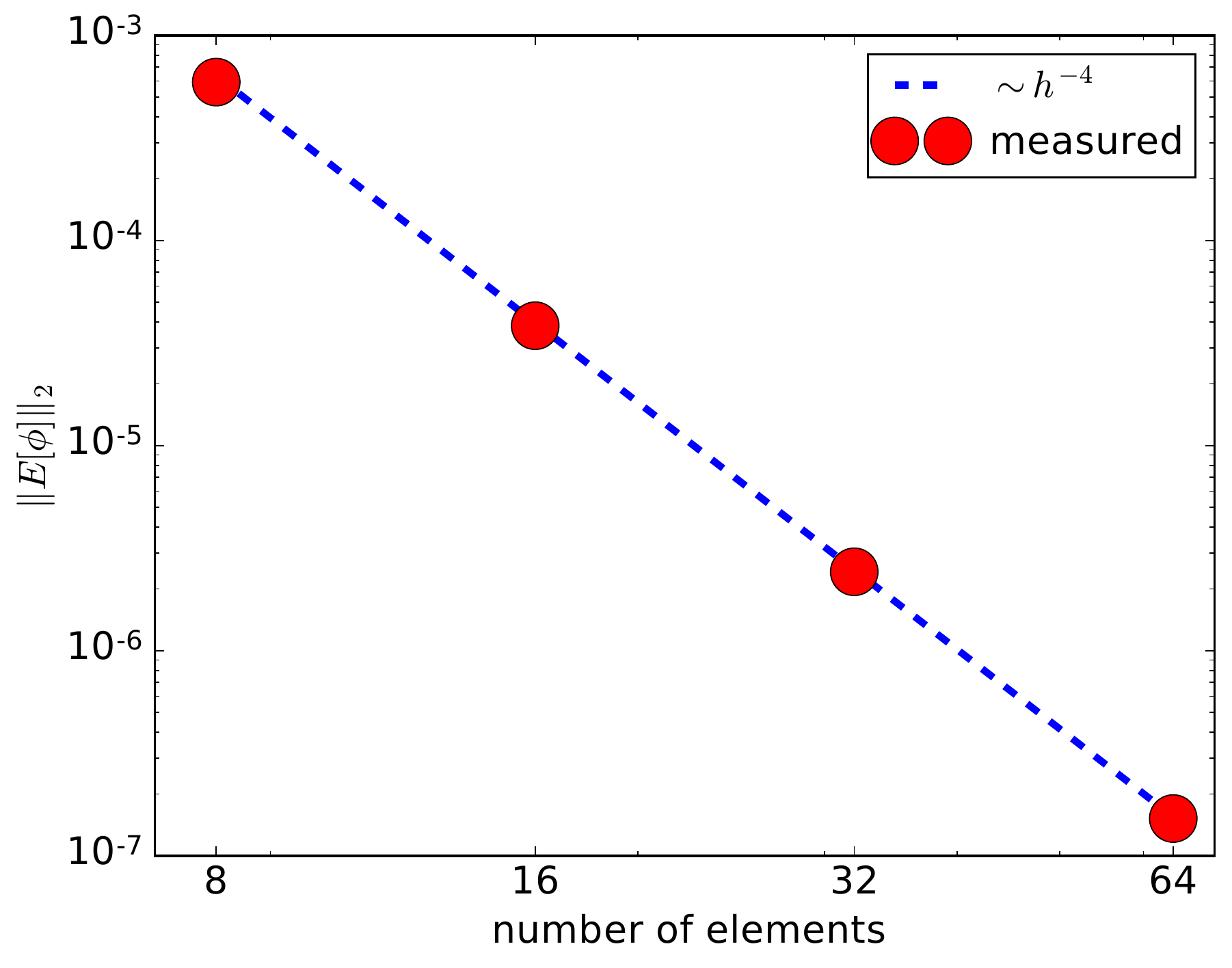}
  \caption{The $L_2$ error $\|E[\phi]\|_2$ of $\phi$ in the
    second-order wave equation \eqref{eq:linear:wave:equation} under
    $h$-refinement for fourth-order elements at time $t=0.75$.}
  \label{fig:we:h-refinement}
\end{figure}

Figure \ref{fig:we:h-refinement} shows the convergence under
$h$-refinement using fourth-order elements. When we double the number
of elements, we halve the element width $h$. Equation
\eqref{eq:wavetoy:convergence:formula:norm2} tells us that we should
see the error scale as $h^{-p}$, or $h^{-4}$ in this case. And indeed
measurements confirm this prediction.

\begin{figure}[tb]
  \centering
  \includegraphics[width=\columnwidth]{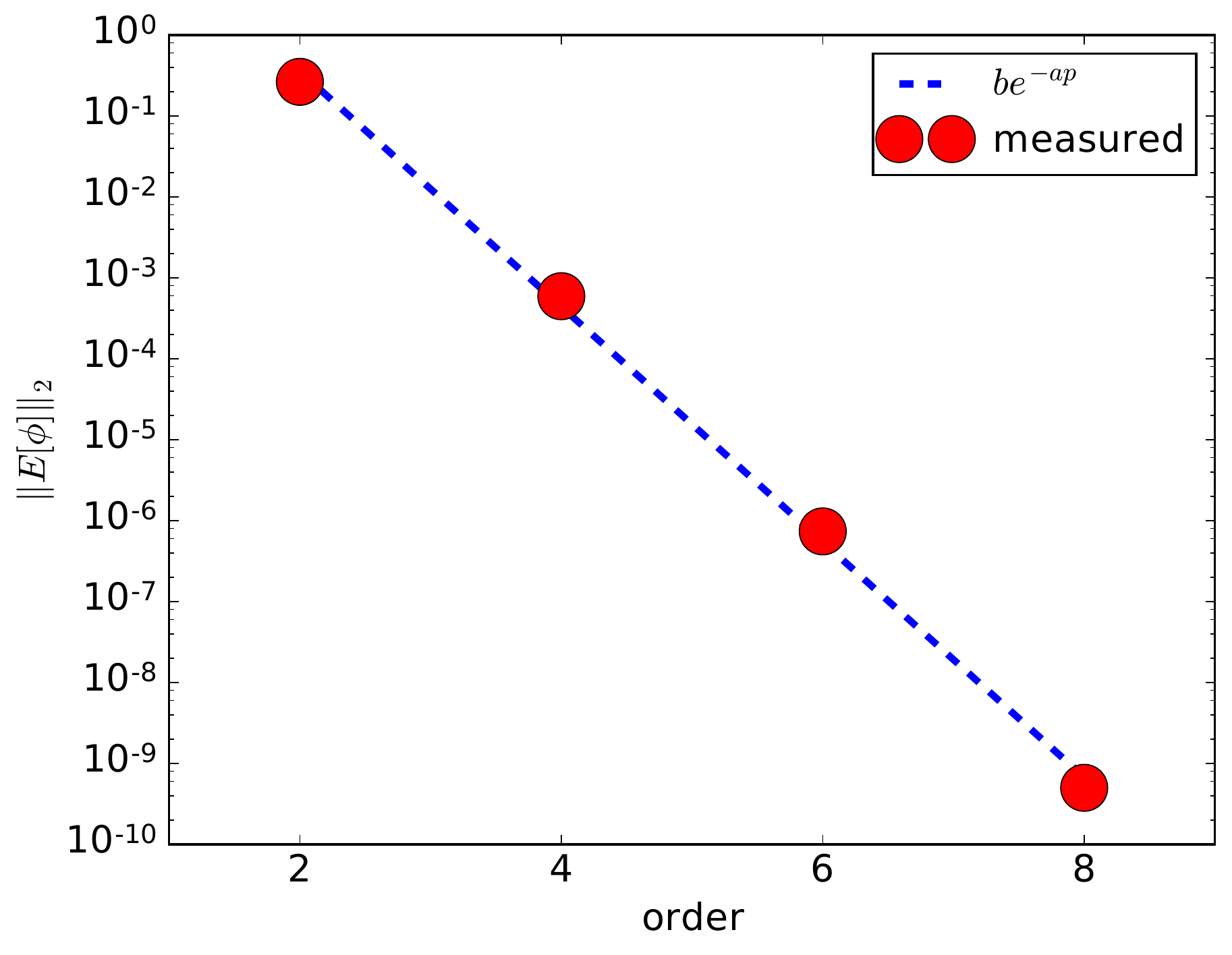}
  \caption{The $L_2$ error $\|E[\phi]\|_2$ of $\phi$ in the
    second-order wave equation \eqref{eq:linear:wave:equation} under
    $p$-refinement with a fixed number of 8 elements at time $t=0.75$.}
  \label{fig:we:p-refinement}
\end{figure}

Figure \ref{fig:we:p-refinement} shows the convergence under
$p$-refinement with a fixed element width $h=1/8$. The convergence
given in equation \eqref{eq:wavetoy:convergence:formula:norm2} now
translates to exponential decay with $p$. This rapid convergence rate is
often called ``spectral'' or ``evanescent'' convergence. Our measured
convergence agrees with this expectation. We measure the convergence
rate to be
$$\|E[\phi]\|_2 = b e^{-a p}$$
for $b \approx e^{5.66}$ and $a \approx 3.34$.

\begin{figure}[tb]
  \centering
  \includegraphics[width=\columnwidth]{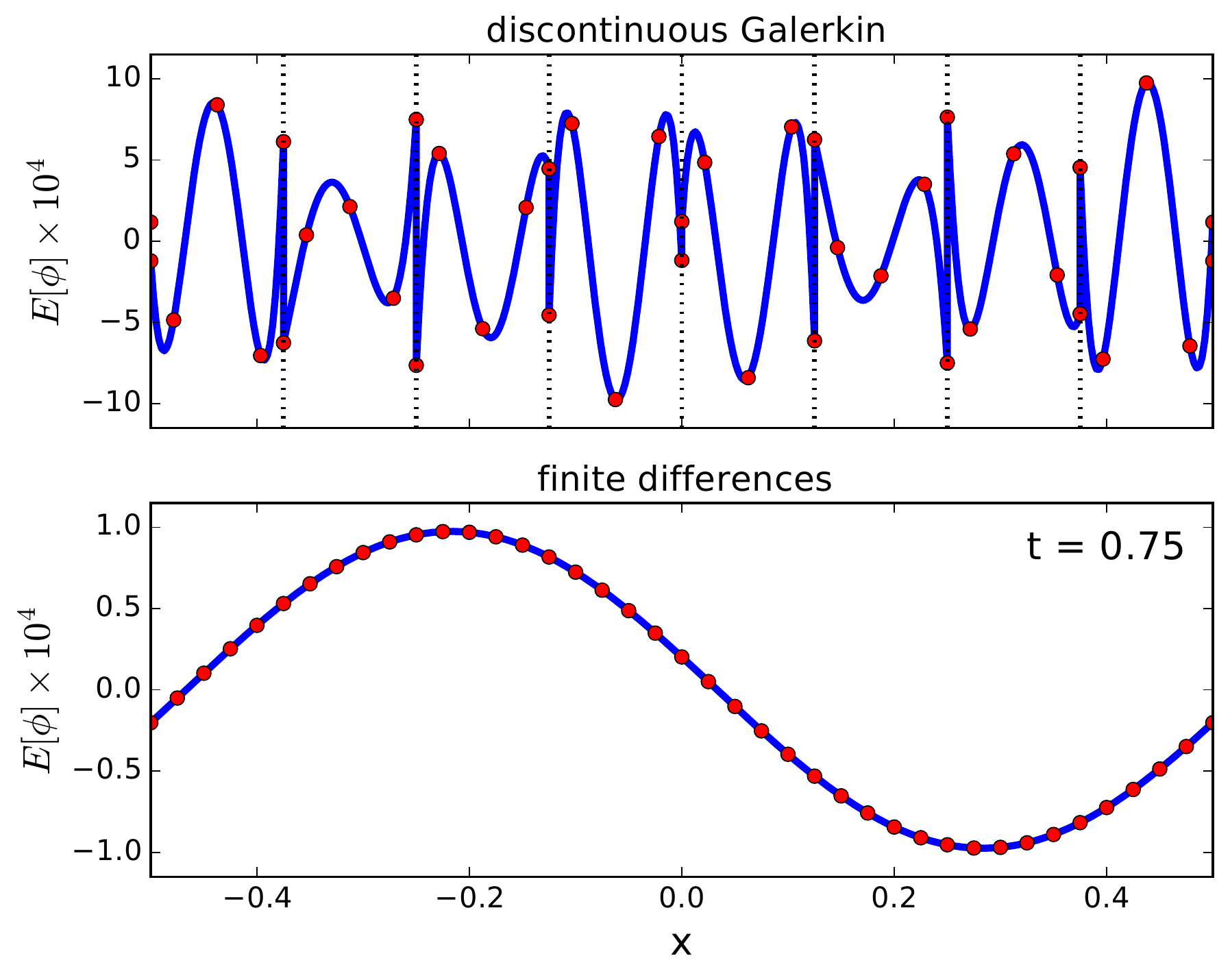}
  \caption{The pointwise errors for the second-order wave equation
    \eqref{eq:linear:wave:equation} with OLDG (top) and finite
    differences (bottom) stencils respectively after 0.75 wave
    periods. These simulations were run with fourth-order stencils and
    40 collocation points (or 8 elements).}
  \label{fig:dg:vs:fd:pointwise}
\end{figure}

\begin{figure}[!tb]
  \centering
    \includegraphics[width=\columnwidth]{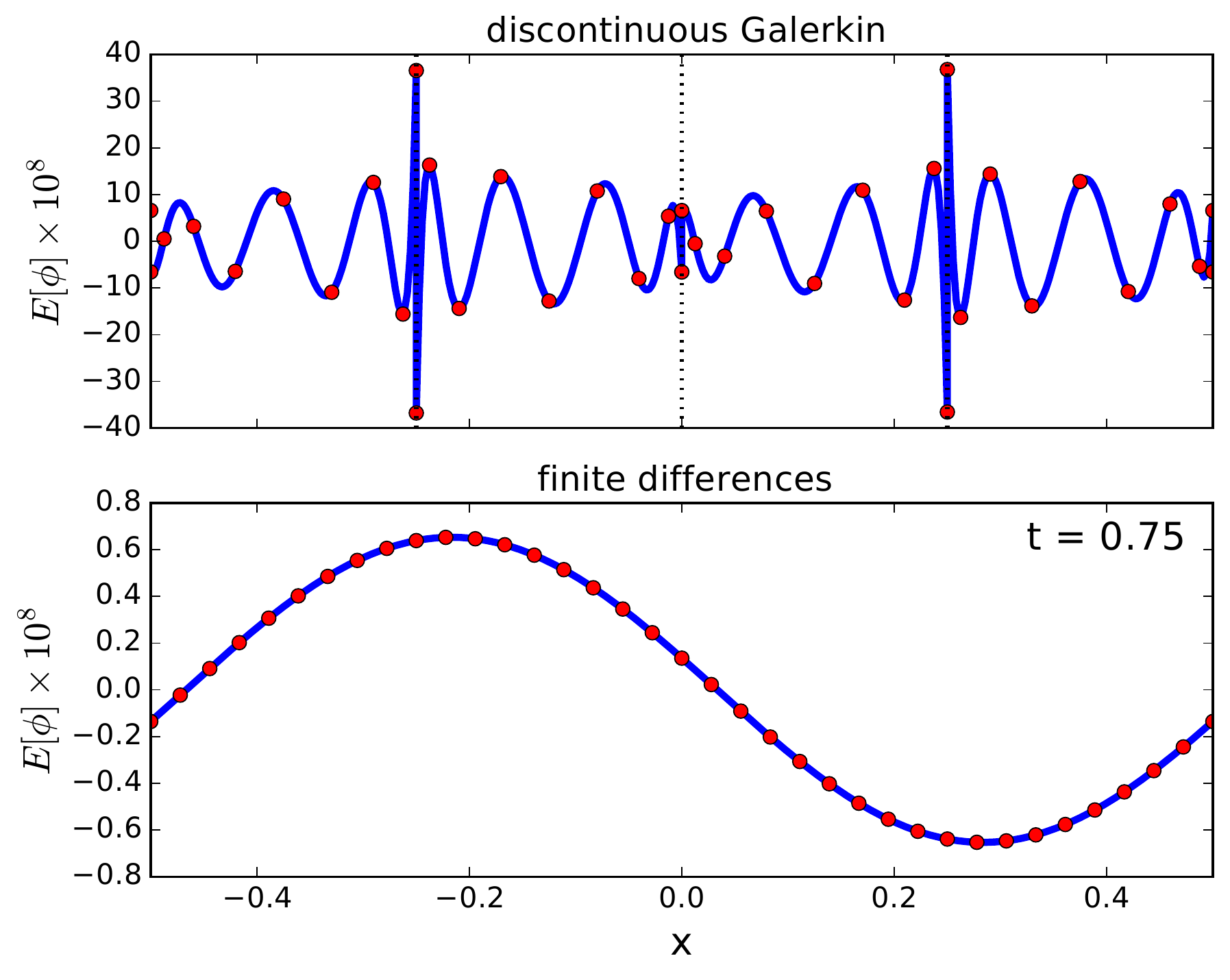}
  \caption{The pointwise errors for the second-order wave equation
    \eqref{eq:linear:wave:equation} with OLDG (top) and finite
    differences (bottom) stencils respectively after 0.75 wave
    periods.
    (Note the differences in scale.)
    These simulations were run with eighth-order stencils and 36
    collocation points (or 4 elements).}
  \label{fig:dg:vs:fd:pointwise:8th:order}
\end{figure}

In figures \ref{fig:dg:vs:fd:pointwise} and
\ref{fig:dg:vs:fd:pointwise:8th:order}, we compare the pointwise error
of our OLDG approach with the pointwise error of a finite differences
scheme. For figure \ref{fig:dg:vs:fd:pointwise}, we use a fourth-order
stencil and 40 collocation points (or 8 elements). For figure
\ref{fig:dg:vs:fd:pointwise:8th:order}, we use an eighth-order stencil
and 36 collocation points (or 4 elements). The curves are generated by
using fourth-order and eighth-order interpolation respectively. In the
OLDG case, this interpolation corresponds to the modal representation
within an
element. The dots are measured values at the collocation points. For
the OLDG stencil, the vertical lines show element boundaries. For
consistency, all simulations were run with the same CFL factor and
time integrator.

For the OLDG stencil, element boundaries are visible by eye as
locations where the function is no longer smooth. We find the
pointwise error for the OLDG stencil is worse than for the finite
differences stencil of the same order and resolution. For fourth-order
stencils, the OLDG error is about ten times worse than the finite
differences error. For eighth-order stencils it is almost fifty times
worse at element boundaries. This error can be mitigated somewhat by
the post-processing technique discussed in section
\ref{sec:postprocessing} below. 

We also find that the error for the OLDG stencil is significantly
higher frequency than the error for the finite differences stencil,
even in the linear case. The high-frequency nature of the error
indicates that artificial dissipation may reduce error, even in the
linear case when it is not required for stability. However, we did not
investigate this possibility. Our experiments indicate that these
traits are roughly generic, although the factor by which the finite
differences error is smaller may depend on the order of the method.

This is a weakness of our OLDG method compared to both traditional
DGFE methods and finite differences methods. Traditional DGFE methods
have significantly more freedom with their numerical flux and they
can, for example, employ upwinding to reduce their error.

However, we emphasize that we have been ``fair'' to finite differences
methods by comparing stencils of the same order. At first glance, one
might assume that a spectral method would have less error than a
finite differences method. However this is only true if the spectral
method is allowed to utilize arbitrarily high-order polynomials. In
the comparisons shown in figures \ref{fig:dg:vs:fd:pointwise} and
\ref{fig:dg:vs:fd:pointwise:8th:order}, we have restricted our
discontinuous Galerkin elements to use polynomials of order fixed to
that of the finite differences stencil. 

\begin{figure}[!tb]
  \centering
  \includegraphics[width=\columnwidth]{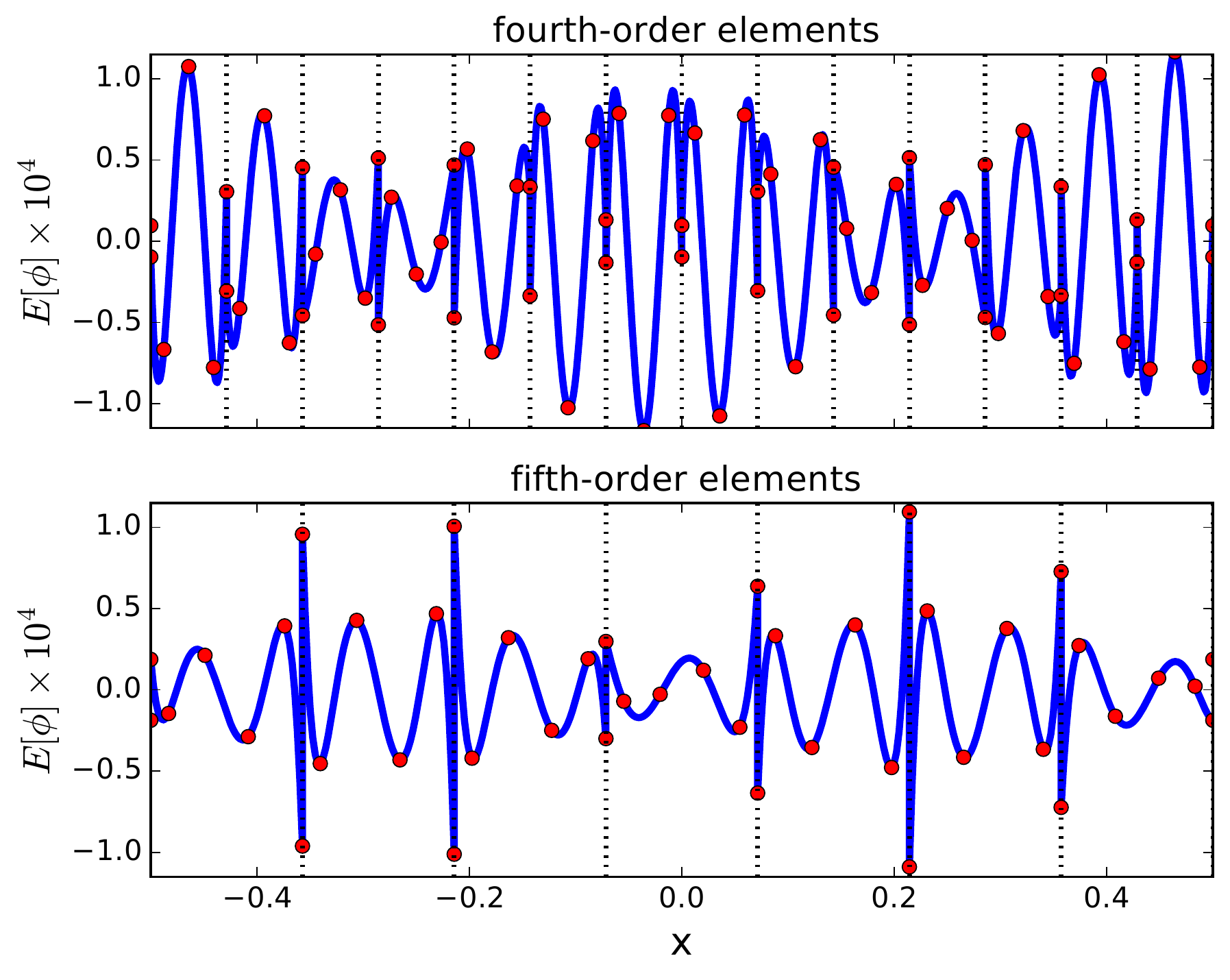}
  \caption{The pointwise errors for the wave equation with fourth-
    (top) and fifth-order (bottom) OLDG stencils respectively after
    0.75 wave periods. The simulation for the fourth-order stencil was
    run with 14 elements and the simulation for the fifth-order
    stencil was run with seven. Note that the errors are comparable to
    the finite differences calculation in figure
    \ref{fig:dg:vs:fd:pointwise}.}
  \label{fig:dg4:vs:dg5}
\end{figure}

In the fixed order case, we need about 14 fourth-order elements (or 70
collocation points) to do as well as the fourth-order finite
differences stencil with 40 points. (See the top panel in figure
\ref{fig:dg4:vs:dg5}.) However, if we vary the polynomial order within
elements, we can get comparable accuracy to the finite differences
stencil at similar computational cost and, as discussed in section
\ref{sec:asymptotics}, significantly improved communication and cache
properties.

The bottom panel of figure \ref{fig:dg4:vs:dg5} shows the pointwise
error for the wave equation with 7 fifth-order elements, or 42
collocation points. The error is comparable to the fourth-order finite
differences case shown in figure \ref{fig:dg:vs:fd:pointwise}. And if
our code were optimized, the computational cost would be similarly
comparable. This example highlights how, even though the pointwise
error for OLDG methods may seem inferior to finite differences in a
``fair'' comparison, they will perform better in realistic situations.

\subsection{Post-processing Element Boundaries}
\label{sec:postprocessing}

The pointwise error shown in figure
\ref{fig:dg:vs:fd:pointwise:8th:order} highlights a conceptual
difficulty with DGFE methods. The ``continuum'' function recovered by
modal representation is not continuous. Rather, it is piecewise
smooth. However, we often know from physical considerations that the
function that we solve for should be smooth over the whole domain. How
then do we recover a continuous function?

Given the collocation points $\phi^k_i$, we solve for the modal
representation, and therefore the polynomial interpolant, within an
element by solving equation \eqref{eq:vandermonde:use}
\begin{displaymath}
  \myvec{\phi}^k = \V^k \myvec{\hat{\phi}}^k
\end{displaymath}
for the modal coefficients $\hat{\phi}^k_i$. However, we can replace
equation \eqref{eq:vandermonde:use} by the following
\begin{equation}
  \label{eq:vandermonde:average}
  A^k \myvec{\phi}^k_{\text{wide}} = \V^k \myvec{\hat{\phi}}^k,
\end{equation}
where $\myvec{\phi}^k_{\text{wide}}$ is a length $P^k+3$ vector
containing the points
$$\{\phi^{k-1}_r\}\cup\{\phi^k_i\}_{i=0}^{P^k}\cup\{\phi^{k+1}_l\}$$
and $A^k$ is a wide operator that maps $\phi^k_{\text{wide}}$ to a
length $P^k+1$ vector which represents $\phi^k$ but with the left- and
right-hand limits of $\phi^k$ at element boundaries mapped to their
average. $A_{ij}^k$ has components
\begin{eqnarray}
  \label{eq:def:A:ij}
  A_{ij}^k &:=& \delta_{ij} + \half \left(\delta_{i,0}\delta_{j,-1} - \delta_{i,0}\delta_{j,0} \right. \\
    &&\qquad\qquad +  \left. \delta_{i,P^k}\delta_{j,P^k+1} - \delta_{i,P^k}\delta_{j,P^k} \right)\nonumber
\end{eqnarray}
for all $i=0,1,2,\ldots,P^k$ and $j=-1,0,1,\ldots,P^k+1$. Here we use
the shorthand introduced in equation
\eqref{eq:shorthand:indexing}. Figure \ref{fig:Ak:example} shows an
example of $A^k$ for polynomial order $P^k=4$.

\begin{figure}[tb!]
  \centering
  \begin{equation}
    \label{eq:A:ij:p4}
    A_{ij}^k = \begin{pmatrix}
      \half & \half & 0 & 0 & 0 & 0     & 0     \\
      0     & 0     & 1 & 0 & 0 & 0     & 0     \\
      0     & 0     & 0 & 1 & 0 & 0     & 0     \\
      0     & 0     & 0 & 0 & 1 & 0     & 0     \\
      0     & 0     & 0 & 0 & 0 & \half & \half \\
    \end{pmatrix}
  \end{equation}
  \caption{The averaging matrix $A^k$ for polynomial order $P^k =
    4$. $A^k$ is independent of element width.}
  \label{fig:Ak:example}
\end{figure}

\begin{figure}[!tb]
  \centering
  \includegraphics[width=\columnwidth]{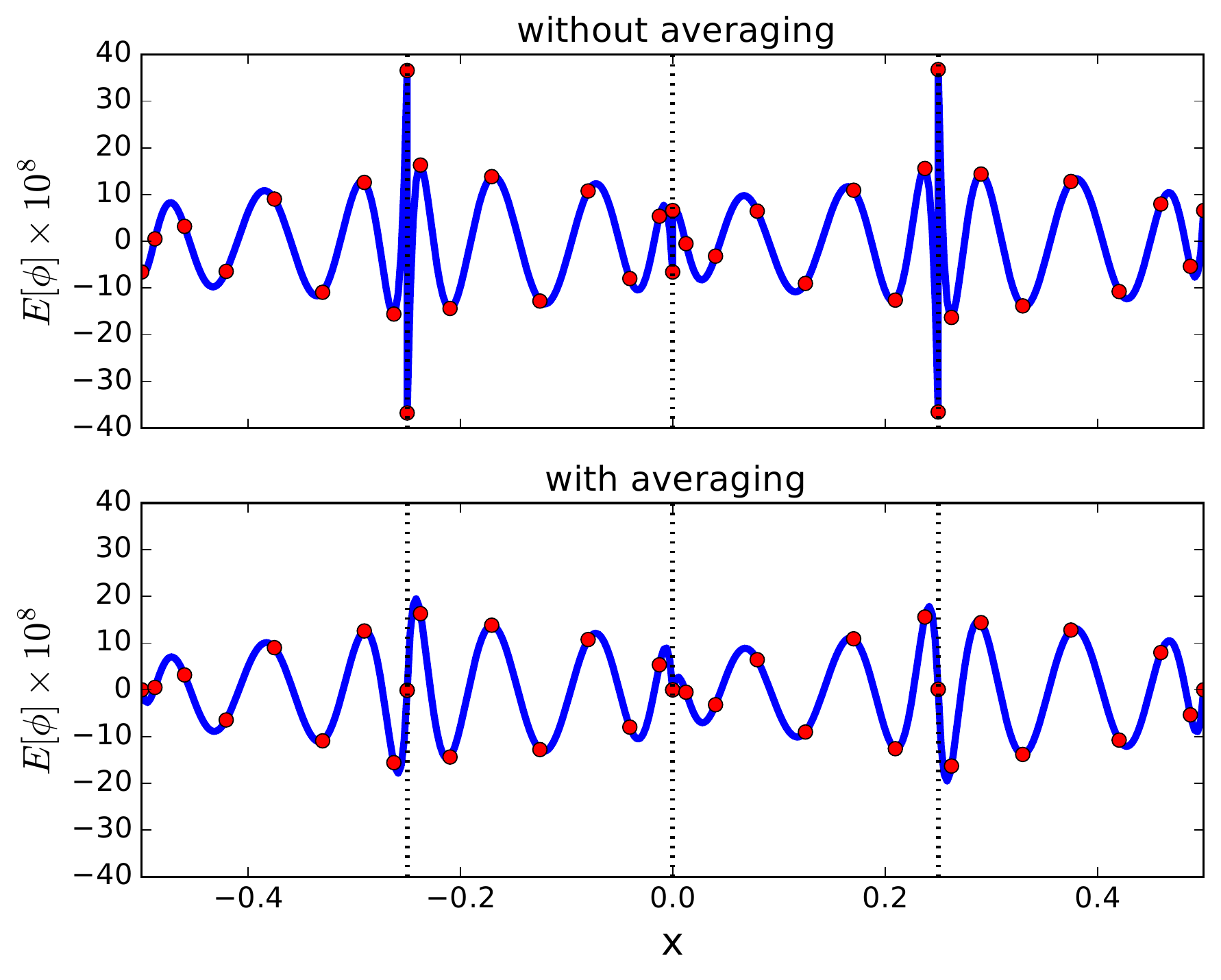}
  \caption{The error in the interpolating polynomial for the
    second-order wave equation \eqref{eq:linear:wave:equation}
    calculated using equation \eqref{eq:vandermonde:use} (top) and
    equation \eqref{eq:vandermonde:average} (bottom)
    respectively. This simulation was run with four eighth-order
    elements. The error is plotted at time $t=0.75$. In this case, the
    maximum pointwise error is reduced by a factor of two.}
  \label{fig:wavetoy:averaging:scheme}
\end{figure}

If we solve equation \eqref{eq:vandermonde:average} to generate our
interpolation, we produce a function that is continuous everywhere,
though it may not be smooth everywhere. Figure
\ref{fig:wavetoy:averaging:scheme} shows the effects of this averaging
scheme on pointwise errors for a simulation run with four eighth-order
elements. As usual, the curves are the interpolated solution and the
points are the collocation points. This procedure reduces pointwise
error near element boundaries, but otherwise does not significantly
change the pointwise error.

\subsection{The Robust Stability Test}
\label{sec:tests:robust:stability}

Our goal, of course, is not to solve the wave equation but to solve
equations as complex as the BSSN system.
To test our scheme's properties in this more
realistic and demanding setting, we perform the Apples-with-Apples
tests developed by the community \cite{AwA1,AwA2}.

The most basic of the Apples-with-Apples tests is the robust stability
test. The robust stability test is an experimental and numerical
analog to the stability condition for hyperbolic systems. It discerns
whether the numerical approximation of our formulation of the
continuum equations is stable to linear perturbations.

We take a constant-time slice of vacuum Minkowski space and introduce
random noise. We choose the amplitude of the noise such that the
Hamiltonian constraint is linearly satisfied. We then evolve the
spacetime and watch the deviation from Minkowski space for various
resolutions. A method passes the test if the deviation grows at most
exponentially in time, such that the \textit{maximum} rate is
\textit{independent of the resolution.}  Practically this means that
the \textit{rate of growth} in the deviation must \textit{not increase}
with increased resolution.

\begin{figure}[tb]
  \centering
  \includegraphics[width=\columnwidth]{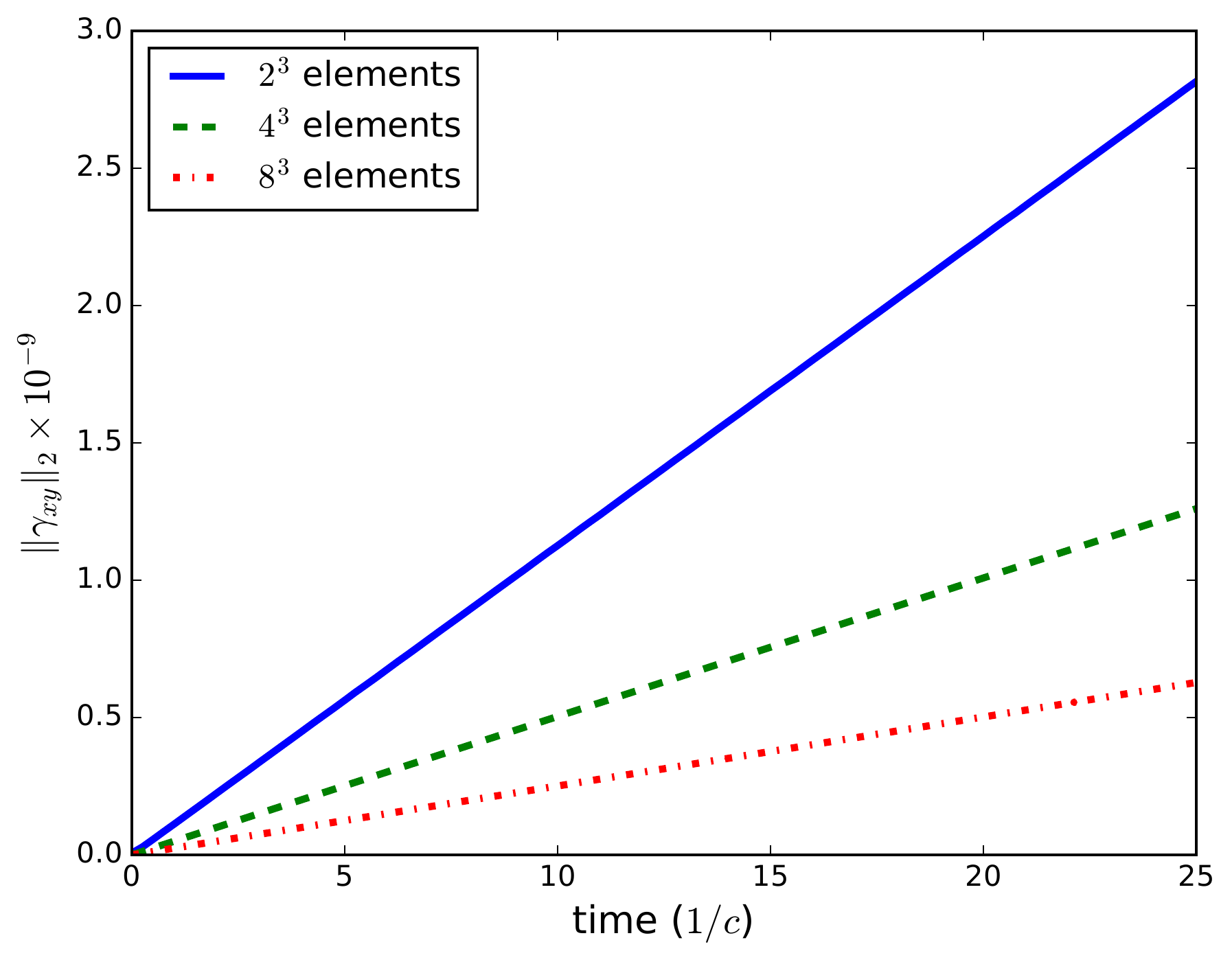}
  \caption{The robust stability test under $h$-refinement. We use
    $4^{th}$-order elements for this test. We find the growth rate to
    be at most linear and to decrease with resolution.}
  \label{fig:robust:stability:h:refinement}
\end{figure}

\begin{figure}[tb]
  \centering
  \includegraphics[width=\columnwidth]{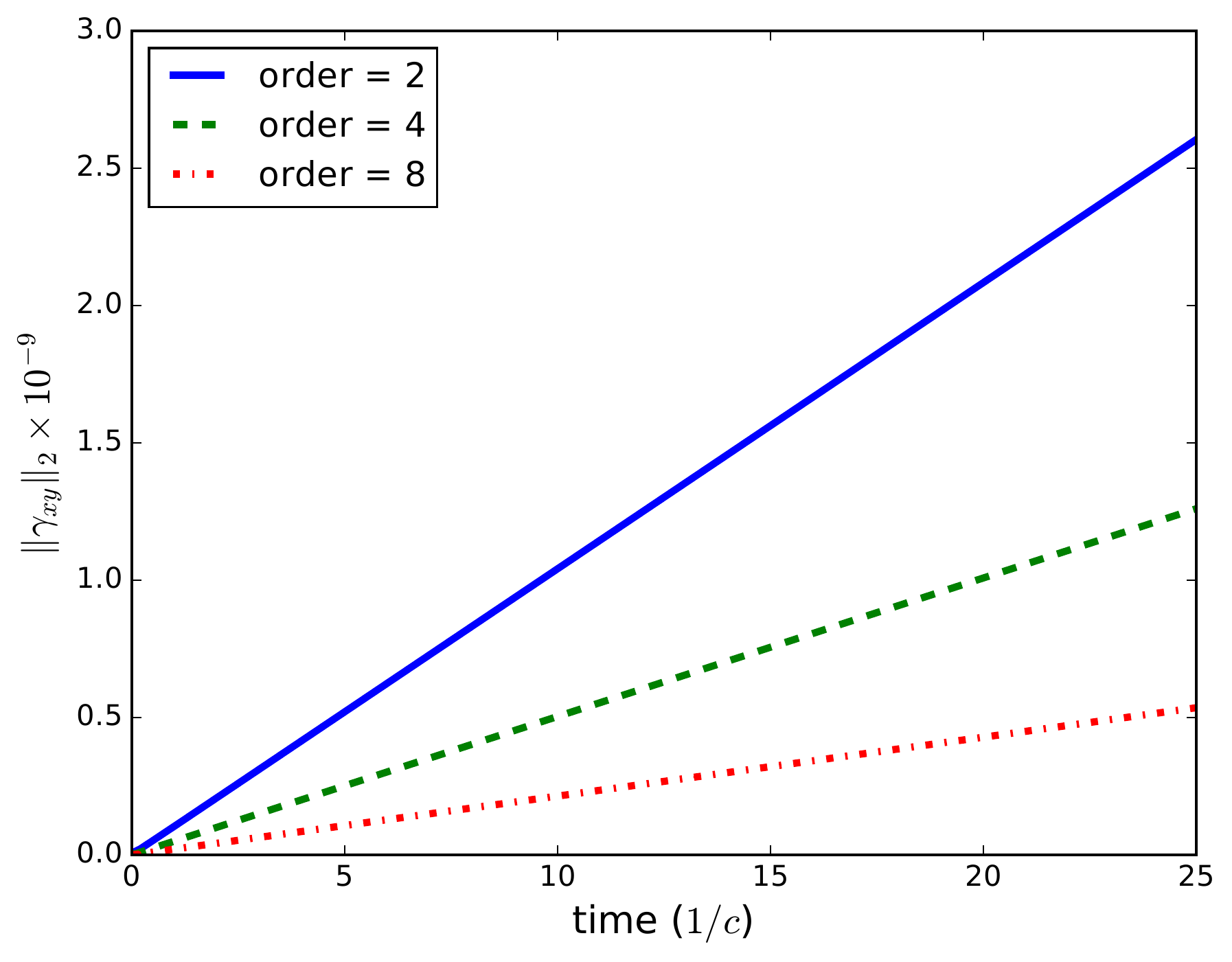}
  \caption{The robust stability test under $p$-refinement.  We use
    $4^3$ elements for this test. We find the growth rate to be at
    most linear and to decrease with resolution.}
  \label{fig:robust:stability:p:refinement}
\end{figure}

Figures \ref{fig:robust:stability:h:refinement} and
\ref{fig:robust:stability:p:refinement} show the results of the robust
stability test under $h$- and $p$-refinement respectively. We plot the
$xy$-component of the spatial metric $\gamma$, which should vanish in
the continuum. Therefore we measure error in the linear regime as a
function of time.  We find our discretization of the Einstein
equations to be stable. By eye, the growth rate is at most linear and
decreases with resolution under both kinds of refinement. For this
test, we use a three-dimensional domain $\Omega = [-0.5,0.5]^3$.

\subsection{Gauge Wave Test}
\label{sec:tests:gauge:wave}

\begin{figure}[tb]
  \centering
  \includegraphics[width=\columnwidth]{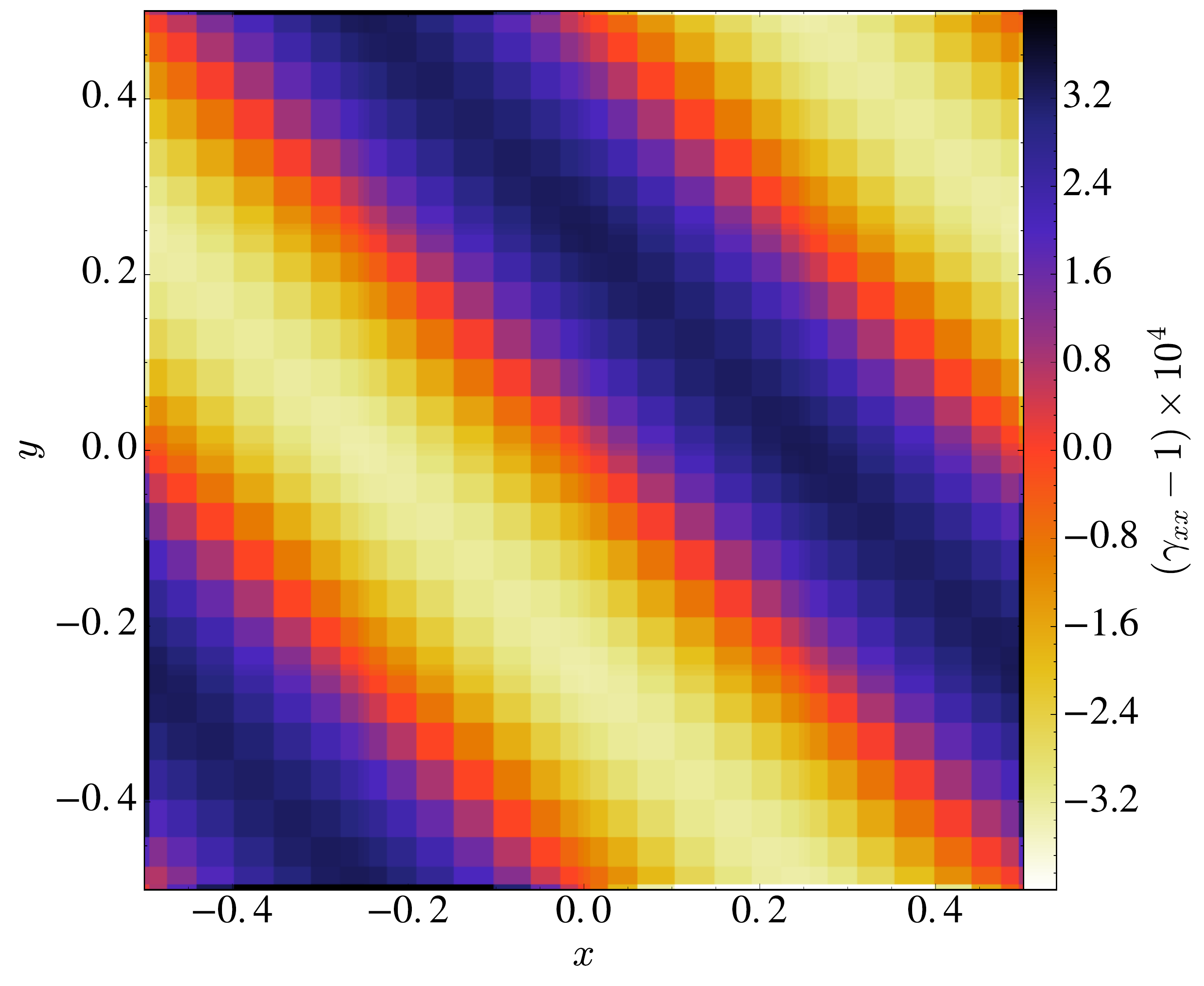}
  \caption{A two-dimensional slice of the three-dimensional gauge wave
    at $t=0$, generated with $8^{th}$-order elements. We plot the
    $xx$-component of the metric as a function of space in the $z=0$
    plane. The amplitude is commensurably smaller since $\gamma_{xx}$
    is a projection of the rotated $\gamma$ onto the $x$-axis.}
  \label{fig:gauge:wave:3D:slice}
\end{figure}

As a test of the accuracy and convergence of our scheme for the BSSN
system, we perform the \textit{gauge wave} test \cite{AwA1,AwA2}. The
gauge wave test is a periodic coordinate transformation of Minkowski
space, which provides a metric with a known analytic solution against
which we can compare. The metric for the one-dimensional gauge wave
test is
\begin{equation}
  \label{eq:def:gauge:wave:metric}
  ds^2 = (1-H) (-dt^2 + dx^2) + dy^2 + dz^2
\end{equation}
with
\begin{equation}
  \label{eq:def:gauge:H}
  H = A \sin\paren{\frac{2\pi(x-t)}{d}},
\end{equation}
for some amplitude $A$ and wavelength $d$. This can be converted to a
three-dimensional wave by rotating about the $y$- and $z$-axes by
$\pi/4$ each. Figure \ref{fig:gauge:wave:3D:slice} shows a
two-dimensional slice of $xx$-component of the spatial metric for the
3D gauge wave. The boundaries of the cells are the collocation points
and the value in the cell is is the average value in that region. We
have chosen a deliberately low resolution to highlight the structure
of the grid. The non-uniform position of the collocation points can be
seen in the varying cell sizes.

In all of our gauge wave simulations, we use an amplitude of $A=0.01$
and a period of $d=1$. In one dimension our domain is the interval
$x\in [-0.5,0.5]$. In three dimensions, our domain is the box
$(x,y,z)\in [-0.5,0.5]^3.$ For the one-dimensional gauge wave, we use
fourth-order elements. For the three-dimensional gauge wave, we use
eighth-order elements. For time integration we use an explicit
fourth-order or eighth-order Runge-Kutta integrator, as appropriate.

\begin{figure}[tb]
  \centering
  \includegraphics[width=\columnwidth]{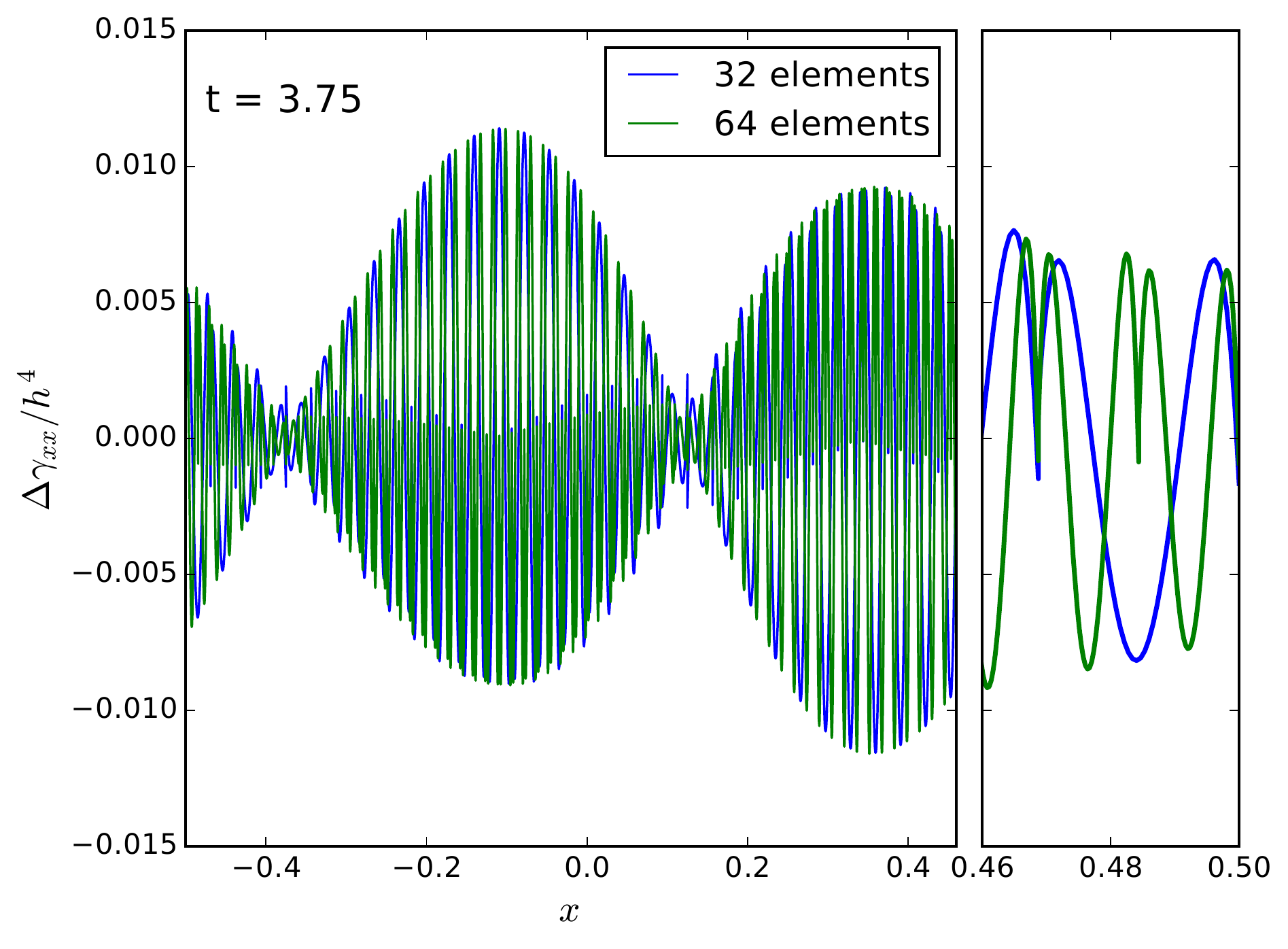}
  \caption{``Pointwise'' convergence of the one-dimensional gauge wave
    under $h$-refinement, with a fixed order of $p=4$, at $t=3.75$. We
    plot the error in the $xx$-component of the metric, rescaled by
    the element width to the $4^{th}$ power. The left pane shows most
    of the domain, while the right panel stretches out the axes so
    that the error is more visible. Since the collocation points do
    not align, pointwise convergence can't be expected. However, the
    fact that the envelopes of the errors align indicates good
    convergence. For the 32 element simulation, the absolute error is
    approximately $\Ord{ 10^{-8}}$. For the 64 element simulation, it
    is approximately $\Ord{10^{-10}}$.}
  \label{fig:gauge:wave:pointwise:convergence}
\end{figure}

Figure \ref{fig:gauge:wave:pointwise:convergence} shows the error in
the $xx$-component of the metric for the one-dimensional gauge wave at
$t=3.75$ light crossing times, rescaled by $h^{-4}$, where $h$ is the
element width for each element. As in figure
\ref{fig:dg:vs:fd:pointwise}, we generate the curves by interpolation
using the modal representation within an element. The curves do not
line up perfectly, but they are all contained within an envelope
function, which converges at fourth-order, as expected. 

\begin{figure}[tb]
  \centering
  \includegraphics[width=\columnwidth]{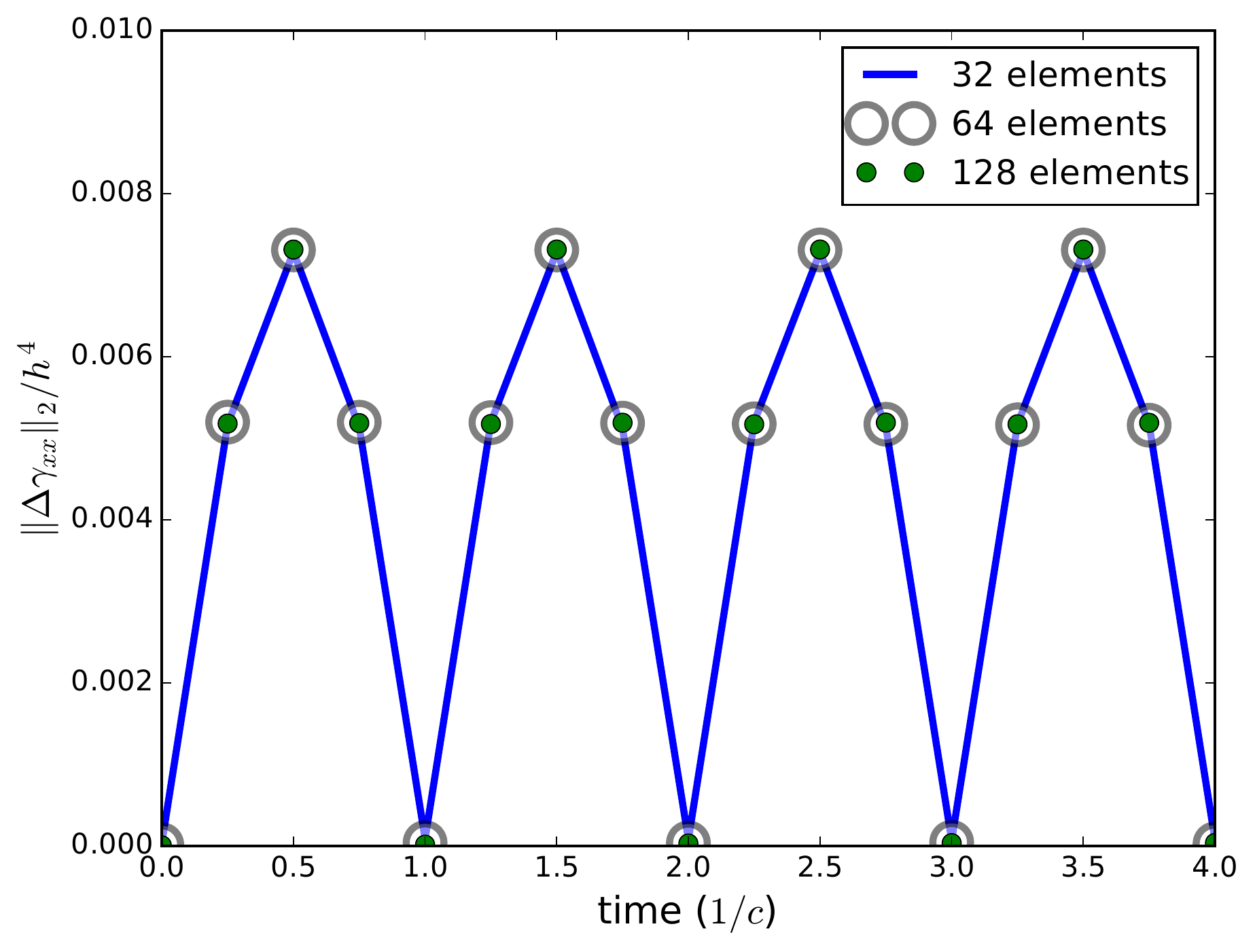}
  \caption{$L_2$-norm of the $xx$-component of the metric for the
    one-dimensional gauge wave, rescaled by $h^{-4}$. These
    simulations were run with a fixed element order of $p=4$. The
    curves align almost perfectly, indicating $4^{th}$-order
    convergence. For the 32 element run, the error is approximately
    $\Ord{10^{-8}}$. For the 64 element run, the error is approximately
    $\Ord{5\times 10^{-10}}$. For the 128 element run, it is approximately
    $\Ord{10^{-11}}$.}
  \label{fig:gauge:wave:L2:convergence}
\end{figure}

Figure \ref{fig:gauge:wave:L2:convergence} shows the $L_2$-norm of the
error for the one dimensional gauge wave as a function of time. We
once again re-scale the error by $h^{-4}$. The fact that the curves
overlap demonstrates fourth-order convergence for the system as it
evolves in time.

\begin{figure}[tb]
  \centering
  \includegraphics[width=\columnwidth]{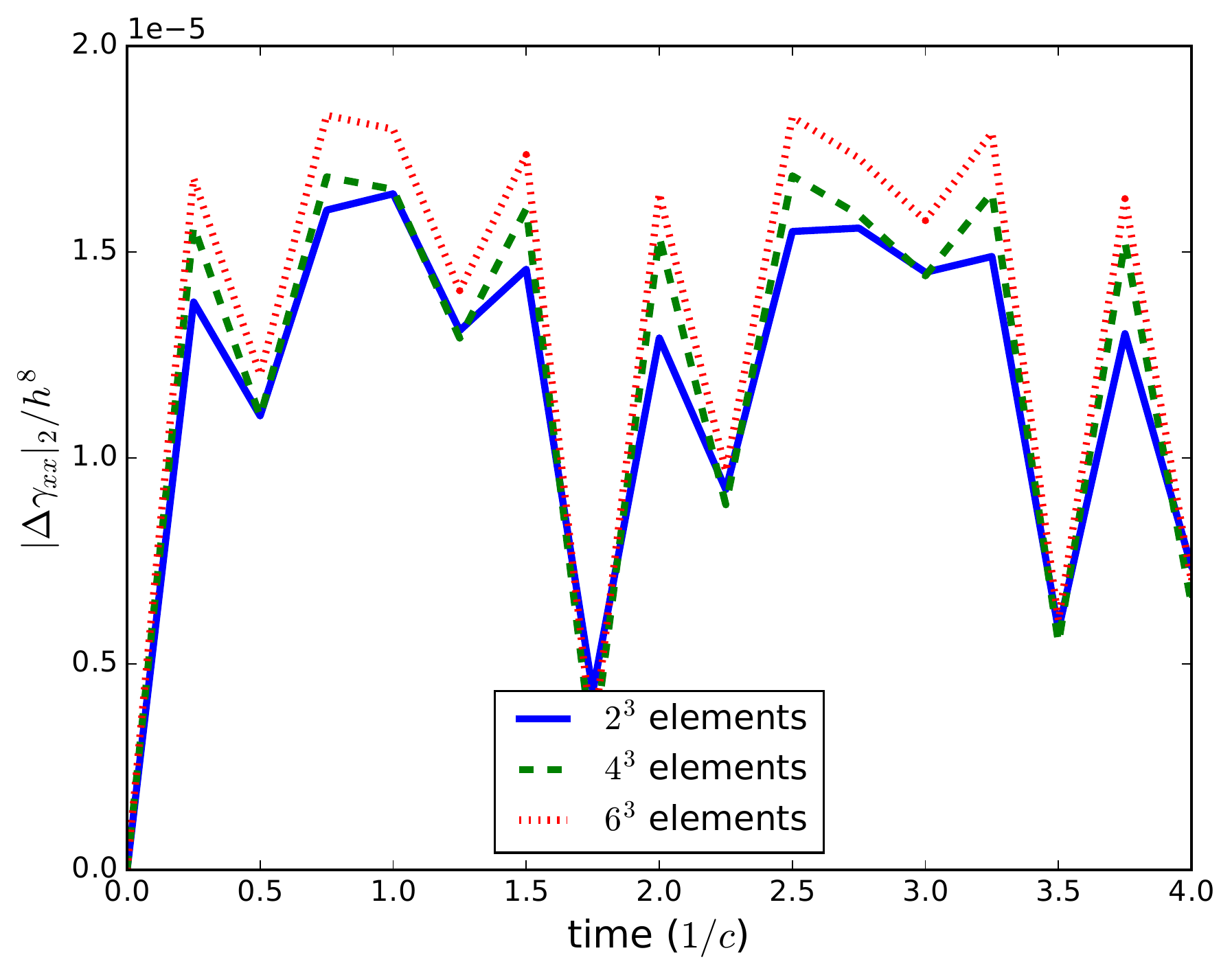}
  \caption{The $L_2$-norm over space of the error of the
    $xx$-component of the metric for the 3D gauge wave, rescaled by
    $1/h^8$.  The curves overlap, showing that the system is
    converging at $8^{th}$-order. This simulation was run with
    $8^{th}$-order elements. For the $2^3$ element run, the error is
    approximately $\Ord{10^{-8}}$. For the $4^3$ element run, the
    error is approximately $\Ord{10^{-10}}$. For the $6^3$ element
    run, the error is approximately $\Ord{10^{-12}}$.}
  \label{fig:dg8:gauge:wave:3D:convergence}
\end{figure}

Figure \ref{fig:dg8:gauge:wave:3D:convergence} shows the $L_2$-norm of
the error for the three-dimensional gauge wave, using eighth-order
elements. We now rescale by $h^{-8}$ and, again, the fact that the
curves overlap demonstrates convergence of the appropriate order under
$h$-refinement. At eighth order, we need very few elements before we
see good convergence.

We note that since the y-axis is rescaled in figures
\ref{fig:gauge:wave:L2:convergence} and
\ref{fig:dg8:gauge:wave:3D:convergence}, it does not represent the
true error. In particular, although the rescaled error is large, the
absolute error is comparable to or better than that in
\cite{AwA1,AwA2}.

\subsection{Gamma Driver Gauge Wave Test}
\label{sec:tests:gamma:driver}

The gauge wave prescription given in equation
\eqref{eq:def:gauge:wave:metric} has a harmonic lapse and vanishing
shift. We would like to test more realistic lapse and shift conditions
in this simplified context, so we seek a generalization of the gauge
wave. In \cite{AwA2}, Babiuc et al. propose the \textit{shifted gauge
  wave}, which generalizes the original gauge wave to include a
nonzero shift.

This nonzero shift is harmonic however and, as discussed in
\cite{AwA2}, the BSSN system performs poorly in this setting. Physical
evolutions of the BSSN system typically use the \textit{Gamma driver}
shift condition of the form \cite{AlcubierreGammaDriver}
\begin{eqnarray}
  \label{eq:gamma:driver:shift}
  \paren{\partial_t - \Lie_\beta}\beta^i &=& B^i\\
  \paren{\partial_t - \Lie_\beta }B^i &=& \alpha^2 \zeta \paren{\partial_t-\Lie_\beta}\tilde{\Gamma}^i - \eta B^i,
\end{eqnarray}
where $\beta^i$ are the components of the shift, $\Lie_\beta$ is the
Lie derivative in the $\beta$-direction, $\zeta,\eta\in\R$ are
constants, and
\begin{equation}
  \label{eq:def:gamma:tilde}
  \tilde{\Gamma}^i = - \partial_j \paren{\psi^4 \gamma^{ij}}
\end{equation}
is the conformally rescaled connection for spatial metric $\gamma$ and
conformal factor $\psi$. The Gamma driver shift is combined with the
\textit{1+log} slicing condition, first developed by Bernstein
\cite{BernseinThesis} and Anninos et al. \cite{anninos1995three}. This
is of the form \cite{AlcubierreConformalDecomp,AlcubierreGammaDriver}:
\begin{equation}
  \label{eq:1+log:slicing}
  \partial_t \alpha = -2\alpha K,
\end{equation}
where $\alpha$ is the lapse and $K$ is the trace of the extrinsic
curvature tensor.

% \begin{figure}[tb]
%   \centering
%   \includegraphics[width=\columnwidth]{bssn_dg4_gauge_wave_1_plus_log_f08_waveform}
%   \caption{The Gamma driver gauge wave. The $xx$-component of the
%     metric as a function position after two wave periods. This
%     solution was calculated using 192 $4^{th}$-order elements.}
%   \label{fig:dg4:gamma:gauge:wave:waveform}
% \end{figure}

Motivated by these observations, we propose a new version of the
shifted gauge wave test, the \textit{Gamma driver gauge wave}, which
tests our method using the gauge conditions typically used with the
BSSN system. We use the same domains and initial conditions as the
gauge wave test, but we impose 1+log slicing and a simplified version
of the Gamma driver shift condition:
\begin{equation}
  \label{eq:gamma:driver:simple}
  \partial_t \beta^i = \zeta \tilde{\Gamma}^i-\eta \beta^i,
\end{equation}
where we choose $\zeta = \eta = 3/4$
\cite{AlcubierreConformalDecomp,AlcubierreGammaDriver}.
% A numerical
% solution to the Gamma driver gauge wave is shown in figure
% \ref{fig:dg4:gamma:gauge:wave:waveform}, where we plot the
% $xx$-component of the metric after two light-crossing times.

We do not know the analytic solution to this system of gauge
conditions, but we can study convergence in this setting by comparing
several coarse resolutions to a fine resolution instead of an analytic
solution. This type of convergence test, which is weaker than
convergence to a known solution, is called a \textit{self-convergence
  test}.

In self-convergence, one must take care to rescale the error by the
correct amount. A system is self-convergent to order $P$ if 
\begin{equation}
  \label{eq:eq:phi:self:convergence:test}
  \frac{(\gamma_{xx})_1 - (\gamma_{xx})_3}{h_1^P - h_3^P} = \frac{(\gamma_{xx})_2-(\gamma_{xx})_3}{h_2^P-h_3^P},
\end{equation}
where $i$ indexes three resolutions, such that $i=1$ is the coarsest
and $i=3$ is the finest. For notational simplicity, we assume all
elements have the same width and order and we therefore suppress the
element index $k$. We also suppress dependence on $x$ and $t$. For
details of where equation \eqref{eq:eq:phi:self:convergence:test}
comes from, see appendix \ref{section:self:convergence:derivation}.

\begin{figure}[tb]
  \centering
  \includegraphics[width=\columnwidth]{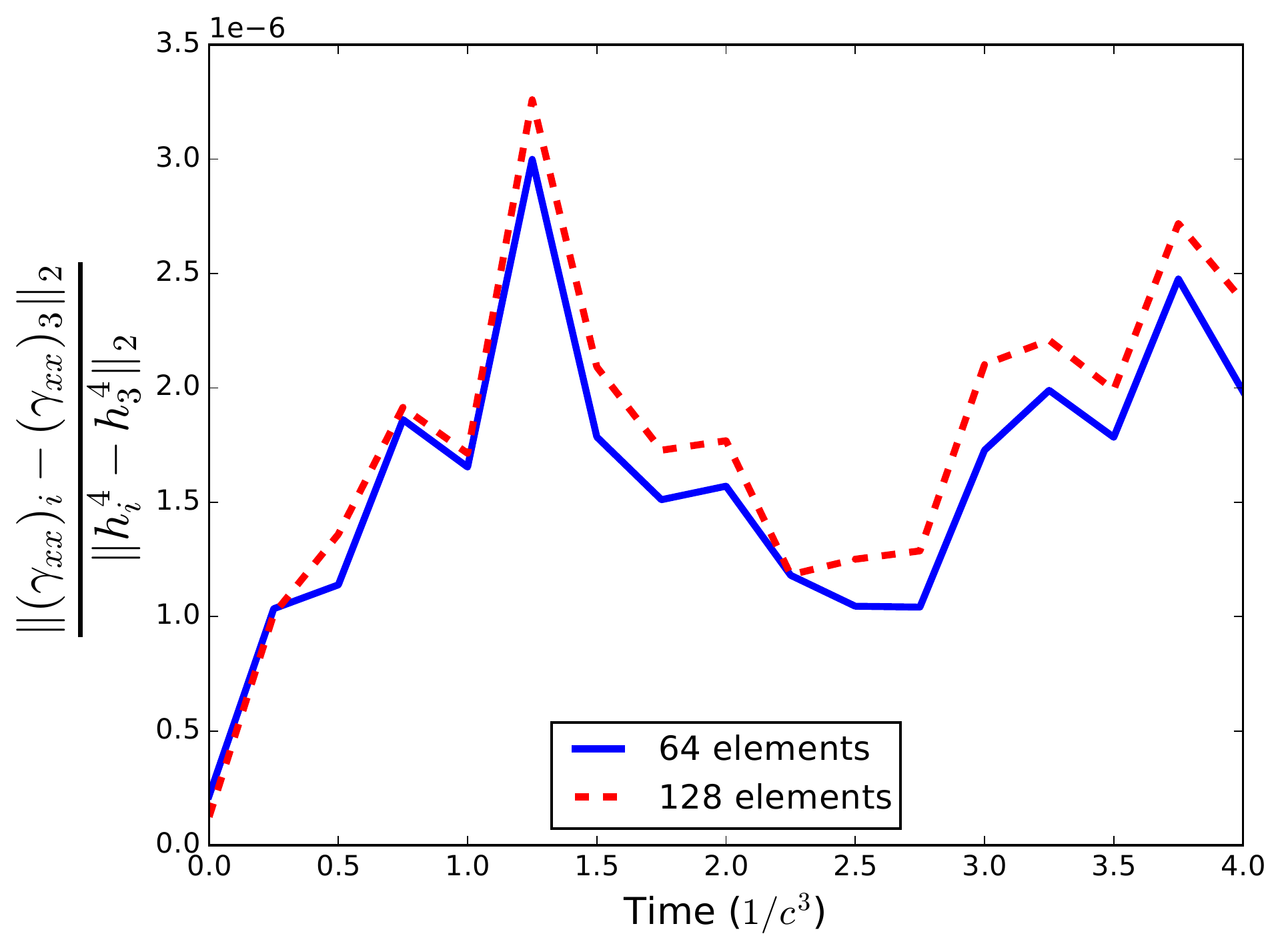}
  \caption{The $L_2$ norm of the difference between two coarse
    resolutions and one fine resolution for the Gamma driver gauge
    wave as a function of time, normalized by an appropriate factor
    based on the number and order of the elements. The fact that the
    curves overlap demonstrates $4^{th}$-order convergence. For this
    test, we use 64, 128, and 192 elements. These differences are
    small, approximately $\Ord{10^{-12}}$ for the comparison between
    64 elements and 192 elements and approximately $\Ord{10^{-13}}$
    for the comparison between 128 elements and 192 elements.}
  \label{fig:dg4:gamma:gauge:wave:l2error}
\end{figure}

Figure \ref{fig:dg4:gamma:gauge:wave:l2error} shows the
self-convergence for the one-dimensional Gamma driver gauge wave with
fourth-order elements. We plot the $L_2$-norm over space of equation
\eqref{eq:eq:phi:self:convergence:test}. Because our implementation is
not yet optimized for performance, the 3D self-convergence test was too
expensive. Therefore we do not perform the
Gamma driver gauge wave test in 3D.

\begin{figure}[tb]
  \centering
  \includegraphics[width=\columnwidth]{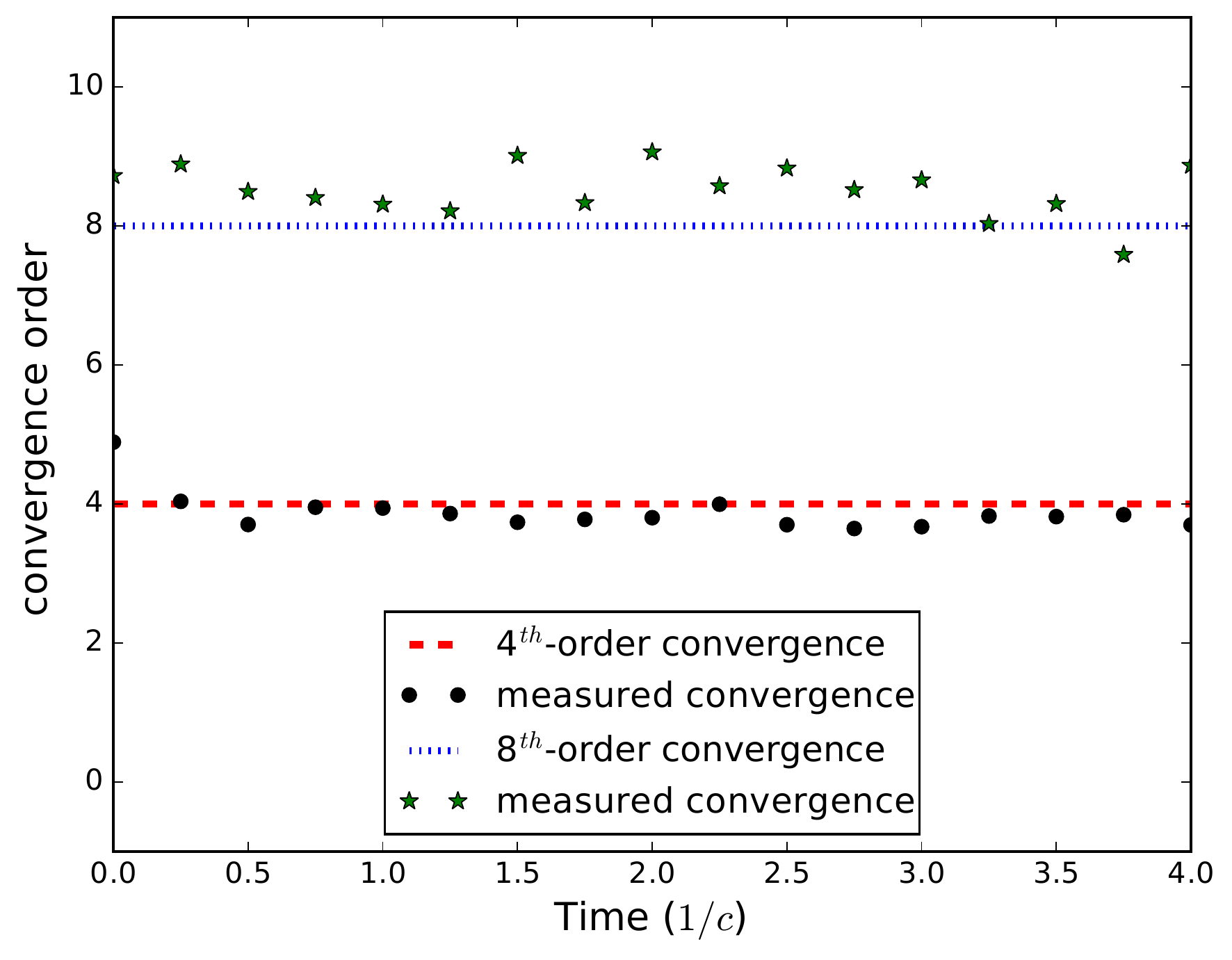}
  \caption{The convergence order for $4^{th}$-order and $8^{th}$-order
    elements as a function of time, as extracted by solving for $P$ in
    equation \eqref{eq:eq:phi:self:convergence:test}.}
  \label{fig:dgx:gamma:gauge:wave:richardson}
\end{figure}

As an additional check, we can treat $P$ in equation
\eqref{eq:eq:phi:self:convergence:test} as a free variable, the
\textit{convergence order}, and solve for it numerically.
If convergence is as we expect, we will recover that
$P$ is the same as the order of the element. Indeed, we can perform
this calculation globally by taking the $L_2$-norm of
$(\gamma_{xx})_i-(\gamma_{xx})_3$ for $i=1,2$ at each timestep and
solving for $P$. We then obtain a measure of convergence as a function
of time. Figure \ref{fig:dgx:gamma:gauge:wave:richardson} shows the
result of this procedure for the Gamma driver gauge wave using both
$4^{th}$- and $8^{th}$-order elements. The measured convergence order
agrees very well with our expectations.

\section{Concluding Remarks}
\label{sec:conclusion}

By performing an LDG discretization at the level of the differential
operator, rather than at the level of the equations, we have developed
a novel DGFE scheme that can be used to discretize arbitrary
second-order hyperbolic equations, in particular also the BSSN
formulation of Einstein's equations. In the process, we have made the
formalism proposed by Hesthaven and Warburton
\cite{hesthaven2008nodal} rigorous and combined it with summation by
parts.

We analyzed and tested our scheme and its stability and accuracy for a
series of standard test problems in numerical relativity, and find the
expected polynomial (for $h$-refinement) and exponential (for
$p$-refinement) convergence. Compared to finite differencing methods,
the solution error is larger when using the same number of collocation
points, but as for other DGFE methods, our OLDG scheme requires
significantly fewer memory accesses and has a significantly lower
communication overhead, and is thus more scalable on current high
performance computing architectures.

Moreover, our focus on the derivative operator allows codes that
currently employ finite differences methods to straightforwardly
replace the finite differences stencil with our OLDG stencil,
converting a finite differences code to a DGFE code. This should
improve the parallel efficiency of such codes.

\section{Acknowledgements}

The authors thank David Radice for many helpful discussions.  The
authors also thank: Federico Guercilena for help with developing the
testing infrastructure; Saul Teukolsky and the SpECTRE collaboration
for discussions on DGFE methods; and Scott Field, Wolfgang Tichy and
Jan Hesthaven for their help with literature search.

The authors acknowledge support from the Natural Sciences and
Engineering Research Council of Canada and from the National Science
Foundation of the USA (OCI 0905046, PHY 1212401). J. Miller
acknowledges the support of the ACM/SIGHPC SC15 Student Travel
Grant. 
Research at Perimeter Institute is supported by the Government of
Canada through the Department of Innovation, Science and Economic
Development and by the Province of Ontario through the Ministry of
Research and Innovation.

We are grateful to the countless developers contributing to open
source projects on which we relied in this work, including Cactus
\cite{Goodale:2002a, Cactuscode:web} and the Einstein Toolkit
\cite{loffler2012einstein,EinsteinToolkit:web,EinsteinToolkit:ascl},
Kranc \cite{Husa:2004ip,Kranc:web}, HDF5 \cite{hdf5}, Python
\cite{rossumPythonWhitePaper}, numpy and scipy \cite{numpy,scipyLib},
Matplotlib \cite{hunterMatplotlib}, and the yt-project
\cite{TurkYTProject}.

This research used computing facilities at the Perimeter Institute,
the Center for Computation \& Technology at LSU, the Shared
Hierarchical Academic Research Computing Network (SHARCNET) and
Compute Canada, the Extreme Science and Engineering Discovery
Environment (XSEDE) (TG-PHY100033), and the Blue Waters
sustained-petascale computing project (PRAC\_jr6).

\appendix

\section{Legendre Polynomials, Collocation Points, and Gauss Lobatto
  Quadrature}
\label{sec:collocation}

The following treatment can be found in standard textbooks such as
\cite{press2007numerical} or \cite{hesthaven2008nodal}. Consider the
domain $\Omega = [-1,1]$ and a function $\psi\in L_2(\Omega)$. We wish
to approximate the continuum function $\psi$ and its
derivatives in a reasonable way. One such approximation is that
\begin{equation}
  \label{eq:legendre:ansatz}
  \psi(x) \approx \psi_{\text{modal}} := \sum_{i=0}^P \hat{\psi}_i \Phi_i(x),
\end{equation}
where $P\in\N$, each $\Phi_i$ is some polynomial basis function such
that $\{\Phi_i\}_{i=0}^\infty$ forms a complete orthonormal basis of
$L_2(\Omega)$, and each $\hat{\psi}_i$ is a constant in $x$ defined by
the projection
\begin{equation}
  \label{eq:legendre:def:psi:hat}
  \hat{\psi}_i = \int_{-1}^1 \psi(x) \Phi_i(x) dx = \braket{\psi,\Phi_i}_\Omega.
\end{equation}
Since the $\Phi_i$'s form a complete basis, equation
\eqref{eq:legendre:ansatz} becomes exact in the limit
$P\to\infty$. The demand that each $\Phi_i$ is a polynomial is
crucial, as we will soon see.

% S[table-format=x.y] fails for multirow tables
% numprint works.
% using it for this table only.
\begin{table}[!t]
  \centering
  \begin{tabular}{| c | n{10}{2} | n{4}{2} |}
    \toprule
    \textbf{Order} $P$ 
    & {\textbf{Collocation Points} $x_i$ }
    & \textbf{Weights} $w_i$\\
    \colrule    
    \multirowcell{2}{2} & 0 & $1.33$\\ 
    [1ex]
    & $\pm 1$ & $0.33$\\
    \colrule
    \multirowcell{3}{4} & 0 & $0.71$ \\
    [1ex]
    & $\pm 0.65$ & $0.54$ \\
    [1ex]
    & $\pm 1$ & $0.1$ \\
    \colrule
    \multirowcell{4}{6} & 0 & $0.49$ \\
    [1ex]
    & $\pm 0.47$ & 0.43 \\
    [1ex]
    & $\pm 0.83$ & 0.27 \\
    [1ex]
    & $\pm 1$ & 0.05 \\
    \colrule 
    \multirowcell{5}{8} & 0 & 0.37\\
    [1ex]
    & $\pm 0.36$ & 0.35 \\
    [1ex]
    & $\pm 0.68$ & 0.27 \\
    [1ex]
    & $\pm 0.90$ & 0.17 \\
    [1ex]
    & $\pm 1$ & 0.03 \\
    \botrule
  \end{tabular}
  \caption{The approximate locations of the abscissas and values of 
    associated weights for Gauss-Lobatto quadrature for several values of $P$.}
  \label{tab:weights:coloc}
\end{table}

We can also approximate $\psi$ in a so-called
\textit{nodal} basis, where we treat it as the polynomial that
interpolates between known points $\psi(x_i)$ for some set of points
$x_i\in \Omega$, $i = 0,1,\ldots,N$, with $N\in\N$. We can write this as
\begin{equation}
  \label{eq:nodal:representation:lagrange:poly}
  \psi(x) \approx \psi_{\text{nodal}}(x) := \sum_{i=0}^N \psi_i l_i(x),
\end{equation}
where $\psi_i = \psi(x_i)$ and $l_i(x)$ are the $P^{th}$-order
Lagrange polynomials,
\begin{equation}
  \label{eq:def:lagrange:poly}
  l_i(x) = \prod_{\substack{0\leq j\leq N\\j\neq i}} \frac{x-x_j}{x_i-x_j}.
\end{equation}
Interpolating polynomials are unique. So in the special case $N=P$,
the two representations \eqref{eq:legendre:ansatz} and
\eqref{eq:nodal:representation:lagrange:poly} in fact describe the
same polynomial.

Suppose $N=P$. Crucially, if equations \eqref{eq:legendre:ansatz} and
\eqref{eq:nodal:representation:lagrange:poly} are different
representations of the same function, there should be a way to
transform between them. This is the \textit{Vandermonde} matrix
defined by its action as a transformation operator
\begin{equation}
  \label{eq:vandermonde:simple:def}
  \psi_i = \sum_{j=0}^P \V_{ij} \hat{\psi}_j.
\end{equation}
From equation \eqref{eq:vandermonde:simple:def}, it is easy to show
that
\begin{equation}
  \label{eq:vandermonde:components}
  \V_{ij} = \Phi_j(x_i).
\end{equation}
In a realistic calculation, we will need both the Vandermonde matrix
and its inverse $\V^{-1}$. Therefore, the matrix should be
well-conditioned.

One basis for which $\V$ is well conditioned is the basis of
\textit{Legendre polynomials}. The Legendre polynomials are solutions
to Legendre's differential equation
\begin{equation}
  \label{eq:legendre:equation}
  \dd{x} \sqrbrace{(1-x^2)\dd{x}\Phi_i(x)} + i(i+1)\Phi_i(x) = 0\ \forall\ i\in\N,
\end{equation}
but they are most easily defined recursively as:
\begin{eqnarray}
  \label{eq:legendre:recursive:definition}
  \Phi_0(x) &=& 1\\
  \Phi_1(x) &=& x\\
  (i+1)\Phi_i(x) &=& (2i+1)x\Phi_i(x) - i\Phi_{i-1}(x)
\end{eqnarray}
for all $i\in\N$. 

In general, we may not be able to compute the integral in equation
\eqref{eq:legendre:ansatz} with perfect accuracy or
efficiency. Therefore, we would like to find a \textit{quadrature
  rule} that allows us to efficiently calculate approximate integrals:
\begin{equation}
  \label{eq:def:quadrature:rule:elementwise}
  \int_{-1}^1f(x) dx \approx \sum_{i=0}^P w_i f(x_i),
\end{equation}
for some set of weights $w_i$ and collocation points $x_i$. Choosing a
quadrature rule involves a choice both of weights $w_i$ and abscissas
$x_i$. There are several good choices for quadrature rules. Because we
are interested in using this discretization for a DGFE method, we want
a rule where the abscissas include the endpoints of the domain
$\Omega$. We therefore use \textit{Gauss-Lobatto} quadrature.
\begin{equation}
  \label{eq:lobatto:quadrature:points}
  x_i \in \{-1,1\}\cup \text{roots}\paren{\Phi_{P-1}'},
\end{equation}
where $\text{roots}\paren{\Phi_{P-1}'}$ is the set of solutions to the
equation
\begin{equation}
  \label{eq:def:root:P}
  \Phi_{P-1}'(x) = 0.
\end{equation}
The weights are then defined as the solutions to the linear system
\begin{equation}
  \int_{-1}^1 \Phi_j(x) dx = \sum_{i=0}^P w_i \Phi_j(x_i)
\end{equation}
for all $j=0,1,\ldots,P$. In table \ref{tab:weights:coloc}, we provide
approximate values for some of the Gauss-Lobatto quadrature points and
their associated weights for different values of $P$. A procedure for
calculating these points precisely can be found in any standard
numerical text such as \cite{press2007numerical}.

By setting the collocation points in equation
\eqref{eq:nodal:representation:lagrange:poly} equal to the abscissas
defined by equation \eqref{eq:lobatto:quadrature:points}, we obtain the
discrete approximation of $L_2(\Omega)$. This representation is exact
for all polynomials of order no greater than $P$.

We are interested in performing this discretization within a DGFE
element
$$\Omega^k = [x^k_l,x^k_r],$$
which is related to the interval $[-1,1]$ by a simple linear
coordinate transformation. This transformation introduces a factor of
$h^k/2$ into each of the weights:
\begin{equation}
  \label{eq:weights:transformed}
  w^k_i = \frac{h^k}{2} w_i
\end{equation}
where the weights must be computed for the element order $P^k$. This
coordinate transformation also introduces a factor of $2/h^k$ into the
modal derivative operator defined in equation
\eqref{eq:modal:derivative}.

\begin{widetext}
\section{Proof That $\xi=1$ Satisfies Summation By Parts}
\label{sec:sbp:proof}

In this section, we prove that the scheme described in section
\ref{sec:methods} satisfies summation by parts if and only if $\xi
= 1$, as asserted in section \ref{sec:methods:sbp}. We begin with the
element-wise derivative operator defined in equation \eqref{eq:def:Dk}:
\begin{displaymath}
  \partial_x \tilde{\psi}^k = D^k \psi^k = \sqrbrace{d^k - \half b^k + \half\xi F^k}\psi^k,
\end{displaymath}
where $d^k$ and $b^k$ are the element-wise differentiation and
boundary operators respectively and the fetch operator $F^k$ produces
information about the boundary of neighbouring elements as defined in
equation \eqref{eq:def:J}. We seek the conditions on it such that
equation \eqref{eq:def:summation:by:parts:omega} holds:
\begin{displaymath}
    \braket{\psi, D\phi}_\Omega + \braket{D\psi, \phi}_\Omega = \braket{\psi,\phi}_{\partial\Omega},
\end{displaymath}
where $D$ is the derivative operator defined on the whole domain as given by equation \eqref{eq:def:D}:
\begin{displaymath}
  D\psi(x) = (D^k\psi^k)(x)\ \forall\ x\in \Omega^k\ \forall\ 1\leq k \leq K.
\end{displaymath}

In terms of the element-wise inner product, we demand that 
\begin{equation}
  \label{eq:element:wise:sbp}
  \sum_{k=1}^{K}\sqrbrace{\braket{D^k\psi^k,\phi^k}_{\Omega^k} + \braket{\psi^k,D^k\phi^k}_{\Omega^k}} = \phi\psi\eval_{\partial\Omega}
\end{equation}
(or some equivalent relation) for all test functions $\phi$ and $\psi$
in $L^2(\Omega)$. For convenience, we define 
\begin{eqnarray}
  \label{eq:def:phiL:psiL}
  \phi_L = \phi(X_l),&\ \phi_R = \phi(X_r),\\
  \text{and }\psi_L = \psi(X_l),&\ \psi_R = \psi(X_r),
\end{eqnarray}
for the boundary elements $\Omega_0 = \{X_l\}$ and
$\Omega_{K+1}=\{X_r\}$. Now, if we plug definition \eqref{eq:def:Dk}
into condition \eqref{eq:element:wise:sbp}, we find that
\begin{equation}
  \label{eq:sbp:expanded}
  \phi\psi\eval_{\partial\Omega} = S_1 + S_2
\end{equation}
where we have split the sum over elements into two sums
\begin{equation}
  \label{eq:def:S1}
  S_1 = \sum_{k=1}^{K}\sqrbrace{\braket{\paren{d^k-\half b^k}\phi^k,\psi^k}_{\Omega^k}+\braket{\phi^k,\paren{d^k-\half b^k}\psi^k}_{\Omega^k}}
\end{equation}
and 
\begin{equation}
  \label{eq:def:S2}
  S_2 =  \half \xi\sum_{k=1}^{K}\sqrbrace{\braket{F^k\phi^k,\psi^k}_{\Omega^k} + \braket{\phi^k,F^k\psi^k}_{\Omega^k}},
\end{equation}
which we will handle separately.

Let us examine $S_1$ first. Recall that $b^k$ is defined in equation
\eqref{eq:def:b:operator} by the relation:
\begin{displaymath}
  w^k b^k = w^k d^k + (d^k)^T w^k,
\end{displaymath}
where $w^k$ is the element-wise weight operator. Therefore, 
\begin{eqnarray}
  \label{eq:w:dmb}
  w^k \paren{d^k - \half b^k} &=& w^k d^k - \half w^k b^k\nonumber\\
  &=& w^k d^k -\half w^k (w^k)^{-1}\paren{w^kd^k+(d^k)^Tw^k}\nonumber\\
  &=& \half\paren{w^k d^k-(d^k)^T w^k},
\end{eqnarray}
and similarly
\begin{eqnarray}
  \label{eq:w:dmbT}
  \paren{d^k - \half b^k}^Tw^k &=& (d^k)^Tw^k - \half \paren{(d^k)^T w^k+w^k d^k}\sqrbrace{(w^k)^{-1}}^Tw^k\nonumber\\
  &=& -\half\paren{w^k d^k - (d^k)^T w^k},
\end{eqnarray}
where we have used the fact that a diagonal matrix is its own
transpose. Therefore,
\begin{equation}
  \label{eq:matrix:anticommute:elemetnwise}
  w^k \paren{d^k - \half b^k} + \paren{d^k - \half b^k}^Tw^k = 0
\end{equation}
and 
\begin{eqnarray}
  \label{eq:inner:product:anticommute:elementwise}
  \braket{\paren{d^k-\half b^k}\phi^k,\psi^k}_{\Omega^k}+\braket{\phi^k,\paren{d^k-\half b^k}\psi^k}_{\Omega^k} &=& \paren{\myvec{\phi}^k}^T\paren{d^k - \half b^k}^Tw^k\myvec{\psi}^k + \paren{\myvec{\phi}^k}^Tw^k \paren{d^k - \half b^k}\myvec{\psi}^k\nonumber\\
  &=& \paren{\myvec{\phi}^k}^T\sqrbrace{w^k \paren{d^k - \half b^k} + \paren{d^k - \half b^k}^Tw^k}\myvec{\psi}^k\nonumber\\
  &=& 0 
\end{eqnarray}
for all $\phi$ and $\psi$ and for all elements $\Omega^k$. Therefore,
every term in the sum $S_1$ vanishes and
\begin{equation}
  \label{eq:S1:solution}
  S_1 = 0.
\end{equation}

We now focus our attention on $S_2$. Recall from equation
\eqref{eq:def:J} that the fetch operator is
\begin{displaymath}
  F^k = b^k (b^{-1}F)^k,
\end{displaymath}
where $(b^{-1}F)^k$ is defined in equations
\eqref{eq:def:b:inverse:J:1} and \eqref{eq:def:b:inverse:J:2} such
that
\begin{eqnarray}
    (b^{-1}F)\psi_l^k&=& \psi^{k-1}_r\nonumber\\
    \text{and }(b^{-1}F)^k\psi^k_r&=&\psi^{k+1}_l.\nonumber
\end{eqnarray}
The fetch operator does the same thing, but selects only the boundary
term, thus making anything else $(b^{-1}F)^k$ does
irrelevant. Furthermore, recall from equation
\eqref{eq:bk:wk:properties} that the product $w^k b^k$ is symmetric
and has unit absolute value. So the sum $S_2$ becomes
\begin{equation}
  \label{eq:S2:soln}
  S_2 = \half\xi \sum_{k=1}^{K}\sqrbrace{\paren{\psi^k_r\phi^{k+1}_l-\psi^k_l\phi^{k-1}_r}+\paren{\phi^{k+1}_l\phi^k_r-\psi^{k-1}_r\phi^k_l}},
\end{equation}
which is a telescoping sum. The $k+1$ terms cancel with the $k-1$
terms, leaving
\begin{equation}
  \label{eq:S2:soln:2}
  S2 = \half\xi \sqrbrace{\paren{\psi^K_r\phi_R + \psi_R\phi^K_r} - \paren{\psi^1_l\phi_L + \psi_L\phi^1_l}}.
\end{equation}
We can interpret 
$$\paren{\psi^K_r\phi_R + \psi_R\phi^K_r}$$
and
$$\paren{\psi^1_l\phi_L + \psi_L\phi^1_l}$$
as twice the average of $\phi$ and $\psi$ evaluated on the right and
left boundaries of the domain respectively. Therefore their difference is a reasonable definition of the product $\phi\psi$ evaluated at the boundary of $\Omega$. And so 
\begin{equation}
  \label{eq:S2:soln:3}
  S2 = (\xi) \paren{\psi\phi}_{\partial\Omega}
\end{equation}
and equation \eqref{eq:sbp:expanded} is satisfied if and only if
$\xi = 1$. $\square$
\end{widetext}

\section{Stability for the Wave Equation}
\label{sec:stability:wave:equation}

Here we use summation-by-parts to demonstrate the stability an OLDG
discretization of the linear first-order-in-time,
second-order-in-space wave equation. This calculation provides an
example of how one demonstrates stability with summation-by-parts.

Consider the second-order-in-space wave equation
\eqref{eq:linear:wave:equation}
\begin{eqnarray}
  \partial_t \phi &=& \psi\nonumber\\
  \partial_t \psi &=& c^2 \partial_x^2\phi\nonumber
\end{eqnarray}
on the domain $\Omega$ subject to appropriate initial and boundary
conditions. Using the OLDG approach, this translates to the
semi-discrete system
\begin{eqnarray}
  \label{eq:discrete:wave:equation}
  \begin{aligned}
  \partial_t \phi^k &= \psi^k\\
  \partial_t \psi^k &= c^2 D^k \pi^k\\
  \pi^k &= D^k \phi^k,
  \end{aligned}
\end{eqnarray}
where $D^k$ is the element-wise wide derivative operator. The
continuum operator $\partial_t$ commutes with the discrete linear
operator $D^k$ so that
\begin{equation}
  \label{eq:discrete:wave:constraint}
  \partial_t\pi^k = D^k \partial_t\phi^k = D^k \psi^k,
\end{equation}
where we have used the equations of motion to remove the time
derivative. We do not use equation \eqref{eq:discrete:wave:constraint}
for evolution. Rather, we treat it as a \textit{constraint} which is
automatically satisfied.

This system admits the \textit{energy norm}
\begin{equation}
  \label{eq:def:discrete:energy}
  \MH= \half\sqrbrace{\braket{\psi,\psi}_{\Omega} + c^2\braket{\pi,\pi}_{\Omega}},
\end{equation}
which is manifestly positive-definite. To show that our discretization
is stable, we show that $\MH$ is non-increasing in time. We
differentiate equation \eqref{eq:def:discrete:energy} to find
\begin{eqnarray}
  \partial_t \MH &=& \braket{\psi,\partial_t\psi}_{\Omega} + c^2 \braket{\pi,\partial_t\pi}_{\Omega}\nonumber\\
  \label{eq:dH:intermediate}
  &=& c^2\braket{\psi,D \pi}_{\Omega} +  c^2\braket{\pi,D\psi}_{\Omega},
\end{eqnarray}
where we have used the discrete equations of motion
\eqref{eq:discrete:wave:equation} and the constraint
\eqref{eq:discrete:wave:constraint}. 

Finally, we integrate by parts to obtain
\begin{eqnarray}
  \partial_t\MH &=& c^2 \sqrbrace{\braket{D\psi,\pi}_{\Omega} - \braket{D\psi,\pi}_{\Omega}} + c^2\braket{\psi,\pi}_{\partial\Omega}\nonumber\\
  \label{eq:dH:is:zero}
  &=& c^2\braket{\psi,\pi}_{\partial\Omega},
\end{eqnarray}
where we have used equation
\eqref{eq:def:summation:by:parts:omega}. The value of this expression
depends on the boundary condition. For a large class of boundary
conditions, including periodicity,
homogeneous Dirichlet ($\psi = 0$) or von Neumann ($\pi = 0$),
or maximally dissipative boundary conditions, this term is either
zero or negative. 

In this case, we have 
\begin{equation}
  \label{eq:eq:dH:le:zero}
  \partial_t\MH \leq 0.
\end{equation}
Then, since $\MH \ge 0$ and $\MH$ is non-increasing, equation
\eqref{eq:discrete:wave:equation} provides a stable scheme. We note
that, although we perform our calculation for a second-order system,
it proceeds almost identically for a fully first-order system.

\section{Convergence for the Wave Equation}
\label{sec:convergence}

Here we present a calculation showing that the scheme described in
section \ref{sec:methods} is convergent for the linear wave
equation. 

\subsection{Strategy}
\label{sec:proof:strategy}

The strategy of our proof is as follows. We use our discontinuous
Galerkin scheme to solve the linear wave equation given arbitrary
initial conditions and compare to the analytic solution. We write both
the analytic solution and the ``numerical'' solution in terms of
element width $h$ and element order $p$ so that we can write the error
as a function of these two quantities. (As usual, we assume that all
elements are the same width and order.)

We use Wolfram Mathematica \cite{Mathematica} to \textit{symbolically}
carry out the OLDG differentiation and Runge-Kutta integration, as
described in section \ref{sec:methods}. In this way, our initial
conditions can be truly arbitrary, and we only need to provide it in
terms of a finite number of arbitrary constants. We have made our
Mathematica code public and placed it in an online repository, where
it can be examined \cite{dgfeSupp}.

\subsection{The Continuum Problem}
\label{sec:proof:continuum}

Consider the one-dimensional domain $\Omega = \R$ and the interval
\begin{equation}
  \label{eq:def:tau}
  \mT = \sqrbrace{0,T}\text{ for some }T>0\in\R.
\end{equation}
We seek functions $\psi(t,x)$ and $\pi(t,x)$ which satisfy the linear
wave equation in its first order in time reduction
\eqref{eq:linear:wave:equation},
\begin{eqnarray}
  \pd{\phi}{t} &=& \psi\nonumber\\
  \pd{\psi}{t} &=& c^2 \partial^2_x\phi\nonumber
\end{eqnarray}
for all $x\in\Omega$ and all $t\in\mT$. Without loss of generality, we
assume that $c=1$.

If $\phi$ is analytic in $x$, then at any time $t$ it can be
well-approximated by a power series
\begin{equation}
  \label{eq:phi:power:series}
  \phi(t=0,x) = a_0 + \sum_{i=1}^N (a_i+b_i) x^i,
\end{equation}
where $a_i$ and $b_i$, $0\leq i\leq N$ are arbitrary constants that
determine the initial profile. Of course, the solution to the wave
equation given this initial condition is known. On the real line,
$\psi$ is a superposition of right- and left-travelling waves that
advect in each direction with speed $c$:
\begin{equation}
  \label{eq:phi:power:analytic}
  \phi(t,x) = a_0 + \sum_{i=1}^N a_i (x+t)^i + \sum_{i=1}^N b_i (x-t)^i
\end{equation}
with time-derivative
\begin{equation}
  \label{eq:psi:power:analytic}
  \psi(t,x) = \sum_{i=1}^N i a_i (x+t)^{i-1} + \sum_{i=1}^N i b_i (x-t)^{i-1}.
\end{equation}
Therefore our initial condition for $\psi$ is given by
\begin{equation}
  \label{eq:psi:power:0}
  \psi(t=0,x) = \sum_{i=1}^{N} i (a_i + b_i) x^{i-1}.
\end{equation}
Note that the initial condition for $\psi$ contains one fewer modes
than the initial condition for $\phi$. Enforcing this at each time
step is equivalent to applying the truncation procedure discussed in
section \ref{sec:truncation}.

\subsection{The Discrete Setup}
\label{sec:proof:setup}

To test the OLDG method developed in section \ref{sec:methods}, we use
it to calculate a numerical approximation to the solution given by
equations \eqref{eq:phi:power:analytic} and
\eqref{eq:psi:power:analytic} with equations
\eqref{eq:phi:power:series} and \eqref{eq:psi:power:0} as initial
conditions. Crucially, we do not want to specify $a_i$ and
$b_i$. Rather we want our solution in terms of them.

For simplicity we break $\Omega$ into a uniform ``grid'' of elements,
all of the same width $h$ and order $p \leq N$. For initial data that
is truly arbitrary, initial conditions \eqref{eq:phi:power:series} and
\eqref{eq:psi:power:0} are accurate up to order $h^N$ and $h^{N-1}$
respectively. To simulate a realistic situation, where the initial
conditions introduce error equivalent to the order of the
discretization scheme, we set $N=p$.

In principle, we have an infinite number of elements since our domain
is the real line. In practice, however, we can examine a finite number
$K$ of elements, spanning some interval $\mI=[-A,A]\subset\Omega$ as
long as that number is sufficiently large so that no information from
elements near the boundary of $\mI$ has time to propagate to elements
near the centre of $\mI$ in time $T$.

We then write the positions of the nodes within elements, $x^k_i$ as a
multiple of the element width $h$, which can be calculated by finding
how many elements away from the origin the element $\Omega^k$ is and
the ``local'' coordinates of $x^k_i$ within $\Omega^k$. We also define
$\phi^k_i$ and $\psi^k_i$ as the restrictions of the fields $\phi$ and
$\psi$ onto the nodes within elements, $x^k_i$.

\subsection{Comparing to the Continuum Solution}
\label{sec:proof:comparison}

Once we define our fields, we integrate them using a $p^{th}$-order
explicit Runge-Kutta scheme and compare to the analytic solution. We
perform this calculation for both second- and fourth-order stencils
and the results seem to be generic. We use an explicit integrator of
the same order as the OLDG stencil we wish to test. Our implementation
of a second-order Runge-Kutta, for example, integrator is given by the
following code.

\lstset{language=Mathematica}
\lstset{basicstyle={\sffamily\footnotesize},
  breaklines=true,
  captionpos={t},
  frame={lines},
  rulecolor=\color{black},
  framerule=0.5pt,
  columns=flexible,
  tabsize=2
}
\begin{lstlisting}
(* The CFL Factor *)
dt = cfl * h; (* factor is arbitrary *)

RK2::usage = "Integrate y via RK2. The state vector is y."
RK2[y_,t_,f_] := Module[{k1,k2,yNew,tNew},
			k1=f[t,y];
			k2=f[t+(2/3)*dt,y+(2/3)*dt*k1];
			yNew = y + dt*((1/4)*k1+(3/4)*k2);
			tNew=t+dt;
			{tNew,yNew}];
\end{lstlisting}

Since the initial conditions are arbitrary, we only need to integrate
by one time step. After integration, we subtract the true solution,
given by equations \eqref{eq:phi:power:analytic} and
\eqref{eq:psi:power:analytic}, from the integrated solution and
calculate the error. Because the wave equation is homogeneous, it is
sufficient to study an element in the center of $\mI$. For
second-order elements (for example), this error is of the form
\begin{eqnarray}
  \label{eq:phi:error:1}
  \phi^k_0 &=& -h^2 \alpha (a_2 - b_2)\\
  \label{eq:phi:error:2}
  \phi^k_1 &=& 0\\
  \phi^k_2 &=& - \phi^k_0
\end{eqnarray}
for $\phi$ and
\begin{eqnarray}
  \label{eq:psi:error:1}
  \psi^k_0 &=& -h\alpha (a_2 - b_2)\\
  \label{eq:psi:error:2}
  \psi^k_1 &=& 0\\
  \psi^k_2 &=& - \psi^k_0
\end{eqnarray} 
for $\psi$, where $\alpha$ is the Courant-Friedrichs-Lewy factor in
this context. For a second-order element, initial conditions
\eqref{eq:phi:power:series} and \eqref{eq:psi:power:0} also have
errors of leading order $h^2$ and $h^1$ respectively, so convergence
is retained. More generally, we find that the error in $\phi$ is of
order $\Ord{h^p}$ and that the error in $\psi$ is of order
$\Ord{h^{p-1}}$.

\section{Making Contact with Standard Discontinuous Galerkin Methods}
\label{sec:lax-friedrich}

%In this section we provide two examples of how our wide derivative
%operator relates to more traditional DGFE formulations. These
%calculations also provides a simple example of how OLDG methods can be
%used to discretize more complicated systems. In appendix
%\ref{sec:lax:friedrich:advection}, we show how our wide derivative
%operator can, for the linear advection equation, be used to recover a
%more standard DGFE formulation. 
%
%In appendix \ref{sec:lax:friedrich:burgers}, we discuss the analogous
%discretization of Burgers' equation. In the case of Burger's equation
%(and all nonlinear systems), there is an ambiguity based on whether or
%not we write the system in manifestly conservative form. In the
%manifestly flux-conservative case, we recover a traditional DGFE
%method with a non-ideal flux. If the system is not written in a
%flux-conservative way, we develop several different schemes.

In this section we provide an example of how our wide derivative
operator relates to more traditional DGFE formulations. This
calculation also provides a simple example of how OLDG methods can be
used to discretize more complicated systems.

% \subsection{The Linear Advection Equation}
% \label{sec:lax:friedrich:advection}

Consider the linear wave equation in first-order, flux-conservative,
form
\begin{equation}
  \label{eq:linear:wave}
  \partial_t \phi + \partial_x f(\phi) = 0,
\end{equation}
where we have introduced the complex variable
\begin{equation}
  \label{eq:wave:variables}
  \phi = \psi + i \pi
\end{equation}
and flux
\begin{equation}
  \label{eq:wave:flux}
  f(\phi) = \paren{\pi + i c^2 \psi}
\end{equation}
for some continuum functions 
$$\psi(t,x) : [0,T] \to \R$$ 
and 
$$\pi(t,x) : [0,T] \to \R$$ 
and constants $T > 0 \in\R$ and $c\in\R$ subject to appropriate
initial and boundary conditions. For simplicity assume the domain is
an interval $\Omega = [X_l,X_r]$.

With the usual choice of Legendre basis functions, this translates to
the semi-discrete system
\begin{displaymath}
  0 = \partial_t \phi^k_i +  D^k \sqrbrace{f(\phi)}^k_i
\end{displaymath}
for all elements $\Omega^k$ and all $i=0,\ldots,P^k$, where $P^k$ is
the order of the element. If we expand the wide derivative operator
$D^k$, we obtain
\begin{equation}
  \label{eq:wave:semi-discrete:1}
  0 = \partial_t \phi^k_i+ d^k f^k_i - \frac{1}{2}b^k\sqrbrace{1 - (b^{-1}F)^k}f^k_i,
\end{equation}
where we have now suppressed the dependence of $f$ on $\phi$.

To obtain the usual representation of a DGFE method,
we must take the inner product with respect to a test function
$\Phi^k_j(x)$, which is its own interpolant. Recall from equation
\eqref{eq:def:vandermonde} that 
$$\V^k_{ij} = \Phi^k_j(x^k_i)$$
and from equation \eqref{eq:nodal:representation:lagrange:poly} that
$$\phi^k(t,x) = \sum_{i=0}^N \phi^k_i l^k_i(x),$$
where the $l^k_i(x)$ are Lagrange interpolants. Finally recall that
derivatives $d^k$ of polynomials of order $P^k$ and lower are exact,
since this is how the narrow derivative operator is defined. We thus
have
\begin{widetext}
\begin{eqnarray}
  \label{eq:advection:inner:product}
   0 &=& \braket{\Phi^k_j,\partial_t \phi^k}_{\Omega^k}
   + \braket{\Phi^k_j,d^kf^k}_{\Omega^k}
   -\frac{1}{2} \braket{\Phi^k_j,b^k\sqrbrace{1-(b^{-1}F)^k}f^k}_{\Omega^k}\nonumber\\
   &=& \sqrbrace{\sum_{i=0}^{P^k}\braket{\Phi^k_j(x),l^k_i(x)}_{\Omega^k}\partial_t \phi^k_i} 
   + \sqrbrace{\sum_{i=0}^{P^k}\braket{\Phi^k_j,\partial_x l^k_i(x)}_{\Omega^k}f_i^k} 
   - \frac{1}{2}\braket{\Phi_j^k,b^k \paren{1-(b^{-1}F)f^k}}_{\Omega^k}\nonumber\\
   &=& \Ma^k \partial_t \myvec{\phi}^k 
   + \St^k \myvec{f}^k
   -  \frac{1}{2}\braket{\myvec{\Phi}^k,b^k \paren{1-(b^{-1}F)f^k}}_{\Omega^k},
\end{eqnarray}
\end{widetext}
where 
\begin{equation}
  \label{eq:mass:stiffness}
  \Ma_{ij}^k = \braket{\Phi^k_j,l^k_i}\text{ and }\St_{ij}^k = \braket{\Phi^k_j \partial_x l^k_i}
\end{equation}
are the \textit{mass} and \textit{stiffness} matrices from standard
discontinuous Gaklerkin methods and where we have suppressed the index
notation to recover a matrix form within each element. 

The last term in equation \eqref{eq:advection:inner:product} still
requires some massaging, however. Recall from equation
\eqref{eq:bk:wk:properties} that $w^k b^k$ is nonzero only on the
boundary and it is always $\pm 1$.  Then we can do away with the
integral over the boundary and recover
\begin{displaymath}
  \Ma^k \partial_t \myvec{\phi}^k + \St (a\myvec{\phi}^k) = \frac{1}{2}\myvec{\Phi}^k\paren{ f^k_- - f^k_+}\eval_{x_l^k}^{x_r^k}
\end{displaymath}
or
\begin{equation}
    \label{eq:advection:2}
  \Ma^k \partial_t \myvec{\phi}^k + \St (a\myvec{\phi}^k) = \myvec{\Phi}^k\sqrbrace{ f^k_- - \half\paren{f^k_- + f^k_+}}\eval_{x_l^k}^{x_r^k}
\end{equation}
where $f_-^k$ and $f^k_+$ are the interior and exterior values of the
flux on an element respectively. In other words $f_-^k$ returns
values on the boundary within an element and $f_+^k$ returns
values on the boundary from neighbouring elements.

We recognize equation \eqref{eq:advection:2} as a standard DGFE 
method in strong form with a simple central numerical flux 
\begin{equation}
  \label{eq:def:central:flux}
  f^*(\phi^k_-,\phi^k_+) = \frac{1}{2} C (\phi^k_++\phi^k_-),
\end{equation}
for the interior and exterior values of $\phi$ at the boundary of
$\Omega^k$. Therefore, in the simplest cases at
least, our scheme matches traditional DGFE methods.\\

\section{Calculating Self-Convergence}
\label{section:self:convergence:derivation}

Here we derive the test for self convergence
\eqref{eq:eq:phi:self:convergence:test} given in section
\ref{sec:tests:gamma:driver}. We follow a procedure first
proposed by Richardson \cite{richardson1911approximate}. Suppose we
are evolving the BSSN equations. Based on equation
\eqref{eq:wavetoy:convergence:formula}, suppose that the error in the
$xx$-component of the metric is of the form:
\begin{equation}
  \label{eq:phi:self:convergence:base}
  (\gamma_{xx})_{i} = \gamma_{xx} + \ME h_i^{P},
\end{equation}
where $i$ indexes three resolutions, such $i=1$ is the coarsest and
$i=3$ is the finest. $\gamma_{xx}$ is the true solution,
$\ME$ is an ``error'' function, and $h$ and $P$ are the element
width and order as usual. For notational simplicity, we assume all
elements have the same width and order and we therefore suppress the
element index $k$. We also suppress dependence on $x$ and $t$.

We now combine formula \eqref{eq:phi:self:convergence:base} for
different values of $i$:
\begin{eqnarray}
  (\gamma_{xx})_1 - (\gamma_{xx})_3 &=& \gamma_{xx} + \ME h_1^P - \gamma_{xx}-\ME h_3^P\nonumber\\
  \label{eq:phi:self:convergence:1}
  &=& \ME \paren{h_1^P - h_3^P}
\end{eqnarray}
and similarly
\begin{equation}
  \label{eq:phi:self:convergence:2}
  (\gamma_{xx})_2 - (\gamma_{xx})_3 = \ME\paren{h_2^P - h_3^P}.
\end{equation}
If we combine equations \eqref{eq:phi:self:convergence:1} and
\eqref{eq:phi:self:convergence:2}, we find equation
\eqref{eq:eq:phi:self:convergence:test}:
\begin{displaymath}
  \frac{(\gamma_{xx})_1 - (\gamma_{xx})_3}{h_1^P - h_3^P} = \frac{(\gamma_{xx})_2-(\gamma_{xx})_3}{h_2^P-h_3^P},
\end{displaymath}
so we can check for self convergence by constructing the left- and
right-hand-sides of equation \eqref{eq:eq:phi:self:convergence:test} and
comparing them. Self-convergence is a weaker statement than
convergence, since it does not guarantee that a numerical solution
converges to the \textit{true} solution. It could, in principle,
converge to something else.

%bibliography
\bibliography{dgfe-for-bssn}
\bibliographystyle{hunsrtnat}

\end{document}